\documentclass[aps,pre,twocolumn]{revtex4}
\usepackage{amssymb,amsfonts,amsmath,bm}
\usepackage[dvipdf]{color}
\usepackage{graphicx}

\usepackage{subfigure}
 \usepackage[T2A]{fontenc}
\usepackage[cp1251]{inputenc}
\usepackage{amssymb, latexsym, amsmath}

 \usepackage{graphicx}

\usepackage{color}
 \allowdisplaybreaks[1]

\begin{document}

\renewcommand{\baselinestretch}{1.2}\normalsize

\title{\large \bf CIRCULATING  MARANGONI  FLOWS  WITHIN  DROPLETS IN   SMECTIC FILMS}

\author{ E.S. Pikina$^{1,2,6}$, M.A. Shishkin$^{2,4}$, K.S. Kolegov$^{2,5}$,  B.I. Ostrovskii$^{2,3}$, and S.A. Pikin$^{3}$. }

\affiliation{$^1$ Landau Institute for Theoretical Physics of the RAS,
142432, Chernogolovka, Moscow region, Russia \\
$^2$ Institute of Solid State Physics of the RAS, 142432 Chernogolovka, Russia \\
$^3$ FSRC "Crystallography and Photonics" of the RAS,
 119333 Moscow, Russia\\
$^4$ NRU Higher School of Economics,
101000,  Moscow, Russia,  \\
$^5$ Astrakhan State University,
 414056 Astrakhan,  Russia\\
$^6$ Oil and
Gas Research Institute of the  RAS,  119333 Moscow, Russia}


\begin{abstract}

We present theoretical study and numerical simulation of Marangoni convection within ellipsoidal  isotropic  droplets embedded in free standing smectic films (FSSF).   The thermocapillary flows are analyzed  for  both isotropic droplets spontaneously formed in FSSF  overheated above the bulk smectic-isotropic transition, and oil lenses deposited on the surface of the smectic film. The realistic model, for which  the upper drop interface is free from the smectic layers, while at the lower drop surface the smectic layering still persists is considered in detail. For  isotropic droplets and oil lenses this leads effectively to a sticking of fluid motion at the border with a smectic shell. The above  mentioned asymmetric configuration is realized experimentally when the temperature of the upper side of the film is higher than at the lower one.
The full set of stationary  solutions for Stokes stream functions describing the  Marangoni convection flows within the ellipsoidal drops were derived analytically.
The temperature distribution in the ellipsoidal drop and the surrounding air was determined in the frames of the perturbation theory.
As a result the analytical solutions for the stationary thermocapillary convection were  derived for different droplet ellipticity ratios and  the heat conductivity of the liquid crystal and air.
 In parallel, the numerical hydrodynamic calculations of the thermocapillary motion in the drops were performed. Both the analytical and numerical simulations predict the axially-symmetric circulatory convection motion determined by the Marangoni effect at the droplet free surface. Due to a curvature of the drop interface a temperature gradient along its free surface always persists. Thus, the thermocapillary convection within the ellipsoidal droplets in  overheated  FSSF is possible for the arbitrarily small Marangoni numbers. The possible experimental observations enabling to check our predictions are proposed.

\end{abstract}

\maketitle


\section{ Introduction}
\label{intro}

A fluid flow within a drop caused by the temperature-dependent surface tension is called the Marangoni convection and was first observed in its classical form by Benard in a process of formation of the characteristic hexagonal convection patterns in flat fluid films \cite{Benard}.
 The onset of thermocapillary convection is determined by a dimensionless Marangoni number, Ma, expressing the ratio of surface tension to viscous forces, which has to reach a certain minimum critical value for instability to occur. The arising convective cells are characterized by a unique critical wavenumber $k_c$, which determines the scale of the nonuniformity in the plane of the film. In general, the formation of different cellular flow regimes, including rolls, hexagonal patterns, hydrothermal waves, etc. have been reported for fluid films of different size and geometry \cite{Koschmieder1974,Davis1987,Koschmieder1992,vanHook1997,Alexeev2005}. An extensive literature on thermocapillary driven flows in fluid films exists and both experimental and theoretical studies are thoroughly reviewed; see for example \cite{Levich1962,Gershuni1972,Getling1991,Roisman2015}.
   The Marangoni phenomenon is important not only for development of fundamental physics of capillarity and wetting; it is frequently encountered in industrial applications, including chemical engineering, food and cosmetic processing, thermal management of micro-fluidic and electronic devices, evaporation related technology \cite{Alexeev2005,Roisman2015,Kolegov2020,Kolegov2021}. The role of Marangoni convection is especially important in thermal processing of electronic and rheological devices    in microgravity conditions where buoyancy effects are negligible
\cite{Orozco,Melnikov2015,Yano2018}.

The past two decades were marked by significant advancements in experimental and theoretical studies of the symmetry and dynamics of Marangoni cellular flows in fluid films of various confinements \cite{Alexeev2005,Roisman2015,Kolegov2020,Nakamura2020,Yoshioka22,Bestehorn2003,Wang2002}.
However, above mentioned  progress largely concerned the Marangoni convection in the systems with a variable flat geometry. In spite of its practical and theoretical significances, thermocapillarity at
curved fluid interfaces has not catched proper attention due to its complexity.
 The convection inside a droplet of a spherical shape \cite{Sasmal1994,Hu2005,Hu2006,Girard2006,Tam2009,Barash2009,Ristenpart2007,Kita2016}
 appears to be  principally different from the conventional Marangoni flows in the plane films.
  This is due to the inhomogeneity, which is imposed on the interface temperature by the curved shape of the drop. Moreover, the spherical geometry of the drop modifies flow patterns, thus affecting the heat and mass transport within the fluid. The thermocapillary flow within the spherical droplet is usually considered as a concomitant process in respect of the main physical phenomena occurring in it; a clear example is the evaporation of a  sessile liquid drop with a pinning contact line in an ambient air \cite{Sasmal1994,Hu2005,Hu2006,Girard2006,Tam2009,Barash2009,Ristenpart2007,Kita2016}. In such a drop the fluid may undergo either outward movement produced by evaporation-driven flow or circulatory motion related to the Marangoni effect. One of the few works that carefully analyzes the effects of Marangoni flows in evaporating sessile drops with the spherical interface is the classical study by Hu and Larson \cite{Hu2005,Hu2006}. In their papers, the authors model convection in a flattened droplet on a partially wetting substrate using both a lubrication analysis and a finite element model (FEM). They found that convective axially-symmetric circulatory motions are occurred driven by a non-uniform temperature distribution at the surface of the droplet which arises from evaporative cooling. In the paper by Tam et al. \cite{Tam2009} a small droplet of water sitting on top of a heated superhydrophobic surface was considered. Similarly to Hu and Larson, an axially-symmetric (toroidal) convection patterns were observed in a spherical drop in which fluid raised along its surface and accelerated downwards in the interior towards the liquid/solid contact point. The internal flow in the drop arises due to the presence of a vertical temperature gradient; this leads to a gradient of  surface tension, which in turn drives fluid away from the contact point along the drop interface. In their work, the authors developed an analytical solution to thermocapillary driven circulatory flow in terms of the Stokes stream functions, which provided a good quantitative agreement between analytical and experimental results.
The affect of Marangoni forces on the evaporation dynamics of the sessile drops was  studied in the theoretical paper by Barash et al. \cite{Barash2009}. The authors identified various dynamic stages of the thermocapillary convection associated with the generation of the array of convective vortices near the surface of a drop and their transformation with time into the single convection vortex.

While the fundamentals of Marangoni convection are well established in the systems with a simple flat geometry, the  analytical description of the thermocapillary flow in fluid drops of ellipsoidal shape is yet not available. In this work we undertake a step in this direction presenting a quantitative description enabling us to account for all  relevant aspects of the Marangoni flows in ellipsoidal droplets, namely, the analytical stationary  and  critical solutions for the Stokes stream functions, the spatial temperature and the velocity distributions for initial stage of the convection. The shape and the axial symmetry of the fluid droplets possessing two spherical interfaces were approximated by an oblate spheroid. Accordingly,  the elliptical coordinate systems were chosen  for the analytical derivations.  The calculations were carried out according to the following scheme:
(i) we derived the equations yielding to approximate the shape of isotropic droplets and lens-like oil inclusions in FSSF  by oblate spheroids. (ii) We   wrote the stationary system of basic equations describing the Marangoni convection in drops in Boussinesq approximation using the ellipsoidal coordinate system. (iii) We solved the  equation describing the thermocapillary motion and derived explicitly the expressions for 2D Stokes stream functions and velocity field corresponding to the convection within the drops. (iy) We solved the thermal conduction equation and find the spatial temperature distribution in the drop using the linear perturbation theory. (y)   We derived and solved the Marangoni boundary condition  and determined the general shape of the   stationary thermoconvection motion as a function of the droplet ellipticity ratio. (yi) Finally, the crossover to the limit of a flat fluid layer was analyzed.

Additionally, the numerical hydrodynamic experiment that models the thermocapillary motion in the ellipsoidal drops was conducted. Both the analytical derivations and numerical simulations predict the axially-symmetric circulatory convection motion within the droplet determined by the Marangoni effect at the droplet free surface. The convection patterns represent either the individual toroidal-like vortices or the series of vortices  distributed within the plane of the drop.

\begin{figure}
    \hskip0.5true cm
   \includegraphics[width=0.9\linewidth]{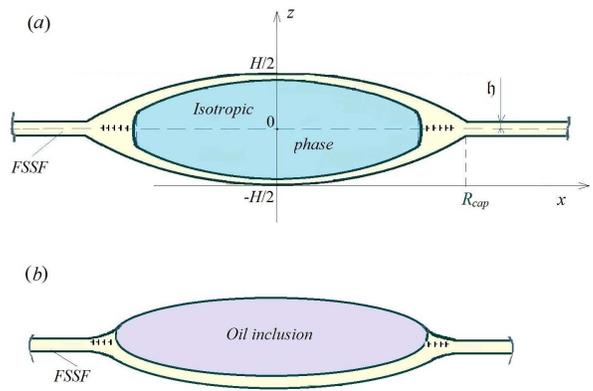}
    \caption{Schematic view  of fluid droplets in free standing smectic films (FSSF): (a) isotropic drops formed in overheated FSSF.   The drop is connected with the FSSF of uniform thickness by a meniscus. The drop has a lens-like shape and is symmetric relative to the horizontal plane. The height and the base radius of the drop are designated as $H$, $R_{cap}$, respectively; (b) oil lenses deposited on the surface of FSSF. In all cases the film thickness $\mathfrak{h}$ is much smaller than the droplet height.      \label{Figure1}}
    \end{figure}

Although  the developed approach is quite general and thus applies to a wide variety of the Marangoni convection problems in ellipsoidal fluid droplets and bubbles, we focus here on two specific cases. First, we consider isotropic droplets spontaneously generated in free standing smectic films (FSSF) heated above the temperature of the bulk smectic-isotropic transition Fig. \ref{Figure1}a.
As a second case we consider the droplets of insoluble fluids (of the type of oil or glycerol) which can be deposited on  overheated FSSF in various ways  \cite{Stannarius08,Dolle2014,Qi16}. For example, the oil vapor can condense at one of the sides of the smectic film thus forming the lens-like oil drops with the lateral diameter of the order of mm \cite{Dolle2014,Qi16}, Fig. \ref{Figure1}b. The FSSF are usually made from the smectic A (Sm-A) and smectic C (Sm-C) liquid crystal materials. The Sm-A phase  consists of a stack of parallel  molecular layers, in which elongated molecules are oriented on average along the layer normal, and exhibit the short-range positional order within the layers. The Sm-C phase differs from the Sm-A phase by a tilt of the long molecular axes with respect to the layer normal. Being stretched on a frame, these materials, due to their layered structure, form free-standing films \cite{Lucht98,Oswald06,Ostrovskii03}  in which the smectic layers align parallel to the two air-film interfaces. The film is attached to its frame via a meniscus, which serves as a reservoir with which the film can exchange matter. The FSSF can also be prepared as bubbles, either connected with an inflation tube or floating freely under microgravity conditions  \cite{Stannarius98,Clark2017,Klopp19}.
In liquid crystals (and thus in smectic films) a free surface usually stabilizes a higher ordered, -  less symmetric,  phases. Due to this the FSSF can in many cases be heated above the bulk smectic disordering temperature without rupturing, and instead show a tendency for the spontaneous nucleation and growth of the isotropic droplets \cite{Schuring02,Dolganovi2019,Pikina2020}. The isotropic droplets have the shape of spherical segments (circular flat lenses), the height of which (of the order up to tens of microns) is about one order of magnitude less than the drop lateral dimension, Fig. \ref{Figure1}a.  Because the thickness of the FSSF is about few molecular layers (approximately 100 nanometers) the height of  isotropic droplet is much larger than the film thickness. Thus, such oblate droplets can be considered as a three-dimensional (3D) fluid objects embedded into quasi-2D smectic film which serves as a frame (substrate) for them.

The occurrence of the thermo-capillary driven macroscopic material transport has been earlier reported in FSSF of certain materials \cite{Godfrey96,Birnstock01,Trittel2019,Stannarius2019}. The linear temperature gradient in  these experiments was applied in the plane of the film, i.e. in the plane of the smectic layers which have a fluid nature. The application of the temperature gradient in the direction along the layer normal in FSSF showing the solid-like elastic response, has not been considered for realization of the Marangoni transport due to a weak permeation in smectic where the molecules are unable to flow through the smectic layers  \cite{Helfrich1969,Lebedev1993,Kleman03}. In our preceding paper \cite{Pikina2021} we have analyzed the possibility of thermocapillary convection within isotropic droplets spontaneously formed in FSSF. The horizontal smectic film with  isotropic droplets formed in it was expected to be heated either from below or from the top, thus creating the vertical temperature gradient  along the FSSF normal. Marangoni forces associated with the temperature dependence of the surface tension induce fluid convection within the isotropic drop. To calculate the Marangoni number for a fluid drop a formal similarity between a drop of a height $H$ and a flat layer of the same thickness was exploited. The validity of this approximation was justified by a small aspect ratio of lens-like isotropic droplets in FSSF. It was shown that along  the lateral drop size about six convection cells (rolls) can be formed. However, the real shape of the drop interface was not taken into account, independently of the fact that a curvature of the drop interface necessarily imposes a temperature gradient along its free surface. This should affect the thermoconvection patterns and mass transport within a drop.

There is another aspect of the Marangoni convection within the ellipsoidal drops which applies to isotropic droplets formed in overheated FSSF. It is well documented that the surface of the isotropic samples heated above the bulk smectic-isotropic temperature is covered by the smectic layers, the amount of which depends on the degree of overheating  \cite{Ocko86,AlsNielsen86,Lucht98}. The similar situation occurs at the interfaces of isotropic droplets in smectic films. Actually, each droplet is connected with the FSSF of uniform thickness via a meniscus the profile and height of which are determined by the set of edge dislocation loops located in its mid-plane \cite{Picano01,Pikina15}, Fig. \ref{Figure1}a.
 In general, a smectic shell covering the drop interface should hinder the development of the Marangoni instability within
  fluid drop - so called sticking effect. This is a delicate moment and is  discussed in more details below, in Sec. II B. This is especially important for isotropic droplets  formed in overheated FSSF where we deal with the smectic layering at both drop interfaces, Fig \ref{Figure1}a.  Concerning the lens-like oil drops deposited on  FSSF, their upper interface is connected directly with the air, Fig. \ref{Figure1}b. Thus, there are no restrictions for development of the Marangoni instability in the oil drops initiated by the surface tension temperature variations at the upper drop interface.

The situation with the asymmetric boundary conditions typical for the oil lenses deposited on FSSF can be also created for the isotropic drops spontaneously  formed in FSSF. To realize this experimentally,  the temperature of the upper side of the drop should be higher than at the lower one. In this case the defects of the layered surface structure,  holes, are preferably formed at the upper (hot) surface of the drop, initiating the growing of the dislocation loops in the direction of the meniscus connecting the droplet with the FSSF. Contrary to this, on the lower (cold) side of the drop the formation of additional layers (islands) is energetically favorable, thus increasing the thickness of the smectic shell (for details see Sec. II B below). As a result, at the upper drop interface the amount of smectic layers continuously diminishes and the surface becomes free, while at the lower drop surface the smectic layering leads effectively to a sticking of isotropic fluid.
This asymmetric configuration is similar to that occurring in oil lenses deposited on FSSF.

The paper is organized as follows. In Sec. II we present  the quantitative description of the  equilibrium shape of the ellipsoidal drops under study. In particular, in Sec. II A we introduce the oblate spheroid coordinates which are systematically used through  all further derivations. The Sec. II B is devoted to the description of the thermal stability of the smectic shell covering isotropic droplets and to the analysis of possibility of thermocapillary motion on the boundary fluid-smectic. The main focus of the remaining sections is a theoretical description of Marangoni flows within  the ellipsoidal drop. In Sec. III we present the basic equations and formulate the  boundary conditions for the ellipsoidal droplets.
 Section IV contains main analytical results for Stokes stream functions describing thermocapillary  flows in ellipsoidal drops in FSSF with asymmetric boundary conditions. The subsections are devoted to calculations of the velocity fields and spatial temperature distributions. Sec. IV E presents the results of numerical simulations of the thermocapillary motion in the drops  in the frame of numerical experiment. Finally, Sec. V gives a concluding discussion. The details of the developed analytical and numerical approaches are described in Appendices A-G and in Supporting Information.

\section{STATEMENT OF THE PROBLEM AND BASIC EQUATIONS}
\label{sec:1}

\centerline{\bf A. Shape of isotropic droplet in FSSF}

\smallskip

Let us discuss first the shape of the isotropic droplets spontaneously formed in overheated FSSF. As it was indicated earlier, these droplets have the shape of spherical segments (caps), Figs. \ref{Figure1}a, \ref{Figure2} \cite{Schuring02,Stannarius08,Clark2017,Pikina2020}.
Due to prolate shape of the drop the inequality $H\ll\,R_{cap}\,$ is usually holds (compare with the designations shown in Fig. \ref{Figure1}) \cite{crw}.
   The drop is connected with the FSSF via a meniscus the shape of which is determined by its dislocation structure.    Initially   the surface of the droplet is covered by a certain amount of the smectic layers. The parameters of the spherical segments of the drop are determined from the condition of minimum of its surface energy under assumption that the volume of the droplet is fixed \cite{Schuring02}. The minimization is usually made by a Lagrange undetermined multipliers method \cite{Schuring02,Pikina2020}, and provides the following relation between the base radius of the cap, $R_{cap}$, and the half -height of the drop, $(H/2)$ \cite{crw}:
  \begin{eqnarray}
   \frac{H}{2}  \approx  \sqrt{\mathfrak{h}^2 + \frac{\gamma - \gamma_{\hbox{\tiny A}} }{\gamma + \gamma_{\hbox{\tiny A}} }\,R_{cap}^2} - \mathfrak{h}
\approx  \sqrt{ \frac{\gamma - \gamma_{\hbox{\tiny A}} }{\gamma + \gamma_{\hbox{\tiny A}} } }\, R_{cap}\, , \
   \label{MinS}
  \end{eqnarray}
where $\mathfrak{h}$ is a half of the film thickness, $\,\mathfrak{h}\,\ll\,H, \,R_{cap}$ and $\gamma$ and $\gamma_{\hbox{\tiny A}}$ are the interfacial tensions between the drop-air and the FSSF-air interfaces, respectively. In accordance with the values of the interfacial tensions the following inequality holds: $H/(2 R_{cap})\ll\,1$. The validity of  Eq. (\ref{MinS}) is confirmed by numerous experimental observations carried out for different smectic materials \cite{Schuring02,Stannarius08,Clark2017,Dolganovi2019}.

\begin{figure}
    \hskip-0.1true cm
   \includegraphics[width=0.8\linewidth]{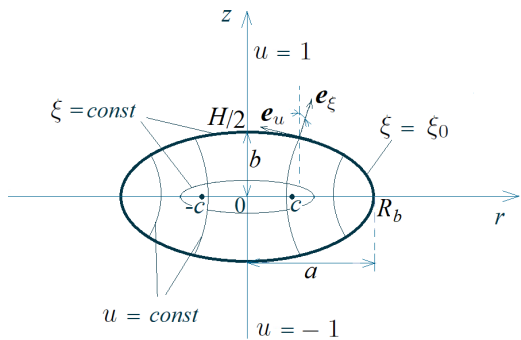}
    \caption{Representation of the frontal profile  of the lens-like isotropic droplet in the approximation of oblate spheroid.   The parameters characterizing the ellipsoidal drop shape  are designated in Fig. \ref{Figure1}.  For convenience, we fixed the zero of the $z$-coordinate axes in the center of the drop (in the middle plane of the FSSF). The parameters: $a, b$ - are the semiaxes of  ellipsoid, $a=c\sqrt{1+\xi_0^2}\equiv R_b$, $b=c\,\xi_0\equiv H/2$, $\xi_0/\sqrt{1+\xi_0^2}=b/a\ll\,1$ (the latter inequality is possible only for $\xi_0 \ll 1$), $c$  is a focus distance (coordinate of the focal point). Here ${\bf e}_{\xi}$ ,   ${\bf e}_{u}$  are the unit vectors in the oblate spheroidal coordinates in their meridional plane. Note that ${\bf e}_{\xi}$  is outward normal vector to the oblate spheroidal surface of constant ${\xi}=\xi_0$,  unit vector ${\bf e}_{\varphi}$ is the  azimuthal unit vector, oriented beyond the page (sheet) plane, ${\bf e}_{u}$ lies in the tangent plane to the oblate spheroid surface and completes the right-handed basis set $\{{\bf e}_{u}, {\bf e}_{\xi}, {\bf e}_{\varphi}\}$.
      \label{Figure02}}
    \end{figure}

In that follows we replace the shape of isotropic droplet in the form of two spherical segments by an ellipsoid (oblate spheroid) characterized by the semiaxes ratio
 $b/a\ll\,1$ (where $b$ and $a$ are a small and a large semiaxis of ellipsoid, respectively), Fig. \ref{Figure02}.
   In doing so we set the $b$ value equal to   $H/2$, while the large semiaxis of the ellipsoid  attains a value
   $a=R_b=H\,\sqrt{1+\xi_0^2}/(2 \xi_0)$.

The validity of the above approximation can be justified by equating the volume of  isotropic drop  in the form of two spherical segments with that of oblate spheroid. Indeed, the volume of an oblate spheroid drop constitutes
\begin{eqnarray}
V_{el}=(4\pi/3)\,a^2\,b\,=(4\pi/3)\,R_b^2(H/2)\, . \ \ \ \label{Vel}
\end{eqnarray}

On another hand the volume of isotropic drop  in the form of  two spherical segments can be written as
 \begin{eqnarray}
V_{cap}=\,\pi\,(H/2)\,(R_{cap}^2+H^2/12)\,+\,\pi\,2\,\mathfrak{h}\,R_{cap}^2\,
\nonumber \\
\approx \pi\,(H/2)\,R_{cap}^2 = \pi\,(H/2)\,(R_b + \delta R)^2\, , \  \label{Vcap}
\end{eqnarray}
where the difference between $R_{cap}$ and $R_b$ is designated as $\delta R $ ($(R_{cap} - R_b)=\delta R \ll R_b$). The approximation for V(cap) in Eq. (3) is valid under assumption $H/(2 R_{cap})\ll\,1$. From the equation $V_{el}=V_{cap}$, we obtain the estimate for $\delta R$:
\begin{eqnarray}
\delta R   \, \approx  \, 0.15 R_b   \,  \ll  \,  R_b  \, . \ \ \ \label{apr}
\end{eqnarray}
Thus, the above approximation of the lens-like isotropic droplets in FSSF by oblate spheroid works well due to their small aspect ratio, $H/(2 R_{cap})\ll\,1$, Fig. \ref{Figure2}. The small deviations of the ellipsoidal cross-section from the initial drop profile can be seen only in the area close to the drop edge.

 On the other hand, the curvature at the end face is much larger than in the upper drop point (their ratio is of the order of $=a^2/b^2$). However, this small area at the drop apex is not affecting the general pattern of the convection motion.
The same geometrical approach was applied to the lens-like oil drops deposited on FSSF, Fig. \ref{Figure1}b. Contrary to inclusions of the isotropic phase, representing the different phase state of the same liquid crystal material, oil is an individual substance  and  has the value of surface tension $\gamma_o$  between the drop-air
interface different from that of the isotropic material. Nevertheless, the same Eq. (1) can be used to describe the shape of the oil lenses.

\begin{figure}
    \hskip-0.6true cm
   \includegraphics[width=0.9\linewidth]{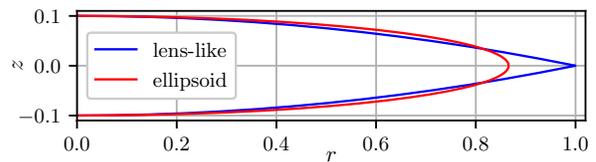}
    \caption{Illustration of the quality of approximation of the shape of the isotropic drop in FSSF by an oblate spheroid.  \label{Figure2}}
    \end{figure}

In accordance with the validity of the approximation of the lens-like drops in FSSF by oblate spheroid the corresponding conventional orthogonal coordinates
$u, \xi,  \varphi\,$ are consistently  employed in further elaborations.  The every point of space is   described by a triple of numbers
($u, \xi,  \varphi\,$),  corresponding to an unique point in the Cartesian coordinates $(x,y,z)$.
The corresponding
orthogonal   system of surfaces consists of
 oblate spheroids formed by  surfaces of constant  $\xi\,$ ($\xi=\xi_0\,$  is the spheroid of the given boundary),
   one-sheeted  hyperboloids of revolution of constant  $\vert u \vert$ (also known as a circular hyperboloid, as the surface generated by a rotation of the hyperbola around the $z$-axis), and planes of $\varphi\,=$ const ($\varphi\,$  is an azimuthal angle), \cite{NLebedev65,NLebedev652,Happel}, Fig. \ref{Figure02}.
The parameter
$\xi_0$ determines the ellipticity ratio and under the reasonable assumption  $\xi_0 \ll 1$ can be written as $\xi_0 = H/(2 c) \approx H/(2 R_b)$. The above parameters are related  to the rectangular coordinates by the  following matrix representation \cite{NLebedev65,NLebedev652,Happel}:
\begin{eqnarray}
  \left(\begin{array}{c}
 x \\ y \\ z
 \end{array} \right)
\, =
\,  \left(\begin{array}{c}
 \,c\,\sqrt{1+\xi^2}\,\sqrt{1-u^2}\,\cos[\varphi]\\
 c\,\sqrt{1+\xi^2}\,\sqrt{1-u^2}\,\sin[\varphi] \\
 c\,u\,\xi\,
 \end{array} \right) \  \
 \label{1elc}
 \end{eqnarray}
where the focus distance $c$ plays a role of a scale parameter and
\begin{eqnarray}
 \ - 1\, \le u \le \, 1 \, , \
0\, \le \xi < \,\infty\, ,
 0\, < \varphi \, \le\,  \,2\,\pi \,\, \qquad  . \  \  \label{1elc1}
\end{eqnarray}

 In turn, the representation of the Lame  coefficients (metric coefficients) in above variables reads:
\begin{eqnarray}
h_{u}^2 =  \Big(\frac{\partial x}{\partial u} \Big)^2 + \Big(\frac{\partial y}{\partial u} \Big)^2 + \Big(\frac{\partial z}{\partial u} \Big)^2 \ , \
 \nonumber \\
h_{\xi}^2 =  \Big(\frac{\partial x}{\partial \xi} \Big)^2 + \Big(\frac{\partial y}{\partial \xi} \Big)^2 + \Big(\frac{\partial z}{\partial \xi} \Big)^2 \ , \
 \nonumber \\
h_{\varphi}^2 =  \Big(\frac{\partial x}{\partial \varphi} \Big)^2 + \Big(\frac{\partial y}{\partial \varphi} \Big)^2 + \Big(\frac{\partial z}{\partial \varphi} \Big)^2 \ , \
  \  \ \label{1metr}
\end{eqnarray}
i.e. the metric coefficients are:
    \begin{eqnarray}
    h_u = c \sqrt{\frac{\xi^2 + u^2}{1-u^2}} \ ,
    h_\xi = c \sqrt{\frac{\xi^2 + u^2}{1+\xi^2}} \ , \
       \nonumber \\
    h_\varphi = c\sqrt{1+\xi^2}\sqrt{1-u^2}
      \  . \ \label{aLamexi}
      \end{eqnarray}

   It is important to note that the values of the parameters $u,\varphi$ at $\xi=0$ describe the points
    ($z=0, \, r= c\sqrt{1-u^2}$) on the mediated circle disk, for which  $\mathbf{r}[\xi=0, u] = \mathbf{r}[\xi=0, -u]$,
     i.e. two points in oblate spheroid coordinates correspond to a single point in real physical  space. This  means that any real physical field
     must be even function of $u$ at $\xi=0$. The same is true for a certain set of conditions for the spatial derivatives of the various physical quantities, the velocity, for example,  at $\xi=0$. On another hand, if the physical field $f[\xi, u]$ splits into a product $f_{\xi}[\xi]f_{u}[u]$, then only the single condition appears: the functions $f_\xi[\xi]$ and $f_u[u]$  should have an  equal evenness.

\bigskip

\centerline{\bf B. Marangoni instability and the smectic   }

\centerline{\bf  layering at isotropic drop interfaces}
\smallskip

\begin{figure}
    \hskip0.5true cm
   \includegraphics[width=0.9\linewidth]{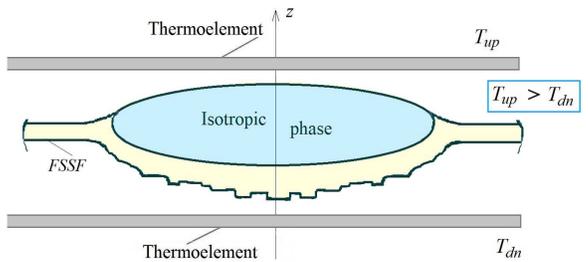}
    \caption{Sketch of the free standing smectic film with isotropic drop embedded in it. The upper surface of the droplet is free of the smectic layers, while the lower interface is covered with a nonuniform smectic shell.The thermoelectric devices above and below the drop are used to
preset a vertical temperature gradient across the drop, $T_{up} > T_{dn}$.    \label{FigSt}}
    \end{figure}

As we indicated earlier the surface of isotropic droplets in FSSF heated above the bulk smectic-isotropic temperature is covered by a certain amount of smectic layers. In view of this, the natural question arises: what might be the reaction of the smectic layering of the drop on the relatively large positive temperature gradient across it ($T_{up} > T_{dn}$, see Fig. \ref{FigSt}). The second question, even more fundamental, is the following: whether the Marangoni instability could develop at the interface between the fluid and the smectic substrate. We remind that smectic state combines a solid like elasticity along the layers normal and  the liquid behavior in the plane of layers -- an absence of the resistance for an applied shear stress \cite{deGennes93}. This second question applies equally to isotropic droplets formed in  overheated  FSSF and to oil lenses deposited on it, Fig. \ref{Figure1}b.

Let us start with the stability of the smectic layering. The amount of smectic layers on the surface of isotropic droplet in FSSF depends on the degree of overheating, initial film thickness, energetics and probability of defect formation, etc. \cite{Lucht98,Picano01,Pikina15}.  As in any layered structure, discrete change of the amount of smectic layers  in the film can only proceed by thermal generation of elementary edge dislocation loops \cite{Kleman03,Turner94}.  This usually occurs in the middle plane of the film and corresponds to the formation either of the surface depletion areas (holes) or surface bulge areas (islands).  The dislocation loops of critical radius can be generated stochastically in the film under favorable conditions and then they are growing at certain rate either in the direction of meniscus producing thinner films, or in the opposite direction -- film thickening  \cite{Oswald97,Oswald03,Oswald06,Ostrovskii04,Pikina15}.

The general approach allowing to calculate the frequency of thermal generation of any type of critical nucleus of energetically favorable defects was proposed by Langer and Fisher \cite{Langer67,freq}. According to our previous findings \cite{Pikina15,Pikina17} the typical value of critical work for nucleation of dislocation  loop of critical   radius   $ R_c \,\sim \, 10^{-8} \div 10^{-7}$ m in the middle plane of overheated smectic film
is of the order  $W_c \sim 10^{-20}$ J. This is smaller then the threshold energy $W_c^* = 2.5 \cdot 10^{-19}$ J, thus indicating that if the temperature of the upper drop surface is sufficiently higher than that at the bottom interface  the large number of dislocation loops appears leading to the thinning of the upper smectic shell of the drop.
The role of the heating protocol (the rate and the waiting period) is essential for the thinning of the smectic layering to proceed.
 This necessarily modifies the amount and the growing rate of the generated dislocation loops.
 Applying the slow preliminary heating \cite{Picano01} it is possible to generate one dislocation loop after another in the smectic shell on the top of a drop and in such layer by layer fashion move away the smectic layers  from the hot half of the drop (see also \cite{Ma}).

 \smallskip

 The situation at the bottom side of the drop is different. In the presence of the positive temperature gradient ($T_{up} > T_{dn}$) the temperature of the meniscus connecting the isotropic drop with FSSF is higher than that at the bottom side of the drop.  For relatively large  temperature difference the energy gain associated with nucleation and growth of dislocation loops of excess smectic layers becomes energetically favorable \cite{Pikina15}. The material, necessary for the smectic shell at the bottom side of the film to thicken, flows from the meniscus surrounding the isotropic drop, thus producing the sequence of islands (bulge areas in the shell), Fig. \ref{FigSt}. This corresponds to the disappearance of a whole set of dislocation loops from the meniscus.  At a certain stage of this process, the activation energy for the formation of such dislocation loops becomes smaller than the threshold energy and the process of formation of islands at the bottom shell of the drop is stopping.
  Such a behavior is in accordance with the formation and movement of islands on the surface of smectic bubble subjected to temperature gradient observed in microgravity experiments at International Space Station \cite{Stannarius2019}

 Now, we turn to the analysis of the possibility of the Marangoni convection in fluid, which is bounded by a smectic shell (substrate). This configuration applies equally to isotropic droplets formed in overheated FSSF and to oil lenses embedded in FSSF, Figs. 1, 4. The Marangoni forces act tangentially at the curved fluid-smectic interface, thus inducing the axially-symmetric flow of smectic material in the plane of smectic layers from the meniscus (hot area) downwards in the direction of the bottom point of the smectic shell (cold area). These fluid motions interfere with each other in the bottom area of the smectic shell, thus producing smectic islands and steps in the film. As a consequence of this process the lamellar structure in the smectic film is strongly destructured in this area, producing the domains in which the orientation of layer normals is inclined relative to the initial fluid-smectic interface. This situation is somewhat similar to a process of collapse of a smectic bubble inflated at the end of a capillary tube \cite{Caillier06}. In both cases the lamellar structure in the area close to the meniscus is strongly destroyed. Below we show that in this case the tangential component of the smectic elastic force compensates the Marangoni forces at the fluid-smectic interface, thus terminating the flow of the smectic material. This means that at the border between the smectic and fluid its tangential velocity turns to zero, which corresponds to the sticking condition.

Indeed, in the invariant form  the smectic elastic tensor can be written as (see for example \cite{Landau7,Lebedev1993})
 \begin{eqnarray}
\sigma^{(el)}_{i k} = B\, (\nabla_n u_{sm} ) n_i n_k  \ , \
\label{sigmel}
\end{eqnarray}
where $B$ is the smectic elastic modulus, corresponding to compression (dilatation) of the smectic layers, $u_{sm}$ is the displacement field of the smectic layers, ${\bf n}$  is the normal to the surface of smectic domains,  $(\nabla_n u_{sm} )=({\bf n} \nabla u_{sm} )$.  The $k-$component of the force ${\bf f}$,  acting on the bottom interface
reads
\begin{eqnarray}
f_k =  B\, (\nabla_n u_{sm} )\,({\bf l}\,{\bf n})\, n_k - \delta p \, l_k  \ ,  \
\label{fk}
\end{eqnarray}
where $\delta p$  is the pressure difference, ${\bf l}$ is the normal to the bottom interface and $({\bf l}\,{\bf n})$ is their scalar product.
Using the condition of  balance of the normal forces
at the internal interface between the smectic and fluid
\begin{eqnarray}
B\, (\nabla_n u_{sm} )\,({\bf l}\,{\bf n})\, n_k l_k - \delta p \, l_k l_k = 0 \ , \
\label{fn}
\end{eqnarray}
and one can derive the pressure difference acting at the interface as
\begin{eqnarray}
\delta p  = B\, (\nabla_n u_{sm} )\,({\bf l}\,{\bf n})^2 \ .  \
\label{dp}
\end{eqnarray}

After substitution of the  expression for $\delta p$ into Eq. (\ref{fk})  the condition of equilibrium of the tangential forces takes the form
 \begin{eqnarray}
  B\, (\nabla_n u_{sm} )\,({\bf l}\,{\bf n})\, ({\bf n}\,{\bf m}) \, - \delta p \, ({\bf l}\,{\bf m})
   \nonumber \\
  =\, B\, (\nabla_n u_{sm} )\,({\bf l}\,{\bf n})\, ({\bf n}\,{\bf m}) = {G}_{Ma} \ , \
  \label{tang}
\end{eqnarray}
where ${\bf m}$ is the  unit vector, tangent to the internal interface ($({\bf l}\,{\bf m})=0$).
 Thus, the  projection of the force ${\bf f}$ on the unit vector ${\bf m}$ is able to compensate the  Marangoni force ${G}_{Ma}$, that corresponds to the condition of  sticking of the fluid at this interface.

In that follows we consider the situation, when the top of the drop is free from the  smectic layers, i.e. it has the free boundary, while the bottom  half of the drop
is in contact with the static smectic substrate (compare with \cite{Stannarius2019}). This asymmetric geometry applies equally to isotropic drops formed in overheated FSSF and to oil lenses embedded in it.

\bigskip

 ******************************************

 \section{GOVERNING EQUATIONS AND BOUNDARY CONDITIONS}
\label{sec:4}

 Normally, surface tension of a liquid is a decreasing function of temperature
\begin{equation}
   \gamma = \gamma_0 -  \varsigma \,T'
      \ , \
      \label{gam}
  \end{equation}
where $T' =  (T_{dr} -  \bar{T})$, $T_{dr}$ is a current drop temperature, $\bar{T}$ is some  constant temperature (far from the drop at $z=0$), and  $ \varsigma\,>0$.
 Below we omit the symbol $'$ for simplification
of the further derivations.
We remind that  only for   the sufficient temperature gradient across the flat fluid film, the
 small temperature variations along the surface initiate change of the surface tension, which in turn cause the fluid to flow and thereby tend to maintain the initial temperature disturbances. Because of viscosity of the liquid the moving surface gives rise to a shear stress which drives a flow in the film interior \cite{Gershuni1972,Koschmieder1992,Getling1991}. As a result the flat  fluid film loses its mechanical stability and the Marangoni convective patterns are developed, as have been shown theoretically by Pearson \cite{Pearson}  using the linear instability analysis.

In this work we present a quantitative description of the Marangoni flows
in ellipsoidal isotropic droplets formed in FSSF based on the formalism of the Stokes stream functions.
Contrary to the flat fluid films,  the mechanical equilibrium within such drops is absent due to their curved shape. Because of the nonuniform temperature distribution the tangential thermocapillary force always exists at the free drop surface (Marangoni force). This leads to a fluid flow along its curved interface, making the thermocapillary flow within the drop thresholdless.

Consider the horizontal FSSF with fluid isotropic droplets  in it, which is placed between two thermoelectric devices.
The film is parallel to the $x-y$ plane, with the layer normal directed along the $z$ axis. The origin of the coordinate frame along $z$ is taken in the center  of the drop. The construction of set-up  allows the heat transfer from the hot plate  to the cold plate placed at the bottom side of the drop. This corresponds to the positive direction of the temperature gradient $ \partial T/\partial z $
  $\,(T_{dn} < T_{up})$, Fig. \ref{FigSt}, and  ensures the absence of the Rayleigh convection in the surrounding air.

The flow is governed by  set of equations, namely, the Navier-Stokes equation, the thermal energy transport equation, the continuity equation for the incompressible fluid and the equation for thermal conduction in the surrounding air
(from (\ref{NSt2}) to (\ref{temp2a})) \cite{Gershuni1972,Getling1991,vanHook1997,Landau6,Falkovich}
\begin{eqnarray}
           \frac{\partial {\bf v}}{\partial t}\,
   \,=\,   -\frac{1}{\rho_0}\,({\bf \nabla} {p})
            +\, \nu\,\mathbf{\nabla}^2 {\bf v}\,-\, \beta\,T\,g\,{\bf e}_z
      \ , \ \label{NSt2} \\
   \frac{\partial {T}}{\partial t}\,+\,({\bf v}\,\mathbf{\nabla})\,{T}  \,=\,\chi\,\Delta T \
        , \    \label{temp2}  \\
      (\mathbf{\nabla} \,{\bf v})  \,=\, 0
      \ , \ \  \label{cont2}\\
   \frac{\partial {T_{a}}}{\partial t}\,  \,=\,\chi_{air}\,\Delta T_{a} \
        , \    \label{temp2a}
\end{eqnarray}
which correspond to the  Boussinesq approximation.   $T_{a} =  (T_{air} -  \bar{T})$, $T_{air}$ is a current temperature in the surrounding air.
In above equations $\beta\,$ is the thermal expansion coefficient,  $\rho_0\,$ is the density of fluid, $\nu = \eta/\rho_0$ is the kinematic viscosity, $\eta\,$ is a dynamical  viscosity coefficient, $g$  is gravitational acceleration and ${\bf v}$  is the flow velocity.  The coefficient of temperature conductivity is designated as
$\chi = \varkappa \,(\rho_0 c_p)^{-1}$,  where $\varkappa\,$ is the thermal conductivity, $c_p$ is a specific heat. In above equations
 the quadratic  over perturbations inertial terms were omitted.
Below we apply a conventional linear perturbation theory to describe the small   deviations of the solutions
in the considered system  from the  zero stationary approximation
($T=T_0+T_1$),  where $\partial T_0/\partial z = A > 0 $, $\vert T_1 \vert \ll T_0$ \cite{Landau6,Gershuni1972}.

 The term $-\,\beta\,T\,g\,{\bf e}_z$   in Eq. (\ref{NSt2}) corresponds to the convective buoyancy  force in the drop.
  Bearing in mind that Marangoni convection at small length scales (i.e. in small size drops we deal with) prevails over the buoyant convection,  we can neglect this  term in Navier-Stokes equation in comparison with the viscous term due to the small Rayleigh number $R=g\beta\,A\,H^4/(\nu\,\chi)\,$ \cite{Gershuni1972}. The relative role of two types of convection can be evaluated  from the comparison of the Rayleigh and Marangoni
\begin{equation}
 \hbox{Ma} = \frac{\varsigma H^2 A}{\chi \eta} \  \
  \label{RMa}
      \end{equation}
numbers \cite{Gershuni1972}, therefore, the Marangoni convection is dominating at drop heights
\begin{equation}
H \ll H_c = \sqrt{\frac{\varsigma}{\rho\,g\beta\,}} \sim 10^4 \mu m \  , \
\nonumber
      \end{equation}
where  the typical values of the liquid crystal parameters \cite{par} are used.
For the ordinary fluids, the transition to buoyancy-dominated convection occurs around a 1 cm which is many orders of magnitude larger than the droplets size considered in our theory \cite{Gershuni1972,Landau6,Lebedev1993}.
    Thus in our case we deal with the pure Marangoni convection, initiated by the gradients of the surface tension at the drop interfaces.

    Because all the coefficients in   Eqs. (\ref{NSt2} --\ref{temp2a})
  are not dependent  on time,   we can  find the stationary solutions of our thermocapillary problem. In this case the left parts of  Eqs. (\ref{NSt2}--\ref{temp2a}), containing terms with time derivatives, vanish.
 The equations (\ref{NSt2}--\ref{temp2a}) are written in  general view with conjunction to the conventional rectangular coordinate $z$.
  To solve the problem of Marangoni convection in ellipsoidal fluid drops it is convenient to rewrite all governing equations (\ref{NSt2}--\ref{temp2a})  and  boundary conditions using the orthogonal oblate spheroid coordinates (see Sec. II A and Fig. \ref{Figure02}).

  Let us start with the formulation of  the boundary conditions for our problem.
  At the surface of the oblate spheroidal drop for fixed $\xi=\xi_0$ the boundary conditions for the fluid velocity components can be written as
                  \begin{eqnarray}
  v_{\xi} = 0 \ \ ({\hbox{at}} \, \xi=\xi_0 ) \ , \
\label{vxibc}
                  \end{eqnarray}
                - that is the condition  of an absence of  flow of the material through the boundary surface of the drop; additionally
                  \begin{eqnarray}
  v_{u} = 0 \ \     \qquad \ ({\hbox{at}} \, \xi=\xi_0, \, u
      \in [ -1, u_0])
             \, , \
             \label{xiubc0}
                       \end{eqnarray}
               which determines the condition  of sticking of a fluid at the bottom boundary surface of the drop (in a contact with the smectic shell). The value of $u_0$
               determines an extension of the boundary with no-slip condition along the drop interface: $u = 0$ corresponds to its termination at the circular edges of the drop, while the positive $u$  indicate the partial overlap of the upper drop interface by the region with sticking  due to a presence of the meniscus, see Figs. \ref{Figure1}, \ref{FigSt}.
                             Next we turn to the boundary conditions for the temperature deviations and the heat fluxes
                 \begin{eqnarray}
  T_{a }\big\vert_{\xi=\xi_0}  = \, T \big\vert_{\xi=\xi_0}
             \, , \ \label{bcT0}
             \\
          T_{a}\big\vert_{\xi\to\infty}  = \, C_{air}\,{c\,u\,\xi} \,
           \, , \ \label{bcT00}
      \end{eqnarray}
              \begin{eqnarray}
     \varkappa\,  \frac{1}{h_{\xi}}\,\frac{\partial T}{\partial {\xi}}\Big\vert_{\xi=\xi_0} =
     \varkappa_{air}\,\frac{1}{h_{\xi}}\,\frac{\partial {{T}_{a }}}{\partial {\xi}}\Big\vert_{\xi=\xi_0}
                 \, , \ \label{bcT2f}
      \end{eqnarray}
       that are the boundary conditions of the equality of the temperature  deviations and the normal heat flux at the air-drop interface.
      It is important that for the system under consideration  $\varkappa=\varkappa_{fluid}\simeq\,0.25\,$W(m K)$^{-1}\gg\,\varkappa_{air}\simeq\,0.026\,$W(m K)$^{-1}$ \cite{Birnstock01}, which  means an almost instant thermal flow inside the drop comparatively to the surrounding air.

Another  class of the boundary conditions for our problem corresponds  to a stress balance
at the surface of the droplet projected both in the normal and tangential directions.
    The hybrid boundary conditions  for the balance of tangential forces $\hat{\sigma}_{\mu \alpha}\,n_{\alpha}$  on two boundaries of the ellipsoidal drop (the free top surface and the bottom one in the contact with FSSF)  are given by  the expressions   (see  Eqs. (\ref{sigmaxi}) from the Appendix B)
\begin{eqnarray}
\sigma^{u\xi} =\eta \Big[ \frac{\partial_u v_\xi}{h_u}+
    \frac{\partial_\xi v_u}{h_\xi} - \frac{v_u}{h_\xi}\frac{\xi}{\xi^2+u^2}
    \Big]
    \nonumber \\
    =  \,  \frac{\partial_u \gamma}{h_u} \qquad ({\hbox{at}} \, \xi=\xi_0, \, u
        \in [u_0, 1])
             \,  , \ \    \label{xiubc}
              \\
                         \sigma^{\varphi \xi} =  0 \ . \  \qquad   \qquad
             \label{xiubc1}
\end{eqnarray}
     The equation (\ref{xiubc})  reflects  the nonuniformity    of the surface tension $\gamma\,$ at the upper drop surface, thus introducing the thermocapillary force, which drives the convection process \cite{Landau6}.
  At this point it is appropriate to note that the zero approximation of the system  of equations (\ref{NSt2} -\ref{temp2a}) coincides with  the zero approximation over the temperature coefficient of surface
tension $\varsigma\,$ (i.e. when $ \varsigma\,=0$, see Eq. (\ref{gam})).
       The detail  analysis of these boundary conditions  is given in Appendix B.

As to the normal stress balance,  it is replaced in our case by the assumption  that the ellipsoidal form of a droplet practically does not change in the process of convection (compare with \cite{Tam2009}).  This assumption is valid because the pressure deviation due to  the nonhomogeneity of the temperature across the drop boundary  is  negligibly small $\delta p/p \sim \delta \gamma/\gamma \sim  10^{-4} \div 10^{-3} \ll 1 $, (see \cite{Ma,par}).


It is convenient to solve the hydrodynamic equations of Marangoni convection for  the axially-symmetric ellipsoidal drops in terms of 2D Stokes stream functions $\psi[u_, \xi]$   \cite{Happel}. By definition, this function determines the instant fluid flow rate divided by $2\pi $ (the half of the total spatial angle). The stream function $\psi\,$  is scaled by $\chi c^2/H$, and thus used in the dimensionless form below.
 According to  \cite{Happel},  the velocity field is related to the  stream function by the following equation written in oblate spheroidal coordinates
 \begin{eqnarray}
      \mathbf{v} = \frac{1}{h_\varphi}\,[{\bf e}_{\varphi} \,\times\, \nabla \psi] \ . \
       \label{vsp}
              \end{eqnarray}
  After substitution of Eq. (\ref{aLamexi}) to Eq. (\ref{vsp}) one  obtains
       \begin{eqnarray}
      \mathbf{v} =  \, - \frac{{\bf e}_{u}}{h_\xi h_\varphi}\partial_\xi \psi \, +\, \frac{{\bf e}_{\xi}}{h_u h_\varphi}\partial_u \psi   \ , \
       \label{bfu}
              \end{eqnarray}
              where ${\bf e}_{\xi}$ and ${\bf e}_{u}$  are the unit vectors along $\xi$ and $u$ axis, respectively.

      To obtain the  dynamic equation for Stokes  stream function it is convenient to introduce the  vorticity
          \begin{eqnarray}
     \vec{\mathbf{\varpi}} = \,[\nabla\,\times\,  \mathbf{v}] \ . \ \
       \label{vort1}
              \end{eqnarray}
      After substitution of Eq. (\ref{bfu})  to Eq. (\ref{vort1}) one arrives to
        \begin{eqnarray}
        \vec{\mathbf{\varpi}}\,  =  \,\frac{{\bf e}_{\varphi}}{h_{\varphi}} \, \hat{\text{E}}^2\psi  \ , \
        \label{vort2}
              \end{eqnarray}
              where
   \begin{eqnarray}
 \hat{\text{E}}^2 \psi =   \frac{1}{c^2\,\big(u^2  + \xi^2 \big)}\,
\Big\{ \,(1  + \xi^2)\,\frac{\partial^2 \psi}{\partial\xi^2}
 +   (1 - u^2)\,\frac{\partial^2 \psi}{\partial u^2}\, \Big\}
     \, . \  \ \ \label{StrF2}
                  \end{eqnarray}
Then   applying the rotor operation to the vorticity $\vec{\mathbf{\varpi}}$  twice  one obtains
 \begin{eqnarray}
    \,[\nabla\,\times\, \,[\nabla\,\times\,  \vec{\mathbf{\varpi}}]\,] = - \,\frac{{\bf e}_{\varphi}}{h_{\varphi}} \, \hat{\text{E}}^2 \,\big( \hat{\text{E}}^2\,\psi \big) =  - \,\frac{{\bf e}_{\varphi}}{h_{\varphi}} \,\hat{\text{E}}^4 \psi\ . \
       \label{vort3}
              \end{eqnarray}
  Applying the rotor operation to both sides of Eq. (\ref{NSt2}), one excludes the pressure $p$,  and using the continuity equation  (\ref{cont2})  and equations (\ref{vort1}), (\ref{vort2}) and (\ref{vort3}) (thus $ [\nabla\,\times\, \nabla^2 \mathbf{v}] \, = -  \,[\nabla\,\times\, \,[\nabla\,\times\,  \vec{\mathbf{\varpi}}]\,]$), expresses the resulting equation through single variable $\psi$ \cite{Happel}.
    The  convective buoyancy  force in Eq. (\ref{NSt2}) was disregarded, as we argued above. In such a way the Navier-Stokes equation in the Boussinesq approximation (\ref{NSt2}) for  the stationary regime   is replaced   by  the following equation for the  stream function
 \begin{eqnarray}
\hat{\text{E}}^2 \,\big( \hat{\text{E}}^2\,\psi \big)
 \,=\,0
    \, .\  \ \  \label{big}
                  \end{eqnarray}

       \smallskip
We note that the boundary condition represented by Eq. (\ref{vxibc}) with account to  Eq. (\ref{bfu}) takes the form:
  \begin{eqnarray}
\psi[\xi_0, u]  \,=\,0 \
    \, . \  \ \ \label{BCS1}
                  \end{eqnarray}
It is important to check the obtained solutions on the absence of singularities, see Sec. II A. In the first place this applies to the components of the velocity field  $v_x, v_z$ and the  vorticity  $\vec{\varpi} $, which should be continuously differentiable functions.

 \bigskip

\section{RESULTS }
\label{sec:5}

\centerline{\bf  A. Stokes stream functions and velocity fields}

In this section we generalize the  formalism developed by Happel$\&$Brenner  \cite{Happel}   to solve Eq. (\ref{big}) for the stream function $\psi\,$. In doing so we first  obtain the solutions of equation $\hat{\text{E}}^2\,\psi\,=0$.
According to definition of the stream function, $\psi=0\,$ along the $z$-axis, i.e. for $u=1$  or  $u=-1$. The solutions for $\psi\,$ can be either symmetrical or asymmetrical over the variable $u$. This means that all solutions of Eq. (\ref{big}) should be proportional to either
$(1 - u^2)\,$ or to $u\,(1 - u^2)\,$, respectively.
In turn, in accordance with the properties of the Legendre polynomials, $P_n(u)$, the
 solutions  of the equation $\,\hat{\text{E}}^2\,\psi\,=0\, $ are proportional to the integrals  from Legendre polynomials, see appendix C, where the straight method of derivation of the solutions of Eq. (\ref{big})  is presented.
 This allowed us to obtain the full set (the linear space)  of solutions of Eq. (\ref{big})
 satisfying all of the above mentioned requirements.
\begin{figure}
\hspace{-0.1cm}\includegraphics[width=1\linewidth]{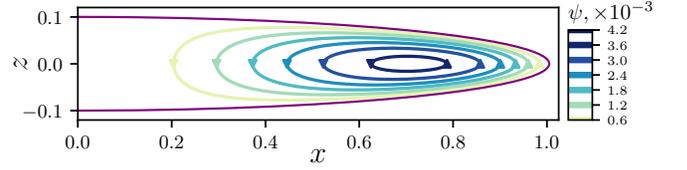}
 \caption{ \baselineskip8pt
  Streamlines corresponding to the basic stream function $\psi_{3}$ for $\xi_0 = 0.1$.; the $\xi_0$ value determines the ellipticity ratio of the droplet and is equal to $H/(2 c) \approx H/(2 R_b)$, see section II A). All lengths are shown in dimensionless form.}
 \label{psi01f}
\end{figure}
\begin{figure}
\hspace{-0.1cm}\includegraphics[width=1\linewidth]{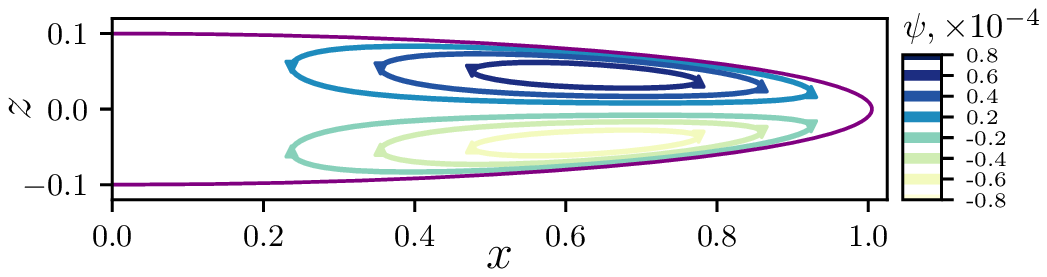}
 \caption{ \baselineskip8pt
 Streamlines corresponding to the basic stream function $\psi_{4}$ for $\xi_0 = 0.1$}
 \label{psi02f}
\end{figure}
\begin{figure}
\hspace{-0.1cm}\includegraphics[width=1\linewidth]{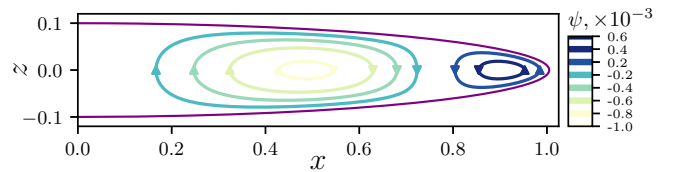}
 \caption{ \baselineskip8pt
  Streamlines corresponding to the basic stream function $\psi_{5}$ for $\xi_0 = 0.1$.}
 \label{psi03f}
\end{figure}

\begin{figure}
\hspace{-0.1cm}\includegraphics[width=0.9\linewidth]{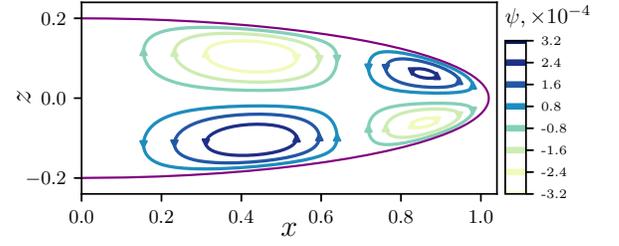}
 \caption{ \baselineskip8pt
 Streamlines corresponding to the basic stream function $\psi_{6}$ for $\xi_0 = 0.2$.}
 \label{psi04f2}
\end{figure}

However, the above mentioned straight method of derivation of the solutions of Eq. (\ref{big}) is too complicated, especially if we  extend it for the large number   of the accounted basic functions. We note that the full solution for the stream function represents the sum over the limited amount of the  basic functions $\{\psi_n[\xi, u]\}$, which  is determined by the number $N_r$. Instead that,  we developed the operator method of the solution of Eq. (\ref{big}) based on the introduction of a set of the  recursive operators, and  the special algebraic technique which allows the stream  functions of different order to interconnect with each other (see Appendices D and E).
  The general expression for the $n$th basic stream function $\psi_n$
can be written as
\begin{eqnarray}
\psi_n[\xi, u]\, = \, \mathcal{F}_n \Big( \mathcal{X}_{n-2}^{(1)}+\frac{-\mathcal{X}_{n-2}^{(1)}[\xi_0]}{\mathcal{X}_{n}^{(1)}[\xi_0]} \mathcal{X}_{n}^{(1)}\Big)
\nonumber \\
 + \, \,\mathcal{F}_{n-2} \Big( \mathcal{X}_{n}^{(1)}
+ \, \frac{-\mathcal{X}_{n}^{(1)}[\xi_0]}{\mathcal{X}_{n-2}^{(1)}[\xi_0]}\mathcal{X}_{n-2}^{(1)} \Big) \  , \ \ \
 \label{GenS}
  \end{eqnarray}
 where the function
 \begin{equation}
    \mathcal{F}_n[u] = \int_{-1}^{u}P_n[u'])\,d{u'} = \frac{P_{n+1}[u] - P_{n-1}[u]}{2n+1} \ , \
    \label{mFn}
\end{equation}
is expressed via the Legendre polynomials $P_n[u]$ of the order $n$,
and
\begin{equation}
    \mathcal{X}^{(1)}_n[\xi] = \frac{\Xi_{n+1}^{(1)}[\xi] - \Xi_{n-1}^{(1)}[\xi])}{2n+1} \ , \
    \label{mXn}
\end{equation}
  where
  $\Xi_{n}^{(1)}[\xi]$ is the  solution of the  equation (\ref{Xieq}), i.e. $\Xi_{n}^{(1)}[\xi]$ is  the $\xi$-depended function multiplier in the solution of the Laplace equation $\,\Delta \phi_{n} = 0\,$ in ellipsoidal coordinates  (see Appendices A, F and G). The above formalism is  applied for $n > 2$.
The  streamlines  for the first four stream functions $\psi_n$ corresponding to  ellipticity ratios  $\xi_0= 0.1$, $\xi_0= 0.2$  are presented  in Figs. \ref{psi01f}--\ref{psi04f2}.
 The lengths $x$ and $z$  are shown in dimensionless form, being scaled by $c = \frac{H/2}{\xi_0} = {a}\,({\sqrt{1+\xi_0^2}})^{-1}$. The number of vortexes along the  long drop semiaxis $a$  for each $\psi_n$ increases with the increase of $n$.
It is  readily seen  that with increasing of  the  ellipticity  ratio  $\xi_0$ the flow pattern remains the same, only  the scale is changed.
The  basic stream functions,  described by Eq. (\ref{GenS}),  are satisfied by the symmetry of the problem and  the condition of the absence of the fluid flow  through the external drop boundary (see Sec. III).

Provided that the stream functions describing the thermocapillary motion are known, the velocity components satisfying the corresponding boundary conditions (\ref{xiubc0},\ref{xiubc}) can be obtained
  \begin{eqnarray}
v_\xi[\xi, u]  =\,\,
  \frac{1}{c^2 ( \sqrt{1 + \xi^2} \sqrt{u^2 + \xi^2})}\,
    \frac{\partial \psi[\xi, u]}{\partial u} \ , \
 \nonumber \\
v_u[\xi, u]  =\,-\,
  \frac{1}{(c^2 \sqrt{1 - u^2} \sqrt{u^2 + \xi^2})}\,
     \frac{\partial \psi[\xi, u]}{\partial \xi}
\ . \ \label{StrL6}
      \end{eqnarray}
  In turn, the velocity components $v_x, v_z$  in oblate spheroid coordinates in accordance with Eqs. (\ref{ez1}--\ref{ez4}) are
  obtained as
                   \begin{eqnarray}
     v_x[\xi, u] =
  \frac{1}{ \sqrt{1 + \xi^2} }\, \big( \xi \sqrt{1 - u^2}
        \, v_\xi
        \nonumber \\
          - \, u \,\sqrt{1 + \xi^2}  \, v_u  \big)\,\cos\varphi
     \ , \ \label{vx1}
                  \end{eqnarray}
    \begin{eqnarray}
     v_z[\xi, u] =
  \frac{1}{ (u^2  +  \xi^2)^{1/2} } \big( u\, \sqrt{1+\xi^2}
        \,v_\xi
        \nonumber \\
           + \, \xi\, \sqrt{1-u^2}  \,v_u  \big)
     \ . \ \label{vz1}
                  \end{eqnarray}

      \begin{figure}
\hspace{-0.1cm}
\includegraphics[width=0.9\linewidth]{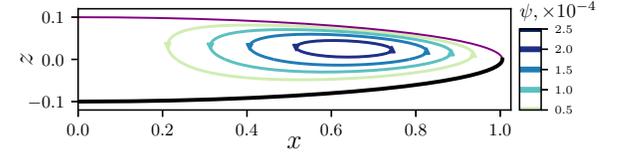}
 \caption{ \baselineskip8pt
 Streamlines corresponding to the first stream function $\psi_{1, st}$,  for $\xi_0 = 0.1$. The bold black line at the bottom part of the drop indicates that the sticking (no slip) conditions are fulfilled at this interface. In contrast to this, the upper part of the drop is free (in contact with the air). }
 \label{psi1St}
\end{figure}


\begin{figure}
\hspace{-0.1cm}
\includegraphics[width=0.9\linewidth]{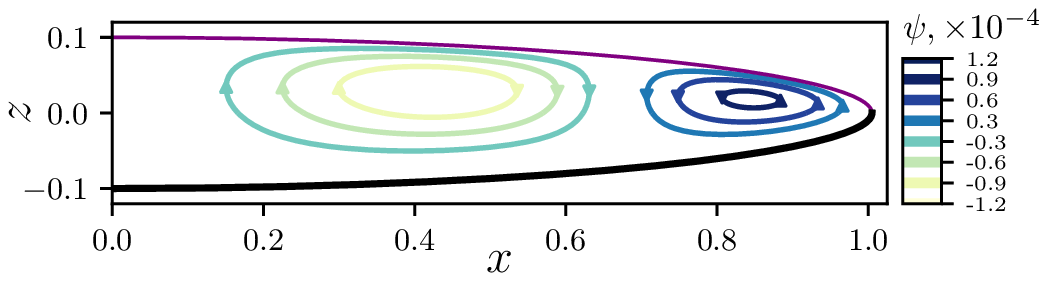}
 \caption{ \baselineskip8pt
 Streamlines corresponding to the first stream function $\psi_{2, st}$,  for $\xi_0 = 0.1$. The bold black line at the bottom part of the drop indicates that at this interface the sticking boundary condition holds. }
 \label{psi2St}
\end{figure}

\begin{figure}
\hspace{-0.1cm}
\includegraphics[width=0.9\linewidth]{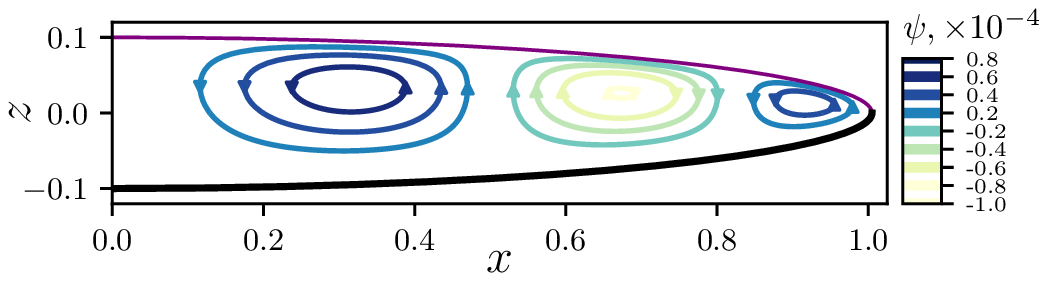}
 \caption{ \baselineskip8pt
 Streamlines corresponding to the first stream function $\psi_{3, st}$,  for $\xi_0 = 0.1$. The bold black line at the bottom part of the drop indicates that at this interface the sticking boundary condition holds. }
 \label{psi3St}
\end{figure}

\begin{figure}
\hspace{-0.1cm}
\includegraphics[width=0.9\linewidth]{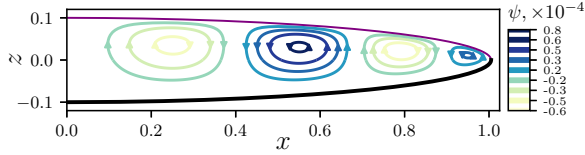}
 \caption{ \baselineskip8pt
 Streamlines corresponding to the first stream function $\psi_{4, st}$,  for $\xi_0 = 0.1$. The bold black line at the bottom part of the drop indicates that at this interface the sticking boundary condition holds. }
 \label{psi3St}
\end{figure}

Next, we  aim  to derive the basic set of the stream functions $\{\psi_{j, st}\}$, explicitly describing the convection flow
 for the asymmetric boundary conditions, i.e. for the case
 of fluid sticking  at the bottom drop surface:  $\partial_\xi \psi_{j, st}(\xi=\xi_0) =0 $ for $u \in(-1,0)$ (i.e. for $u_0=0$), and designation  $..._{st}$ indicates this property.
To solve this problem it is optimal to  represent the set of basic stream functions $\{\psi_j[\xi, u]\}$  (see Eq. (\ref{mFn})) and corresponding tangential velocities  $v_{u, j}$ as  sets of odd and even functions. This allows us to obtain the correct view of the new basic stream  functions satisfying the nonsymmetric boundary conditions.
Moreover, these stream functions provide a continuous variation through the points  of contact between the free and bounded by the smectic layers  surfaces of the drop, (for details of the calculation procedure see Appendix E).
The streamlines corresponding to the three first stream functions $\psi_{j, st}[\xi, u]$, satisfying the  boundary conditions of sticking at the bottom interface, are shown in  Figures \ref{psi1St}--\ref{psi3St}. The number of vortexes along the long drop semiaxis $a$  for each
 $\psi_{j, st}$ is equal to $j$.

    \smallskip

      \centerline{\bf B. Temperature  distribution  }

   \smallskip

 Now, we can turn to the   couple  of governing equations of Marangoni convection within ellipsoidal fluid drops describing  the thermal energy transport inside the drop  and the  thermal conduction in the surrounding air, Eqs. (\ref{temp2}) and (\ref{temp2a}).
 We need to know the  temperature  distribution   to obtain the general stationary solution of the thermocapillary convection inside the drop.
The  conventional linear perturbation theory is applied  again  to solve the system of equations (\ref{NSt2})--(\ref{temp2a}) in the stationary
regime.
  In doing so the approximations $T = T_0 + T_1$ and $T_a = T_{0\, air} + T_{1\, air}$ are used, where  $T_0$ is the solution for the case, the fluid motion is absent,
     and $T_{1}$ is the   small temperature deviation.
   This results in the following equations for temperature deviations:
\begin{eqnarray}
         \Delta T_0 = 0 \ , \
    \label{T01} \\
     \Delta T_{0\, air} = 0 \ , \
    \label{T0air} \\
   \chi  \Delta  T_1 = v_z \partial_z T_0
   \ , \
    \label{T1} \\
         \Delta T_{1\, air} = 0 \ . \
    \label{T1air}
\end{eqnarray}
The full set of solutions of the above Laplace equations in oblate spheroid coordinates is given in Appendix A.
As a first step we calculate the  temperature distribution $T_0$. In accordance with the symmetry of the problem the stationary heat flux far away from the drop (in the surrounding air) is directed along the z-axis: \\
$T_{0\, air} \to  C_{air}\,{c\,u\,\xi} $, where $C_{air}$ is a uniform temperature gradient across the drop, $z={c\,u\,\xi}$.
According to   Appendix A the regular  kernel of the Laplace operator can be expressed as $P_n[u]\Xi_n^{(1)}[\xi]$.
 The corresponding heat flow  is regular at  the point $u=\xi=0$.
 To find the temperature  distribution  in the air it is convenient to use
    the linear combinations of  $\Xi_n^{(1, 2)}$, damped at $\xi\to+\infty$, which are designated below as $\Xi_n^{(a)}$.
Thus,  the solution   in the air  has the form

\begin{eqnarray}
 T_{0 \,air} \, = \,  C_{air} \,c
      \underbrace{u \,\xi}_{P_1 \, \Xi_1^{(1)}} + \sum^{\infty}_{n=1} \alpha_n\, P_n[u]\,\Xi_{n}^{(a)}[\xi]
    \label{solT1}
\end{eqnarray}

The  temperature  distribution $T_{0}$  inside the drop can be written  (with account for the regularity condition of the heat flow  at
the  point   $u=\xi=0$) as:
\begin{eqnarray}
T_{0} =  \sum^{\infty}_{n=1} \beta_n P_n[u] \Xi_{n}^{(1)}[\xi] \ . \
\label{solTd0}
\end{eqnarray}

The solutions of Eqs. (\ref{solT1}) and (\ref{solTd0}) should  satisfy the boundary conditions (\ref{bcT0}) and (\ref{bcT2f})  at $\xi=\xi_0$,  because the Legendre's polynomials $\{P_n\}$ form an orthogonal basis, these conditions have to hold for any value of variable  $u$.

\smallskip

 $  \ n=1$ : $\ \begin{cases}
    \beta_1 \xi_0 = C_{air}\,  \xi_0 + \alpha_1\, \Xi_1^{(a)}\vert_{\xi=\xi_0} \  , \\
    \varkappa\, \beta_1 =  \varkappa_{air}\,C_{air} \,c \,+ \varkappa_{air}\,\alpha_1 \,\partial_\xi \Xi_1^{(a)}\vert_{\xi=\xi_0} \, ,
        \end{cases} \to $

\begin{eqnarray}
\to \beta_1 = c \,\underbrace{C_{air} \,\kappa\, \frac{1-\xi_0 (\ln{\Xi_1^{(a)}})'\vert_{\xi=\xi_0}}{1- \kappa \xi_0 (\ln{\Xi_1^{(a)}})'\vert_{\xi=\xi_0}}}_{A} = c\,A,
  \label{BCTtb} \\
   \alpha_1 = \,-\, C_{air}\,  \frac{(1 - \kappa )}{( \Xi_1^{(a)}/\xi_0 - \kappa\,\partial_\xi\, \Xi_1^{(a)})\vert_{\xi=\xi_0} }   \ , \ \
 \label{BCTt}
\end{eqnarray}
where $\kappa = \varkappa_{air}/\varkappa$ is a relative heat conductivity.
\medskip

 $  \ n>1$ : $\ \begin{cases}
    \beta_n \xi_0 \, \Xi_n^{(1)}\vert_{\xi=\xi_0} =  \alpha_n\, \Xi_n^{(a)}\vert_{\xi=\xi_0} \  , \\
    \varkappa\, \beta_n \,\partial_\xi \Xi_{n}^{(1)}\vert_{\xi=\xi_0} =  \varkappa_{air}\,\alpha_n \,\partial_\xi \Xi_{n}^{(a)}\vert_{\xi=\xi_0} \ ,
    \end{cases}  \ \to \ $
\medskip

    $\ \to  \ \beta_n = 0, \alpha_n = 0 \quad \hbox{for} \  n>1 $ .
\medskip

In that follows we  present all the variables  in  dimensionless form , using appropriate scaling relations,  symbol $\widetilde{...}$ designates the dimensionless  variables, correspondingly.
 All lengths are scaled by $c = \frac{H/2}{\xi_0} = \frac{a}{\sqrt{1+\xi_0^2}}$,  velocities by $\chi/H$ (i.e. $\mathbf{v} = \Tilde{\mathbf{v}} \, \chi/H$),
 time by  $c/v=\frac{H^2/2}{\xi_0\,\chi}$,  and temperatures by $H A$ (i.e. $T = \Tilde{T}\,  A\, H$) \cite{Koschmieder1974,Gershuni1972}.
Then for the dimensionless temperature distribution  $\widetilde{ T_0}$  we  obtain
\begin{equation}
\widetilde{ T_0} =  \frac{u \xi}{2 \xi_0} \ . \
   \label{T0f}
\end{equation}
Let stress that the temperature distribution   $\widetilde{ T_0}$ is necessary to find the solution for the stream functions describing the main contribution to the stationary   thermocapillary convection within the drop.

   In turn, the dimensionless equation for the temperature distribution $\widetilde{T}_1$ reads:
\begin{equation}
  \widetilde{\Delta} \widetilde{T}_1 = \, \frac{c^2}{H^2} \, \widetilde{v}_z \,  =  \,\frac{1}{4\xi_0^2} \, \widetilde{v}_z \ . \
       \label{EqT1}
\end{equation}
 Below we omit the symbol $\widetilde{...}$ for simplicity.

It is convenient to rewrite Eq. (\ref{T1}) for the distribution of the temperature deviation $T_{1}$ inside the drop in terms of the stream function $\psi$:
\begin{equation}
       [\partial_\xi (1+\xi^2)\partial_\xi + \partial_u (1-u^2)\partial_u]\, T_{1} = \frac{( u \partial_u \psi - \xi \partial_\xi\psi)}{4\xi_0^2}\,  \, . \,
 \label{Tnin}
\end{equation}
The right side of Eq. (\ref{Tnin})  can be decomposed over the Legendre's polynomials in the form  $\sum_n f_n[\xi]P_n[u]$. In such a way we can find the temperature response to each $f_n[\xi]P_n[u]$ in view of $T_n[\xi]P_n[u]$.
 Thus, for each functional coefficient $T_n(\xi)$ we obtain the following equation
\begin{equation}
    \partial_\xi(1+\xi^2)\partial_\xi  T_n - n(n+1) T_n = f_n[\xi] \ . \
    \label{Tn}
\end{equation}
It is easy to check  that the right part of Eq. (\ref{Tn})  can be  presented as a linear combination
$f_n = \sum_k \mathcal{W}_{n,k}^{(1)}\,\Xi_k^{(1)}$, where $\mathcal{W}_{n,k}^{(1)}$ is expansion coefficient. This allows us to write the response of the  functional coefficient $T_n(\xi)$ to each $\Xi_k$ as
\begin{eqnarray}
    T_{n, k}[\xi] = \underbrace{\frac{\mathcal{W}_{n,k}^{(1)}}{k(k+1)-n(n+1)}}_{W_{n,k}^{(1)}}\Xi_k^{(1)}[\xi]
    \, , \   \  n \ne k\ . \
    \label{Tnk}
\end{eqnarray}
The condition $n \ne k$ is always  valid here due to the specific form of the expression for the velocity component $v_z$  (see Appendix G).

To derive the full solution for the temperature deviations $T_1$ within the drop it is necessary to add  to a partial solution
$\sum_k W_{n,k}^{(1)}\,\Xi_k^{(1)}[\xi]$ of Eq. (\ref{Tnin}) the homogeneous solution $ {\Lambda_n^{(1)}}\Xi_n^{(1)}[\xi]$,
that is
\begin{eqnarray}
    T_{1} =  \sum_n \, \Big\{ \sum_k W_{n,k}^{(1)}\,\Xi_k^{(1)}[\xi]\, + {\Lambda_n^{(1)}}\, \Xi_n^{(1)}[\xi]\Big\}\, P_n[u] \ . \
    \label{Tdr1}
\end{eqnarray}

 In the next step it is necessary to take into account  the continuity of heat and the heat flux at the boundary of the drop ($\xi=\xi_0$), where the solution for the first order temperature correction  outside the  drop can be written as (see Appendix A)
  \begin{eqnarray}
  T_{1\, air} \, = \,\sum_n \, \Lambda_n^{(a)} \Xi_n^{a}[\xi]\,P_n[u] \ . \
       \label{Ta}
   \end{eqnarray}
In such a way we obtain the system of equations
         \begin{eqnarray}
      \begin{cases}   {\Lambda_n^{(a)}} \Xi_n^{(a)}\vert_{\xi=\xi_0} = \big\{ \sum_k W_{n,k}^{(1)}\Xi_k^{(1)}  + \Lambda_{n}^{(1)} \Xi_n^{(1)}\big\}_{\xi=\xi_0} \ , \ \label{bcT1} \\
             {\kappa} \Lambda_n^{(a)} \partial_\xi \Xi_n^{(a)}\vert_{\xi=\xi_0} = \big\{\sum_k W_{n,k}^{(1)}
        \partial_\xi \Xi_k^{(1)} + \Lambda_{n}^{(1)}\partial_\xi\Xi_n^{(1)} \big\}_{\xi=\xi_0} \, ,  \label{bcT2}
        \end{cases}
   \end{eqnarray}
from which we obtain
 \begin{eqnarray}
\Lambda_{n}^{(1)} =  \left\{\frac{ \kappa\, \frac{\sum_k W_{n,k}^{(1)}\Xi_k^{(1)} }{\Xi_n^{(a)}} -  \frac{\sum_k W_{n,k}^{(1)} \partial_\xi \Xi_k^{(1)}}{\partial_\xi \Xi_n^{(a)}}}{\frac{\partial_\xi\Xi_n^{(1)}}{\partial_\xi\Xi_n^{(a)}} - \kappa\frac{\Xi_n^{(1)}}{\Xi_n^{(a)}}}\right\}_{\xi=\xi_0} \,\ . \
\label{Wn}
\end{eqnarray}

Thus, we derived the distribution of the temperature  deviations $T_{1}$ inside the drop. This allows us to find the general stationary solution for the stream functions and to analyze  its stability relative to the increase of the initial temperature gradient.

    \smallskip

      \centerline{\bf C. General stationary solution. }

   \smallskip

 Now, having in hands the analytical expressions for the temperature distribution within the ellipsoidal drops, we can  solve explicitly the Marangoni boundary conditions (\ref{xiubc},\ref{xiubc1}).
 After substitution of the components   $v_u, v_\xi$ from Eqs. (\ref{bfu}, \ref{StrL6})  to Eqs. (\ref{xiubc})
       and taking into account that at the drop boundary  $v_\xi = 0$ ($\xi=\xi_0$),
                        we  obtain the general expression for Marangoni boundary condition at the free surface of the drop
                    \begin{equation}
  -\,  \frac{h_u}{h_\xi} \partial_\xi \frac{\partial_\xi \psi}{h_\xi h_\varphi h_u}= -{\hbox{Ma}}\, \frac{\partial_u T}{h_u}  \ \  ({\hbox{at}}  \,  u     \in [u_0, 1]  )\ . \
\label{BCT0}
\end{equation}
In view of Eq. (\ref{BCT0}) the system of Eqs. (\ref{xiubc},\ref{xiubc1}) can be rewritten in the form
                   \begin{eqnarray}
 &      \Big\{ 2\, \xi \, \partial_\xi \psi  \, - \,  (u^2  +  \xi^2)\, \partial^2_\xi \psi   \,
 \qquad  \qquad  \qquad \qquad  \nonumber
   \\  \hskip-0.95true cm & =   -\,\hbox{Ma}\, \frac{(u^2  +  \xi^2)^{3/2}} {\sqrt{1  +  \xi^2 }} ({1 -  u^2})  \partial_{u}T \Big\}_{\xi=\xi_0}
            \,   ({\hbox{at}}  \, u
        \in [u_0, 1])
             \,  , \qquad   \label{bc24xi}
              \\
  & \qquad \partial_\xi \psi \, = 0  \qquad \qquad \qquad  ({\hbox{at}} \, \xi=\xi_0, \, u
      \in [ - 1, u_0])
             \ , \qquad   \qquad     \label{bc241xi}
                            \end{eqnarray}
   where Ma is the Marangoni number, defined earlier in Eq. (\ref{RMa}). Estimating contributions to the right part of Eq. (\ref{bc24xi}) and taking into account the  solutions for ($\psi, T_0, T_1$), we obtain that for the  typical  parameters of the system $\hbox{Ma}\ll\,10^3$. In this case the contribution from the temperature deviation $T_1$ in solving of the Marangoni boundary condition is negligibly small comparatively to that from  $T_0$ and can be omitted.
  Hence, the full  stationary solution for the stream function in the main approximation can be written either as expansion over initial basic functions $\psi_i$
 \begin{eqnarray}
\psi \, = \,\sum_{i=3}^{N_r} \, c_i \, \psi_i \ , \ \ \
 \label{tpsi}
  \end{eqnarray}
  or over the basic functions $\psi_{i, st}$, satisfying to the condition of sticking at the bottom surface:
   \begin{eqnarray}
\psi \, = \, \sum_{i=3}^{N_r} \, c_{i, st} \, \psi_{i, st}\ , \ \ \
 \label{tpsist}
  \end{eqnarray}
  where $N_r$ - is the number of  basic functions used in the summation.
Now,  we have two ways to find the stream function $\psi$.

The first one is straight: to  substitute expansion (\ref{tpsi}) in the system (\ref{bc24xi}-\ref{bc241xi}) and  to find the set of $c_i$, satisfying the following system of equations:
  \begin{eqnarray}
        \big(\partial_\xi-\frac{1}{2\xi_0}(\xi_0^2+u^2)\,\partial_\xi^2 \big){\psi}\vert_{\xi=\xi_0}
         =\underline{\sum_j \hat{O}_{ij}^{free} \mathcal{F}_i[u] c_{j}}
      \nonumber \qquad   \quad
      \\
\hskip-0.5true cm     = - \hbox{Ma}\, \frac{(\xi_0^2+u^2)^{3/2}}{2\xi_0\sqrt{1+\xi_0^2}}\, \frac{(1-u^2)}{2} = \underline{r[u]} \, ,\    ({\hbox{at}}  \, u
        \in [u_0, 1])\ , \ \
      \label{1way}
     \\
      \sum_{i,j} \hat{O}_{ij}^{stick}c_j\mathcal{F}_i[u] \, = 0  \  , \ \ ({\hbox{at}} \,  u
      \in [ -1, u_0])
             \, , \ \  \qquad   \qquad  \quad  \label{2way}
                             \end{eqnarray}
       where we introduced an expansion of the left part of the Marangoni condition (\ref{1way}) over  set of functions $\mathcal{F}_i[u]$  using the matrix representation  $\hat{O}_{ij}^{free}$   at the free boundary of a drop for $\xi=\xi_0$.
       To obtain the right part of  Eq. (\ref{1way})  we substituted the expression (\ref{T0f}) for the temperature distribution $T_0$  to the right part of the first equation in the system (\ref{bc24xi}).
       The matrix $\hat{O}_{ij}^{free}$  describes  the action of the operator $(\partial_\xi-\frac{1}{2\xi_0}(\xi_0^2+u^2)\,\partial_\xi^2)$  on  the expansion (\ref{tpsi}) for the stream  function $\psi$ at $\xi=\xi_0$. In turn, matrix  $\hat{O}_{ij}^{stick}$  describes  the action of the operator
        $\partial_\xi $  on  the expansion (\ref{tpsi}) for  the stream  function $\psi$ at $\xi=\xi_0$.
      In   Eq. (\ref{1way}) we have introduced two new definitions (underlined as $\underline{...}$), which are used   to  simplify
        the further derivations.

 \begin{figure}
    \hskip-0.5true cm
   \includegraphics[width=0.75\linewidth]{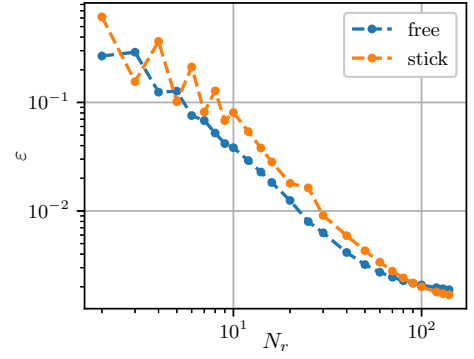}
   \caption{Illustration of the convergence of the expansion procedure for the stream function in dependence on the number of the basic functions.  Dependence of the relative error $\varepsilon$ in deviations  of norms $E[\{c_j\}]^{free}$, $E[\{c_j\}]^{stick}$ from zero   ($\hbox{Ma} = 1$, $\xi_0 = 0.1$). \label{er}}
    \end{figure}
       \begin{figure}
    \hskip-0.5true cm
    \hskip-0.0true cm
   \includegraphics[width=0.9\linewidth]{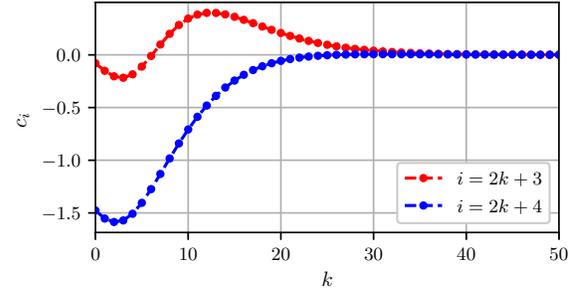}
  \caption{ Coefficients in decomposition of the stream function  $\psi$ for even and odd basic solutions
 ($\hbox{Ma} = 1$, $\xi_0 = 0.1$, $u_0 = 0.1$). Illustration of the convergence of the expansion of the solution for $\psi$ over  basic functions in dependence on their number $i$  as a function of a current integer  number $k$. Both even and odd basic solutions are shown; the convergence is reached for $k$ values about 25. \label{ci1}}
    \end{figure}
    \begin{figure}
    \hskip-0.5true cm
   \includegraphics[width=0.75\linewidth]{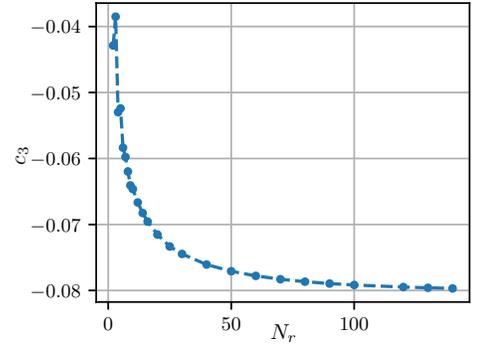}
  \caption{Convergence of the coefficient $c_3$ of the expansion of the final stream function over the basic functions in dependence on their number ( $\hbox{Ma} = 1$, $\xi_0 = 0.1$). \label{c3n}}
    \end{figure}
\begin{figure}
    \hskip-0.5true cm
   \includegraphics[width=1.05\linewidth]{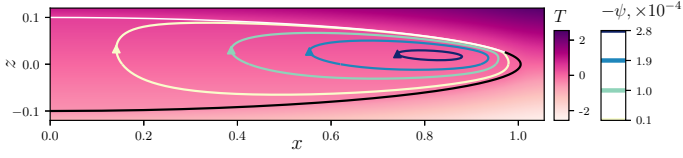}
                  \caption{Streamlines and temperature distribution for the generalized  thermocapillary convection within the drop
                   ($\xi_0 = 0.1$, $\kappa =0.2$,  $u_0=0.25$, Ma=1, $ N_r = 110$;  $\kappa = \varkappa_{air}/\varkappa$ is a relative heat conductivity); the color scale for the temperature is characterized by a more saturated light-purple color in a hot areas). The bold black line at the bottom part of the drop indicates that at this interface the sticking condition holds.  \label{sl1}}
                       \end{figure}
    \begin{figure}
    \hskip-0.5true cm
   \includegraphics[width=1.05\linewidth]{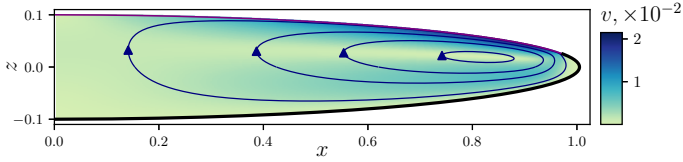}
    \caption{ Modulus of velocity for the generalized thermoconvection flow within the drop ($\xi_0 = 0.1$,
    $u_0=0.25$, $\kappa =0.2$,  Ma=1, $ N_r = 110$). The dimensional values of $v$ can be deduced using the corresponding scaling parameter
     $\chi/H = 4\cdot 10^{-3}$ m s$^{-1}$. The bold black line at the bottom part
of the drop indicates that at this interface the sticking condition holds.  The circulation period $\Delta t$ for certain velocities can be determined from the integral over the closed trajectory of the motion  as shown  in the Discussion section below. In dimensionless form $\Delta t$ values are 124, 151, 206 and 484 as counted off from the center of the vortex to its periphery, respectively (from dark streamline to light one in Fig. \ref{sl1}). The dimensional values of $\Delta t$ can be obtained using the corresponding scaling parameter $\Delta t = ({(H^2/2)}{(\xi_0\,\chi)^{-1}})\, \Delta {\widetilde{t}}$,  ${(H^2/2)}{(\xi_0\,\chi)^{-1}}$, which is about  0.0125 s for the typical geometrical and material characteristics of the drop.  In seconds they constitute: 1.55 s, 1.89 s, 2.575 s, 6.05 s, respectively. \label{sl1-1}}
    \end{figure}
      \begin{figure}
    \hskip-0.3true cm
   \includegraphics[width=1.05\linewidth]{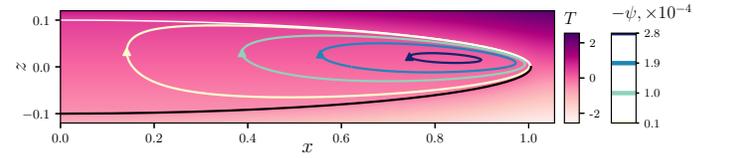}
   \caption{Streamlines   and temperature distribution (color map)  for the generalized movement of thermocapillary convection  ($\xi_0 = 0.1$,    $u_0=0$, $\kappa =0.2$, Ma=1, $ N_r = 110$).
     The bold black line at the bottom part
of the drop indicates that at this interface the sticking condition holds.  \label{sl2-1}}
    \end{figure}
 \begin{figure}
    \hskip-0.5true cm
   \includegraphics[width=1.05\linewidth]{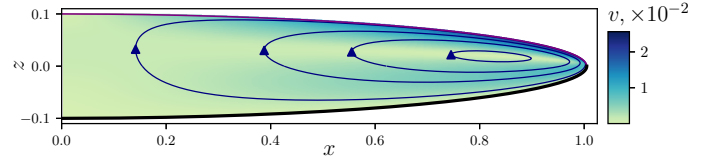}
    \caption{  Modulus of velocity for the generalized thermoconvection flow within the drop ($\xi_0 = 0.1$,
    $u_0=0$, Ma=1, $ N_r = 110$). The bold black line at the bottom part
of the drop indicates that at this interface the sticking condition holds. The circulation period $\Delta t$  for certain velocities in dimensionless form is 135, 156, 206 and 489 as counted off from the center of the vortex to its periphery (from dark streamline to light one  in Fig. \ref{sl2-1}). The dimensional values of $\Delta t$  constitute 1.69 s, 1.95 s, 2.58 s, 6.11 s, respectively.\label{sl2}}
    \end{figure}

We note that  the  irrationality $(\xi_0^2+u^2)^{3/2}$ is present in the  right part   of Eq. (\ref{1way}), i.e. in the ${r[u]}$, but this  irrationality is absent in the functions $\{\mathcal{F}_i[u]\}$ (see expression (\ref{mFn})), determining  the dependence of the left part of this equality on $u$.
This means, that the full solution of  the system  (\ref{1way}) is the infinite series. However,  we are able to obtain only  the finite approximation  for this solution.
The series is breaking  once the following convergence criterion is
satisfied:   the sum of the norms $E[\{c_j\}]^{free},  E[\{c_j\}]^{stick}$ of deviations  of the  equations (\ref{2way}) from zero,
  \begin{eqnarray}
   \hskip-0.5true cm  \begin{cases}
    E[\{c_j\}]^{free} = \int_{free} \frac{d{u}}{1-u^2} (\hat{O}_{ij}^{free}c_j\mathcal{F}_i[u] - r[u])^2
         \\
     = \hat{O}_{ij}^{free}\hat{O}^{f}_{km} c_j c_m \; \underbrace{\int_{free}\frac{d{u}}{1-u^2}\mathcal{F}_i\mathcal{F}_k}_{F_{ik}^{free}} \; \\ -\, 2\hat{O}_{ij}^{free}c_j\;\underbrace{\int_{free}\frac{d{u}}{1-u^2}r[u]\mathcal{F}_i[u]}_{R_i} +\,  \int_{free}\frac{(r[u])^2}{1-u^2}d{u} \ , \ \\
     E[\{c_j\}]^{stick} = \int_{stick}(\hat{O}_{ij}^{stick}c_j\mathcal{F}_i[u])^2 \frac{d{u}}{1-u^2}
            \\
      = \hat{O}_{ij}^{st}\hat{O}_{km}^{st} c_j c_m \hat{F}_{ik}^{st} \ . \
      \label{1wNorm}
 \end{cases}
             \end{eqnarray}
should  be minimal for the obtained $N_r$-measured set of $c_{i}$, i.e.
\begin{eqnarray}
   \partial_{c_{\alpha}} E[\{c_{j}\}]   = 2\underbrace{\hat{O}^{free}_{k\alpha}\hat{F}^{free}_{ki}\hat{O}^{free}_{ij}}_{M^{free}_{\alpha j}}\,
   c_j\,  - 2 \underbrace{\hat{O}_{i\alpha}^{free} R_i}_{V_\alpha}
        \nonumber  \\
        + 2 \underbrace{\hat{O}^{st}_{k\alpha}\hat{F}^{st}_{ki}\hat{O}^{st}_{ij}}_{M_{\alpha j}^{st}}\, c_j =  0 \,
    \,  \to (\hat{M}^{st}+\hat{M}^{free}) \vert{c}\rangle = \vert{V}\rangle \ , \
      \label{2wNorm}
\end{eqnarray}
where $ \vert{c}\rangle $ and $\vert{V}\rangle $  are  designations of the corresponding columns.
The above expressions (\ref{1wNorm}) and (\ref{2wNorm}) can be  essentially simplified with account to expression (\ref{Fik}).

          \smallskip

\begin{figure}
    \hskip-0.5true cm
    \includegraphics[width=0.95\linewidth]{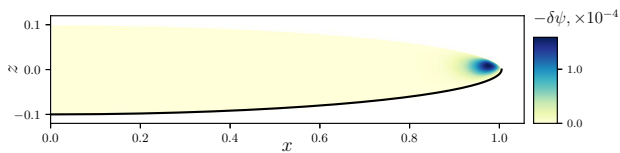}
    \caption{Difference  of streamlines $\psi\,$  for the generalized thermoconvection flow within the drop   for $u_0=0$ and $u_0=0.25$, ($\xi_0 = 0.1$, $\kappa =0.2$, Ma=1, $ N_r = 110$). The bold black line at the bottom part
of the drop indicates that at this interface the sticking condition holds. \label{slf1}}
    \end{figure}

Applying the above procedure  the main approximation for the stream functions with the given accuracy of  determination   are obtained
(relative deviations of norms $E[\{c_j\}]^{free}$, $E[\{c_j\}]^{stick}$ from zero are about $10^{-3}$), see Figs. \ref{er}--\ref{c3n}.
 Additionally, an improved representation of the velocities and temperature distribution corresponding to the stationary thermocapillary convection within the ellipsoidal drop in dependence on the values of $\xi_0$, $\kappa$ and  $u_0$ are calculated,  Figs. \ref{sl1}--\ref{sl2}.
The deviations $ E[\{c_j\}]^{free}$ and $E[\{c_j\}]^{stick}$ converge to zero when the number of the basis functions $N_r$ increases ($N_r\to\infty\,$, see Fig. \ref{er}).
It is important that the differences in the velocity distribution in the Marangoni vortex
 between the geometries with $u_0=0$ and $u_0 \ne 0$ (when  the bottom drop surface with the sticking boundary condition partly overlaps the circular edge of a drop)
 show up themselves only in a butt  end region of the drop, and do not affect the convection motion in the main volume of the drop, see Fig. \ref{slf1}.

 The second way  to obtain the full stream function is to use the set of functions $\psi_{i, st}$, see Eq. (\ref{tpsist}). It is possible for $u_0=0$. It is clear  that the functions $\psi_{i, st}$  automatically satisfy the sticking  boundary condition ($\partial_\xi \psi \, = 0 $) on the bottom surface of a drop - the second equation in a system of equations (\ref{bc241xi}).
Therefore, the only we need is to resolve the first equation in  a system (\ref{bc241xi}). The  Marangoni boundary condition  at the top (free)  surface reads
\begin{eqnarray}
     \big(\partial_\xi-\frac{1}{2\xi_0}(\xi_0^2+u^2)\partial_\xi^2 \big){\psi}\vert_{\xi=\xi_0}
            =\underline{\sum_{i,j}  \hat{O}_{ij}^{free} \mathcal{F}_i c_{j, st}}
      \nonumber \\
     = \,  -\, {\hbox{Ma}}\, \frac{(\xi_0^2+u^2)^{3/2}}{2\xi_0\sqrt{1+\xi_0^2}}\, \frac{(1-u^2)}{2}
     = \underline{r[u]} \, . \,
     \label{Mar2}
\end{eqnarray}
Similarly to that was done earlier, we substituted the expression (\ref{T0f}) for the temperature distribution $T_0$  to the right part of the first equation in  the system (\ref{bc241xi}).
Again,  we need to find the constants $c_{i, st}$ of expansion of the full stream function over basic functions $\psi_{i, st}$.
By analogy with the previous case
 we find the finite approximation  of this solution.
The series is broken  once the following convergence criterion is
satisfied:   the norm of deviation of the Eq. (\ref{Mar2}) from zero,
\begin{eqnarray}
    E[\{c_{j, st}\}] = \int \Big(\sum_{i,j}   \hat{O}_{ij}^{free} \mathcal{F}_i c_{j, st} - r(u)\Big)^2 \frac{d{u}}{1-u^2}  \ , \ \
  \label{Normst}
\end{eqnarray}
should be minimal for the  $N_r$-measured set of $c_{i, st}$,
(details of calculation of $ E[\{c_{j, st}\}]$ are presented in Appendix F).
The above deviation $ E[\{c_{j, st}\}]$ converges to zero when the number of the basic functions $N_r\to\infty $.
  The results obtained for the stream functions and velocity distributions within the drop by a second method of calculations are closely the same as shown in    Figs.  \ref{sl1}--\ref{sl2}.


\medskip

\centerline{\bf D. Stability of  the  stationary solution.}

 \centerline{\bf  Crossover to the limit of the plane fluid layer }

After the stationary solutions for the thermocapillary flows within isotropic fluid droplets in FSSF are determined
 $(\psi=\psi_1+\dots, T = T_0 + T_1 + \dots )$, the natural question about  stability of these solutions relative to the increase of the initial temperature gradient (i.e. increasing of the Marangoni number,  $\hbox{Ma}$) arises. To answer this question let us imagine that the hydrodynamic characteristics of the system in certain moment slightly deviate from those of the stationary solution. Our aim is to  trace the evolution of these deviations with time.

The partial solutions of the Eqs. (\ref{NSt2}-\ref{temp2a}) can be written  in the form of normal pertubations which have exponential  dependence on time \cite{Koschmieder1974,Gershuni1972,Landau6,Lebedev1993,Falkovich},
   \begin{eqnarray}
  {{v}}_{\mu} \propto\,\exp[\,\lambda\,t ]      \ , \
  \label{sol1}\\
 {T}_1 \propto\,\,\exp[\,\lambda\,t  ]      \ , \
  \label{sol2}
  \end{eqnarray}
   where exponent $\lambda\,$ determines the time character of perturbation evolution. The normal  perturbations with the negative sign of the real part of $\lambda\,$  are decaying, while the   perturbations with the positive sign of the real part of $\lambda\,$ correspond to the  growing fluid motions.
  Then, the stationary solutions for the thermocapillary motion are stable if the condition  for all solutions $Re[{\lambda}]<0$ is fullfilled. However, the spectrum of perturbations depends on the value of the Marangoni number, Ma.
While for a small Ma values all $\lambda_i$ possess the negative sign of the real part of lambda, starting from the some larger Ma the perturbations with the positive $Re[{\lambda}]$ arrive. Thus, the margin of stability of thermocapillary flow is determined by a minimal Marangoni number $\hbox{Ma}_c$ for which the normal perturbation $(\psi_c, T_c)$  reaches the zero   $Re[{\lambda}]$ value  for a first time.
To find the stability limit we introduce the general  expansions  $(\psi_\Sigma = \psi + \psi_c, T_\Sigma = T + T_c)$  and substitute them in  Eqs. (\ref{NSt2})--(\ref{temp2a}) in order to  make a linearization procedure  over $(\psi_c, T_c)$.
  We note that the equation  $\hat{E}^4 \psi_c = 0$ over $\psi$  is  linear initially,  thus, the basis of  the solution for $\psi_c$ remains the same. This means that we use the same  expansion $\psi_c = \sum_j c_{cj} \psi_{j, st}$  over  $\{ \psi_{j, st} \}$, which  is a set  of basic functions,  corresponding  to  the sticking  conditions at the bottom surface of a fluid drop, see Sec. IV A and Appendix F).

The  equation $\Delta T_\Sigma = \frac{c}{H}\, (\mathbf{v_\Sigma} \nabla )\, T_\Sigma$ for the temperature distribution after linearization takes the form
    \begin{eqnarray}
        \Delta T_c
                = \frac{c}{H}\, (\mathbf{v}_c \nabla)\, T_0   \ , \
         \label{Tc}
    \end{eqnarray}
    where the terms $\frac{c}{H}\,(\mathbf{v}_c  \nabla) \,T_1$ and $\frac{c}{H}\,(\mathbf{v}_1 \nabla)\, T_c$  are omitted due to a  higher order of smallness.
     Thus, for the case $\hbox{Ma} \ll 10^3$,  Eq. (\ref{Tc})
    is coincides with Eq. (\ref{T1}) which allows the temperature amendments of the first order to be calculated as a response to a set
     $\{c_{cj} \psi_{j, st}\}$; in such a way  we obtain an expansion $T_c = \sum_j c_{cj} T_{1\,j}$.

At the next step  it is necessary  to find  the critical  $\hbox{Ma}_c$ and  corresponding vector $\{c_{cj}\}$,
in order to satisfy Marangoni boundary condition (\ref{BCT0})  at $\xi=\xi_0$:
\begin{eqnarray}
  \Big\{(u^2  +  \xi^2)\, \partial^2_\xi \psi_c  \,-\,  2\, \xi \, \partial_\xi \psi_{c} \, \Big\}_{\xi=\xi_0}
 \qquad \qquad \nonumber \\
  =    -\,\hbox{Ma}\,  \Big\{\frac{(u^2  +  \xi^2)^{3/2}} {\sqrt{1  +  \xi^2 }}\, ({1 -  u^2\,})  \,\partial_{u} T_{c} \Big\}_{\xi=\xi_0}\ , \
\label{BCn}
\end{eqnarray}
where in accordance with  expression  (\ref{Tdr1}) and  with  account  to equality (\ref{duPn})  each $\partial_{u}T_{j}$ can be written as
\begin{eqnarray}
   \partial_{u}T_{j}  = -\, \sum_n \Big\{ \sum_k \big( W_{n,k}^{(1)} \big)_{j, st}\, \Xi_k^{(1)}\, + \big(\mathbf{W_n^{(1)}} \big)_{j, st}\,  \Xi_n^{(1)} \Big\}\,
   \nonumber \\
 \times\,  \frac{n(n+1)}{1-u^2}\mathcal{F}_n[u]\, . \ \qquad \qquad
\label{Tcj}
\end{eqnarray}
For simplification we designate the right part of the equality (\ref{BCn}) as $\hbox{Ma}_c \sum_j \,r_j[u]\, c_{cj}$.
In such a way the equation (\ref{BCn}) can be rewritten  as

\begin{figure}
    \hskip-0.5true cm
    \includegraphics[width=0.65\linewidth]{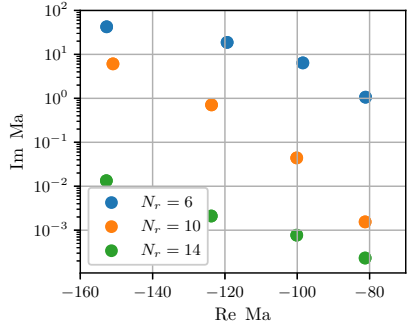}
  \caption{Illustration of the fact, that image part of the Marangoni number, Ma, decreases with the increase of the number $N_r$ of the accounted basic functions (four lowest values of Ma spectrum for each $N_r$ are presented, $\xi_0=0.1$, $\kappa = 0.2$).
    \label{MaSp}}
    \end{figure}

 \begin{widetext}

\begin{equation}
    (\partial_\xi-\frac{1}{2\xi_0}(\xi_0^2+u^2)\partial_\xi^2 ){\psi}_c\vert_{\xi=\xi_0}
          = \,\sum_{i,j}  \hat{O}_{ij}^{free} \mathcal{F}_i  \, c_{cj} = \hbox{Ma}_c \sum_j \,r_j[u] \, c_{cj} \ . \
      \label{eqMa}
\end{equation}

Similarly to the previous section, we  are searching for the solutions,  for which  the norm of deviation from the equation  (\ref{eqMa})  turns to zero
\begin{eqnarray}
    \mathcal{L} =\sum_{jm} c_{c j} c_{c m}  \Bigg[\underbrace{\sum_{i,l} \int \hat{O}_{ij}^{free} \mathcal{F}_i \hat{O}_{lm}^{free} \mathcal{F}_l\,\frac{d{u}}{1-u^2}}_{\hat{M}^{(0)}_{jm}} - \, \hbox{Ma}_c\, \underbrace{\int \Big\{\sum_{i} \hat{O}_{ij}^{free} \mathcal{F}_i  \, r_m[u]  + \sum_{l} \hat{O}_{lj}^{free} \mathcal{F}_l r_j[u] \Big\}\,\frac{d{u}}{1-u^2}}_{\hat{M}^{(1)}_{jm}}\;
    \nonumber \\
    +  \, \hbox{Ma}_c^{2} \underbrace{\int r_j[u] r_m[u] \frac{d{u}}{1-u^2}}_{\hat{M}^{(2)}_{jm}}\Bigg] \,= \,0\ . \qquad \qquad \qquad
     \label{qwF}
\end{eqnarray}
\end{widetext}
 The vector $ \{c_{c j}\} $ and  the corresponding  minimal critical value  \hbox{Ma}$_c$
   can be determined from the equation (\ref{qwF}) for $\mathcal{L}$.
This equation can be rewritten in  the following form
\begin{equation}
{(\hat{M}^{(0)}} - \hbox{Ma}_c \,{\hat{M}^{(1)}} + \hbox{Ma}_c^{2}\,  {\hat{M}^{(2)}}) \; \vert{c_c}\rangle = 0 \ , \
\label{QEP}
\end{equation}
where  the components of  matrixes $\hat{M}^{(0)}, \hat{M}^{(1)}, \hat{M}^{(2)}$  are determined in expression (\ref{qwF}) as   the  interlinear (footnote) designations.
Thus, our problem is reduced to the quadratic eigenvalue problem \cite{QEP}. The
Standard method of its  solution
  is a reduction to the generalized eigenvalue problem (see Appendix G).
 However, there is a certain complication in its solution: the obtained Ma, are complex numbers, containing real and imaginary parts, due to the irrationality in the left part of Eq. (\ref{BCn}).
  To overcome the above problem we have used one of the properties of our system lying in the fact that an increase of the number of the basic functions $N_r$ leads to the diminishing of the image part of Ma$_c$; it
  turns to zero when $N_r\to\,\infty\,$. In practice,  we are searching for the solution depending on the number $N_r$ of  the basic  functions, for which the image part of Ma would be less than $10^{-3}$. The illustration of the  progress  in  these calculations are shown in Fig. \ref{MaSp}.
The negative
value of the critical Marangoni number Ma$_c$ is not accidental. This means  that
for the case under consideration -- the temperature of the
free drop surface is higher than that at the bottom surface with the
sticking (no-slip) conditions, the stationary thermocapillary convection
is stable.

      \begin{figure}
    \hskip-0.5true cm
    \includegraphics[width=0.75\linewidth]{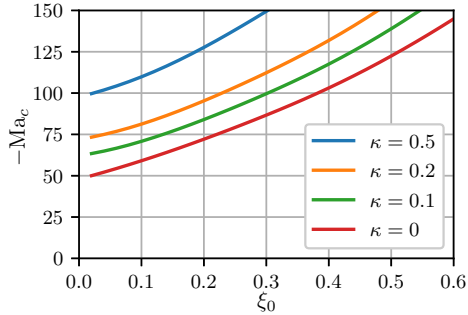}
  \caption{Phase diagram for the critical Marangoni number Ma$_c$ as a function of the ellipticity ratio $\xi_0$ for different values of the relative heat conductivity $\kappa\,$.
    \label{MaPh}}
    \end{figure}

  \begin{figure}
    \hskip-0.5true cm
    \includegraphics[width=1.05\linewidth]{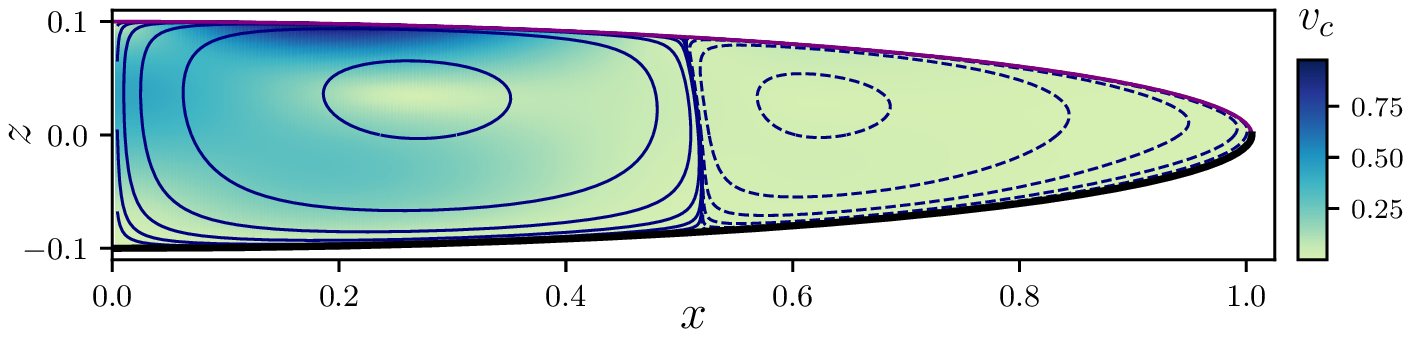}
  \caption{Critical  Marangoni flows in the ellipsoidal drop ($\xi_0=0.1$, $\kappa = 0.2$, Ma$_c=-81.2$). The dashed and solid lines indicate the opposite direction of the fluid velocity in the neighboring vortexes. The bold black line at the bottom part
of the drop indicates that at this interface the sticking condition holds. \label{lim1}}
    \end{figure}
    \begin{figure}     \hskip-1.02true cm
    \includegraphics[width=1.1\linewidth]{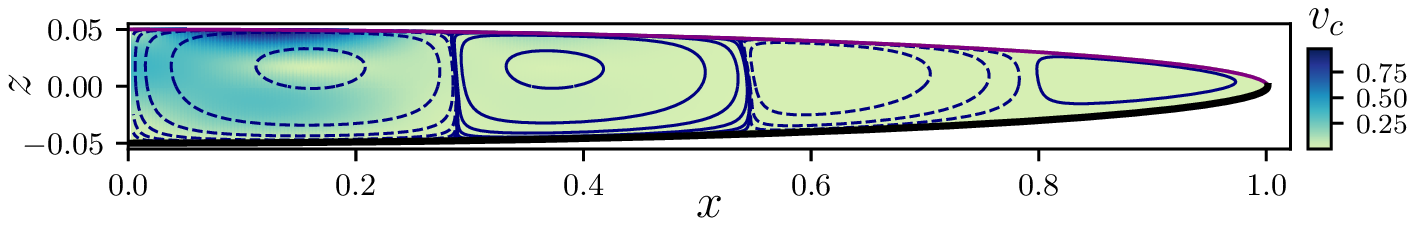}
  \caption{  Critical  Marangoni flows in the ellipsoidal drop ($ \xi_0=0.05$, $\kappa =0.2$, Ma$_c=-75.9$). The dashed and solid lines indicate the opposite direction of the fluid velocity in the neighboring vortexes.  The bold black line at the bottom part
of the drop indicates that at this interface the sticking condition holds. \label{lim2}}
    \end{figure}
           \begin{figure}     \hskip-0.8true cm
    \includegraphics[width=1.1\linewidth]{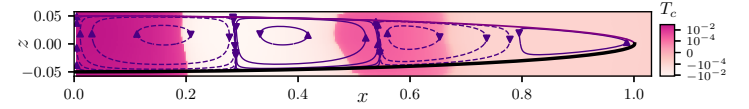}
  \caption{ Streamlines   and temperature distribution (the color scale for the
temperature is characterized by a more saturated light-purple color in hot areas)  for the  critical Marangoni flows in the ellipsoidal drop ($\xi_0=0.05$, $\kappa =0.2$, Ma$_c=-75.9$).  The dashed and solid lines indicate the opposite direction of the fluid velocity in the neighboring vortexes.  The bold black line at the bottom part
of the drop indicates that at this interface the sticking condition holds.\label{lim3}}
    \end{figure}

  As was already mentioned in  Introduction, the vast  amount of papers devoted to Marangoni convection is focused on a rather simple case of a flat fluid films. Our preceding paper on thermocapillary convection within isotropic droplets in FSSF exploited the same approximation.  The formalism of the stream functions which we successfully applied to study the Marangoni convection in ellipsoidal drops provides a unique possibility to investigate a crossover from the lens-like drop to a flat fluid layer. This is made by reducing of the droplet ellipticity ratio $\xi_0$ that leads in the limit $\xi_0 \to 0$ to the case of a flat fluid film.
 The phase diagram for Ma$_c$ as a function of the ellipticity parameter $\xi_0$ for different values of the relative heat conductivity  $\kappa $ is shown  in  Fig. \ref{MaPh}.
The modulus of the obtained  value Ma$_c$  $ \approx - 50 $  for $\kappa \to 0 $ matches well with the
 case  Ma$_c$  $ \sim 50 $ described in the literature  for a flat fluid film placed on a heat-insulating hard substrate with the  sticking conditions  \cite{Gershuni1972}.
 The results of  calculations of  streamlines and velocities for the  negative Ma$_c$ values are presented in  Figs. \ref{lim1}--\ref{lim3}.
 According to our calculations with diminishing of the ellipticity ratio $\xi_0$ the amount of vortices in the direction of the axial drop cross section progressively increases, see Figs. \ref{lim1}--\ref{lim3}. The number of basic functions, which is necessary for an accurate convergence of the calculation procedure, is estimated as $N_r \sim \xi_0^{- 1}$.
These calculations are in good agreement with the results of our preceding paper \cite{Pikina2021}, where the formation of about 6 convection cells (rolls) along the lateral drop size was predicted.

\smallskip

 \centerline{\bf E. Numerical experiment  }

\smallskip
To get further insight about Marangoni flows within ellipsoidal isotropic droplets embedded in FSSF we obtained the numerical results which took the real shape of the drops and their material and transport properties into account. The details of the numerical experiment are presented in Supporting Information.
In short, to simulate the thermocapilary flow within the drops we used the cylindrical coordinates $(r, \phi, z)$. The maximum droplet radius in the horizontal plane and the maximum droplet height in the vertical plane are indicated as $R_{in}$ and $h_d$, respectively.  In the problem under consideration, the transfer of mass and heat does not depend on the angular coordinate $\phi$. This allows us to consider the hydrodynamic problem as a two-dimensional and proceed with the numerical calculations in the coordinates $(r, z)$. In such formulation,
the geometry of the lens-like droplet is described by an ellipse with a semi-major axis $a = R_{in}$, and a semi-minor axis $b = 0.5 h_d$ (see, Suppl. Information).

The hydrodynamic flows in the droplet  are described as follows. The Navier-Stokes and the continuity equations for the incompressible fluid, as well as the heat transfer equation  are written in the cylindrical coordinates. The stream function $\psi$ in this case satisfies the relations $\partial \psi/ \partial z = r u$ and $\partial \psi/ \partial r = -r v$, where $u$ and $v$ are  the horizontal and the vertical components of the liquid flow velocity, respectively. The vorticity is introduced in cylindrical coordinates as $\omega= \partial u/ \partial z - \partial v/ \partial r$. At the end, the mathematical model consists of three equations which are solved with respect to three variables:  $\omega$, $\psi$  and temperature $T$. The above equations are accompanied by a set of initial and boundary conditions, also written in cylindrical coordinates.

To solve numerically the Marangoni convection problem the commercial package FlexPDE Professional Version 7.18/W64 3D was used \cite{Liu2018}. The mathematical algorithm is based on the Galerkin finite element method with application of the modified iterative Newton-Raphson method \cite{Yamamoto2006}. The time intervals in the program are generated automatically in order to minimize the calculation error. To secure the solution reliability the special attention to the mesh convergence was paid (see, Suppl. Information).

 The calculations were performed for the time $t_\mathrm{max}=20\, t_\mathrm{rel}$, Ma$=1$ and correspond to the stationary regime ($t_{rel}$ is the heat relaxation time in the air due to the thermal conductivity, see, Suppl. Information). This time is enough for the system to reach the stationary state. The later is achieved due to a fact that the thermal conductivity dominates over the convective heat transfer for the considered values of the material and geometric parameters, of the drops and their environment. According to  numerical calculations, the maximum velocity of the convective transfer is $v_\mathrm{max}\approx 10^{-5}$ m/s. Then the convective transfer time can be evaluated as $t_\mathrm{conv}=R_\mathrm{in}/v_\mathrm{max}\approx 10$ s. To estimate the heat transfer time determined by thermal conductivity we use the value of the fluid thermal conductivity $\chi_l=\kappa_l/ (c_l \rho_l)\approx 4\times 10^{-8}$ m$^2$/s to obtain the corresponding time $t_\mathrm{cond}=R_\mathrm{in} h_d/ \chi_l\approx 0.05$ s. It is readily seen that $t_\mathrm{cond} \ll t_\mathrm{conv}$, thus confirming our initial claim of predominance of the thermal conductivity. The results of the numerical calculations presented below are obtained for the time $t=t_\mathrm{max}$.

\begin{figure}[!htb] \includegraphics[width=0.6\linewidth]{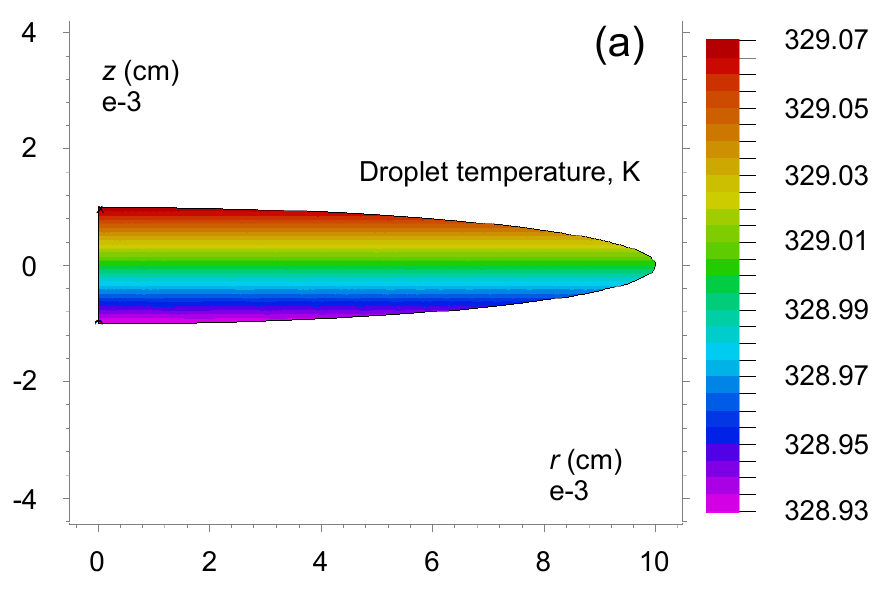} \\

\includegraphics[width=0.6\linewidth]{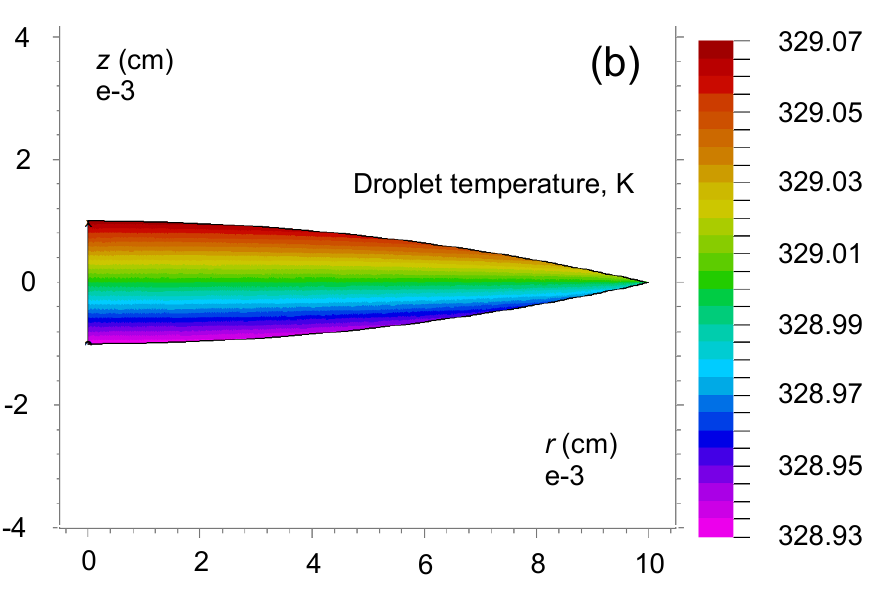}
\includegraphics[width=0.6\linewidth]{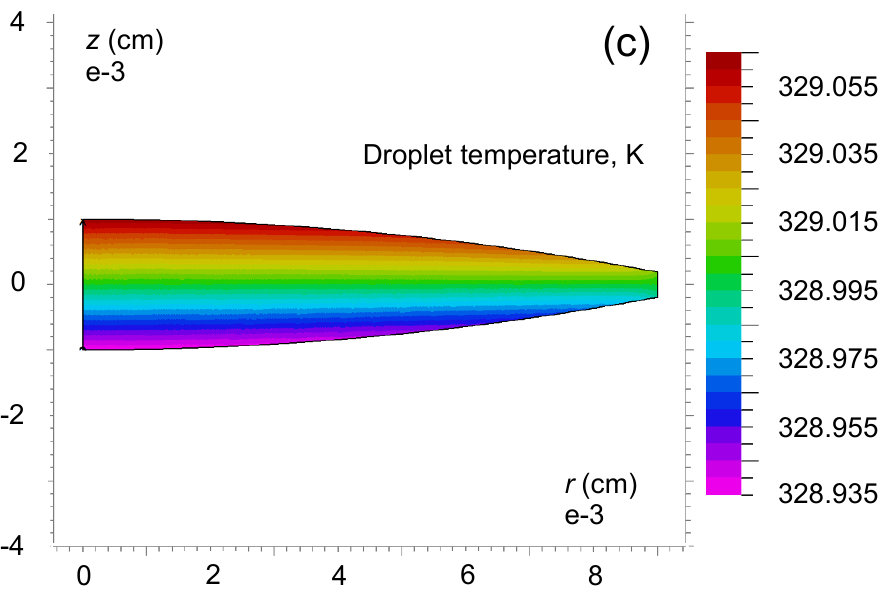} \\ \caption{\label{DrTemp} Temperature distribution within the ellipsoidal isotropic droplets;  the smectic shell  is in  contact with the bottom surface of  isotropic droplet (a, c, e). The droplet shapes:  ellipse (a), a biconvex lens (c), and a lens with the cut end at the edge (e).} \end{figure}

\begin{figure}[!htb] \includegraphics[width=0.7\linewidth]{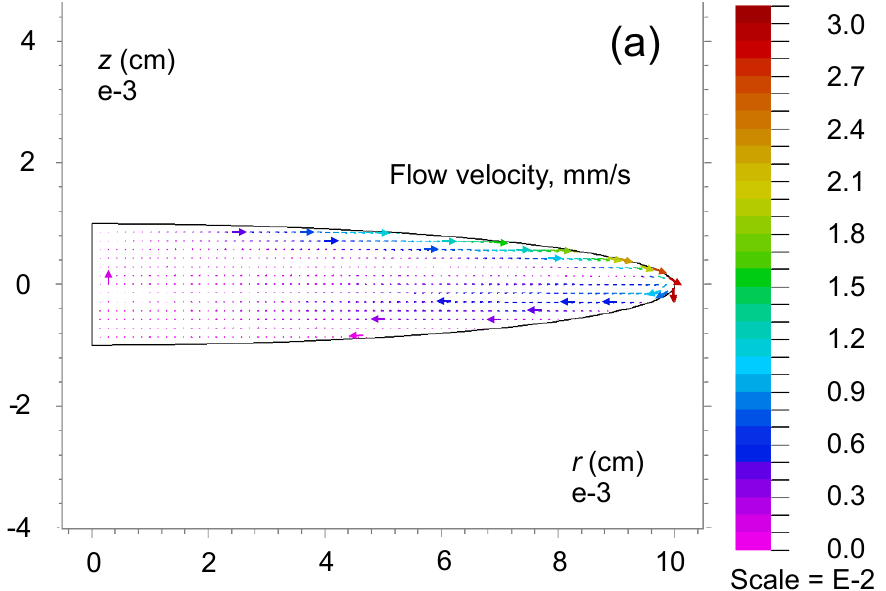} \\

\includegraphics[width=0.7\linewidth]{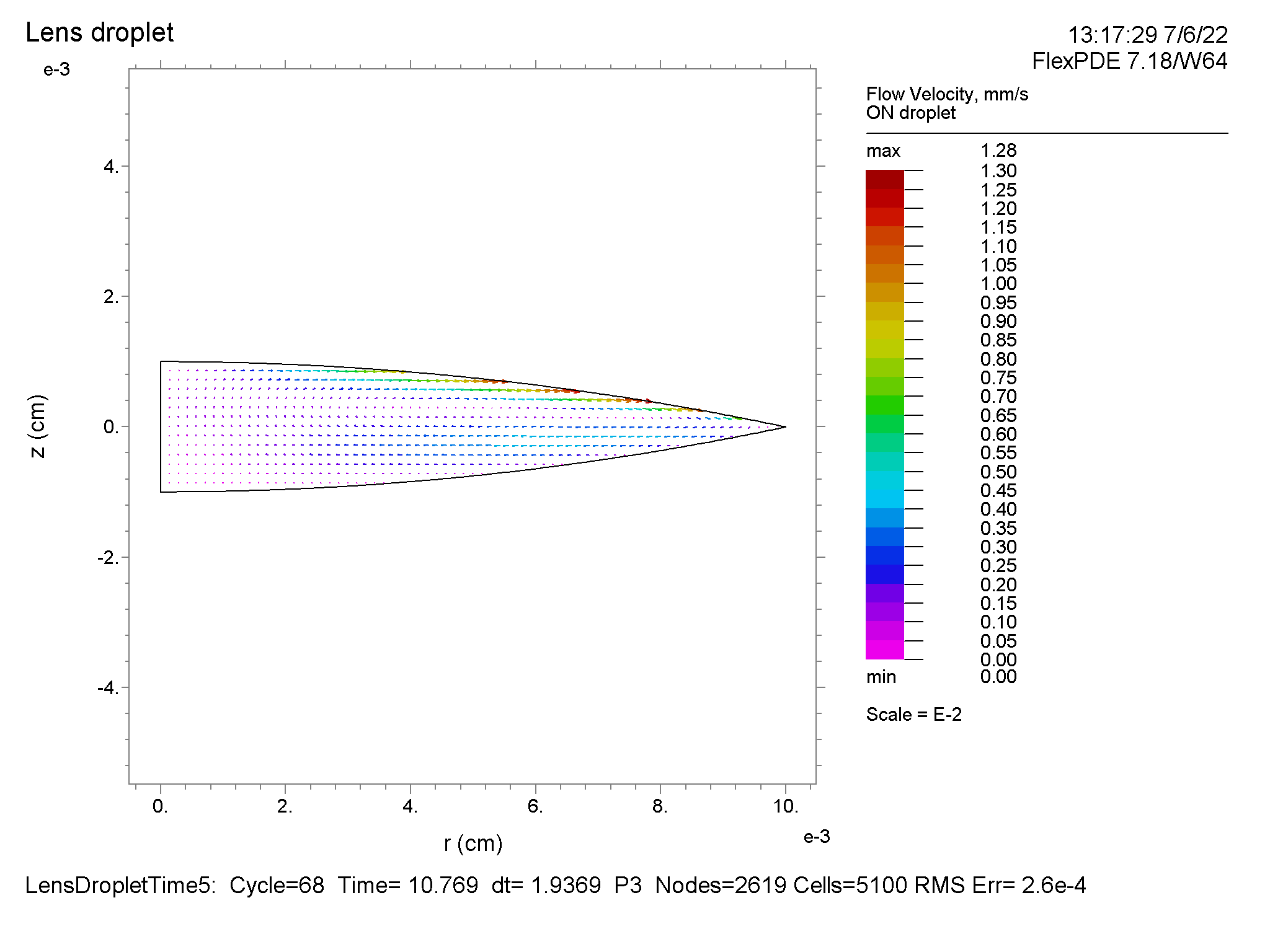}
\includegraphics[width=0.7\linewidth]{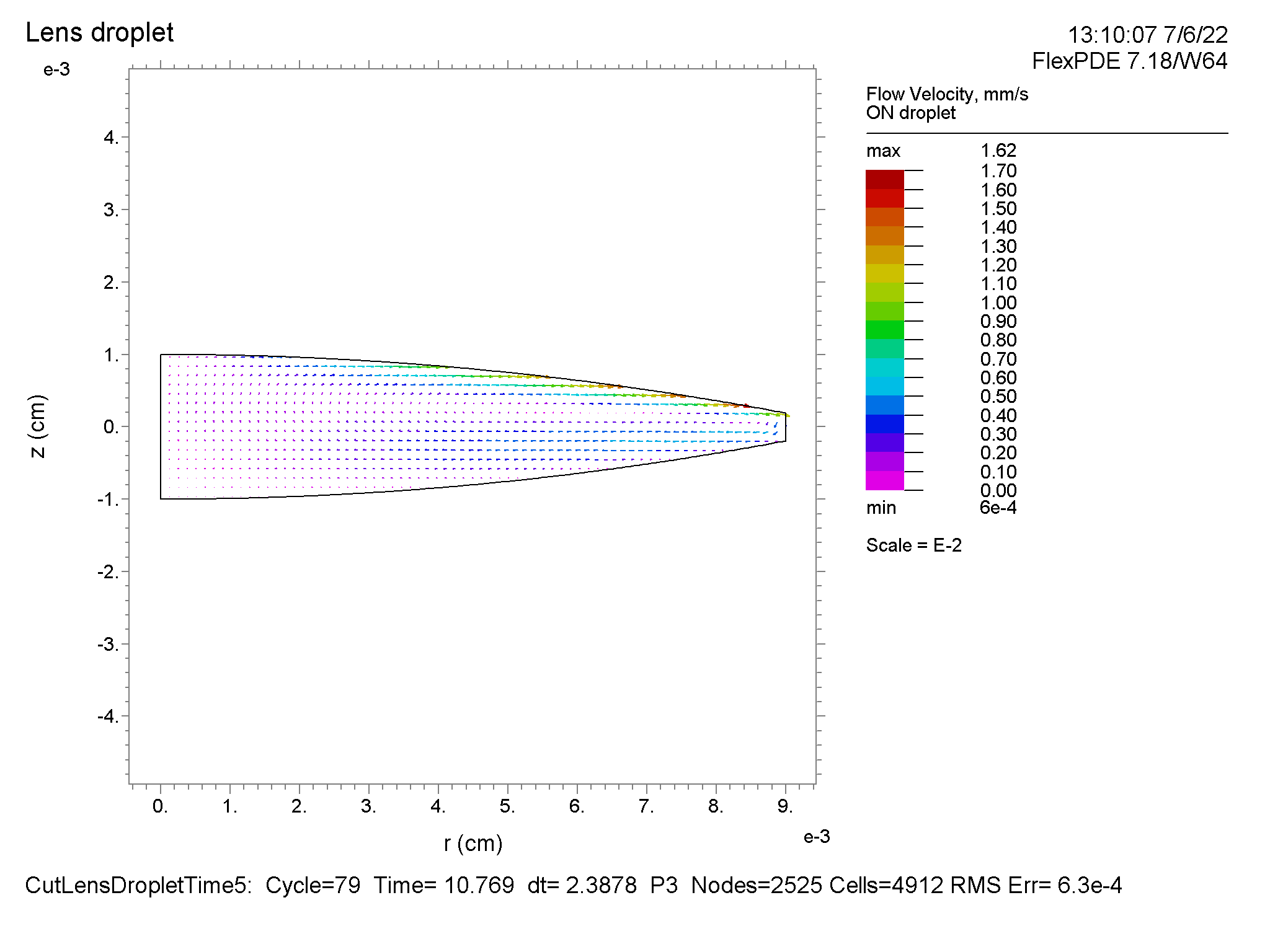} \\ \caption{\label{fig:FlowVel} Distribution of the fluid flow velocity in within the ellipsoidal droplets; the smectic shell  is in  contact with the bottom free surface of isotropic droplet (a, b, c). The droplet shapes:  ellipse (a), biconvex lens (b), and a lens with the cut end at the edge (c).}
\end{figure}

In our numerical calculations we considered the case for which the smectic shell (substrate) is in contact with the lower surface of isotropic droplet, while the upper interface is free, Figs. \ref{Figure1}, \ref{FigSt}. This situation is realized for $T_\mathrm{up}>T_\mathrm{dn}$ (see Sec. II B), so  the values $T_\mathrm{up}=334$ K and $T_\mathrm{dn}= 324$ K were chosen for further calculations. By default, the Marangoni boundary condition (Suppl. Information,  Eq. (S.7)) was used for the free (upper) surface of the droplet. Contrary to this, the boundary no-slip (sticking) condition (Suppl. Information, Eq. (S.8)) was used for the case the smectic shell (substrate) was in immediate contact with the lower surface of a drop and for the cut end of the fluid lens.
 The corresponding temperature distribution is shown in Fig. \ref{DrTemp}. In accordance with our calculations the shape of the droplet does not show any significant effect on the temperature distribution in the droplet for the values of the parameters used.
 In Fig.~\ref{fig:FlowVel} the distribution of the fluid flow velocity in droplets is shown for different types of droplet shapes.

\begin{widetext}

\begin{figure}[!htb]		
\includegraphics[width=0.4\linewidth]{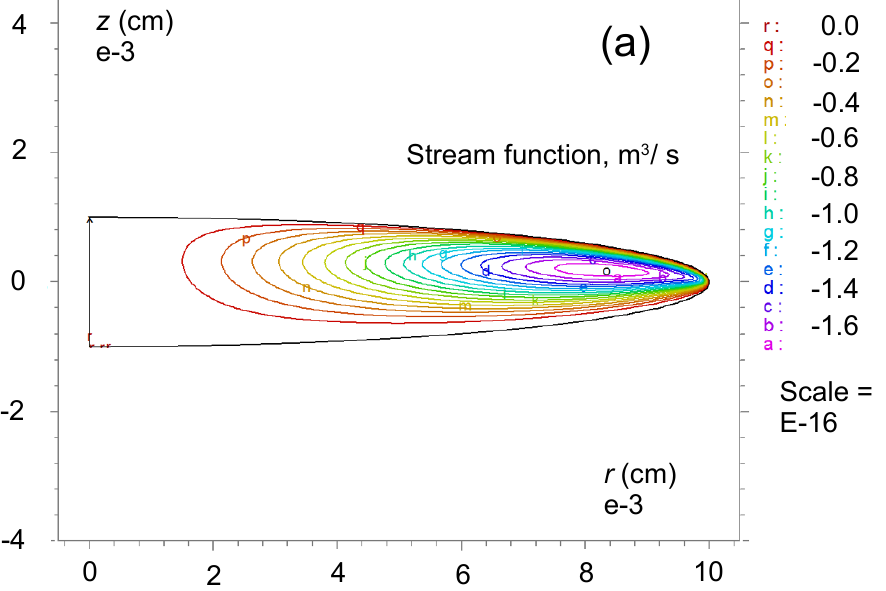}
\includegraphics[width=0.4\linewidth]{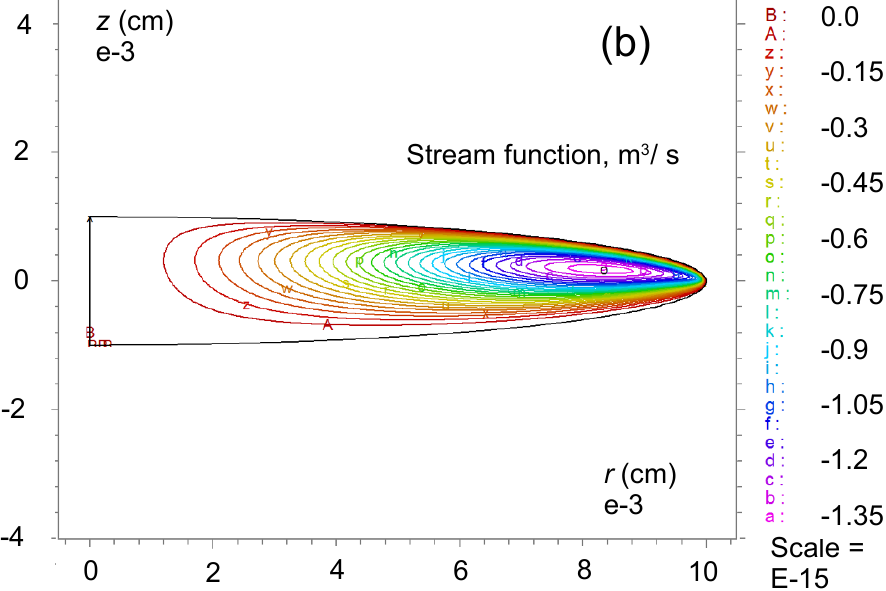}\\

\includegraphics[width=0.4\linewidth]{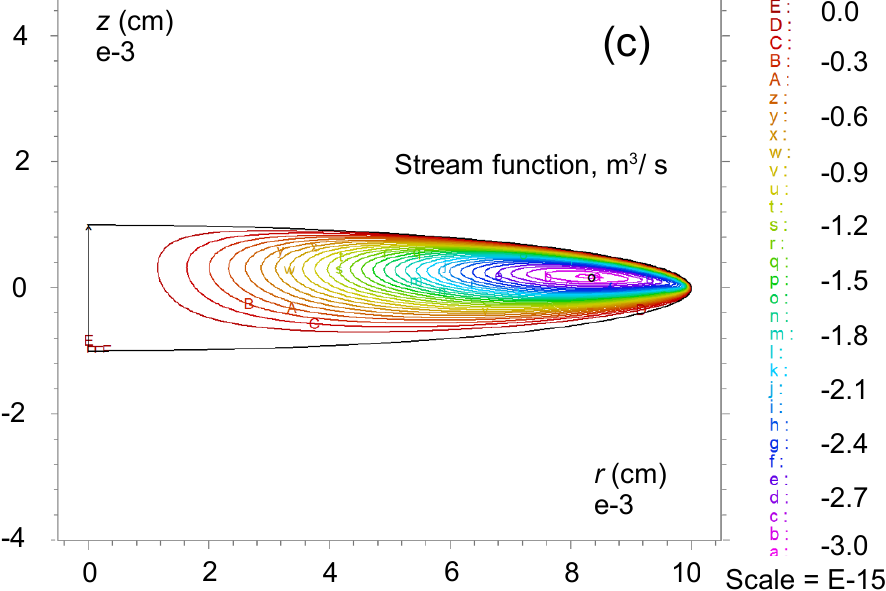}
		\caption{\label{StrF1}
			Stream functions for the ellipsoidal drop with a smectic film on the bottom free surface:  (a) $t=0.1t_\mathrm{rel}$, (b) $t=0.5t_\mathrm{rel}$  and (c) $t=20t_\mathrm{rel}$.}
	\end{figure}

   \end{widetext}

According to Fig.  \ref{DrTemp},   the temperature decreases along the upper  (hot) surface of the droplet and  conversely increases at its  lower (cold) surface when going in the direction from the symmetry axis to the edge of the droplet along the radial coordinate $r$.  Again, there are no noticeable quantative differences in the temperature distribution along the radius $r$ for the different shapes of the droplet.

The fluid flow circulating in the $(r,z)$ plane is directed clockwise in the case under consideration (Fig. \ref{fig:FlowVel}). Because  $T_{up} > T_{dn}$, the Marangoni flow is directed along the free surface of the droplet from the hot area to the cold area, (i.e.  from the area of the low surface tension to the area of the high surface tension).  It is important that the no-slip (sticking) condition at the boundary between the smectic and the isotropic liquid slows down the thermocapillary flow.

The corresponding stream functions are presented in Figs. \ref{StrF1} for several consecutive time points. There are no qualitative differences in the plots of the stream function $\psi$ between different time periods. We observe one axially symmetric vortex in the ellipsoidal drop that is in agreement with our analytical results,  Figs. \ref{sl1}--\ref{sl2}. The quantitative differences are due to  a fact that the flow velocity gradually increases until it reaches a stationary state. We conclude that numerical results for Marangoni convection within ellipsoidal droplets in FSSF are in good accordance with that obtained by analytical methods in Sec. IV A - D.

\section{ DISCUSSION}

We have developed a hydrodynamic theory of the Marangoni flow in the axially-symmetric ellipsoidal fluid droplets on the basis of the formalism of the Stokes stream functions. This approach was applied to ellipsoidal isotropic drops spontaneously formed in overheated FSSF and to droplets of insoluble fluids (of the type of oil or glycerol) deposited on it. The asymmetric geometry, for which the upper drop interface is connected with the air, while it's bottom surface is in contact with the static smectic  layering  was considered. This situation can be realized experimentally when the temperature of the upper side of the film is higher than at the lower one.   Due to the nonuniform temperature distribution the tangential Marangoni force always exists at the free drop surface. This leads to a fluid flow along its curved interface, which is possible for the arbitrarily small Marangoni numbers. The thermocapillary flow occurs along the free surface of the droplet from the hot area to the cold one, leading to the formation of the individual torroidal-like vortices within the drop. Our calculations indicate that the no-slip (sticking) condition at the boundary between the smectic and the isotropic fluid slows down the circulatory Marangoni flow.

There is another point related to the implementation of the no-slip (sticking) boundary conditions at the interface between the fluid and the smectic substrate. In Sec. II B we have shown that for the model of the fluid drop and its environment used in our theoretical analysis, the tangential component of the smectic elastic force compensates the Marangoni force at the fluid-smectic interface. This hinders effectively the flow of the smectic material and leads to a sticking of fluid motion at the border with a smectic shell. In principle, there is another possibility for the smectic motion - so called, permeation, i.e. the flow of the material through the smectic layers \cite{Lebedev1993,deGennes93}, which is usually disregarded due to its low velocity. Our estimations indicate that this is indeed the case; in the limit of the small value of the permeation constant $\lambda_p \simeq 10^{-16}$ m$^2$ Pa$^{-1}$ s$^{-1}$  the permeation velocity can be written as $v_{perm} \sim  {\mathbf{D}}_T\,A\,T^{-1}$  \cite{Lebedev1993,deGennes93}, where ${\mathbf{D}}_T$ and $A$ are the thermodiffusion coefficient and the temperature gradient across the drop,respectively. Using the typical values  ${\mathbf{D}}_T\sim\,10^{-10}$ m$^2$s$^{-1}$, $A = 10^4$ K m$^{-1}$ and $T = 300$ K we obtain $v_{perm} \sim \,10^{-9}$ m s$^{-1}$. According to our calculations the maximum velocity of the convection flow is $v_{max} \sim \,10^{-5}\div\,10^{-4}$ m s$^{-1}$.  Thus,  $v_{perm} \ll v_{max}$,  that confirms our initial assertion that permeation process in smectics is too slow.  We conclude that permeation in smectics can not provide the Marangoni transport at the fluid-smectic interface justifying our assumption of the sticking conditions at this interface.

One of the direction of our research is a study of the stability of the stationary solutions for the thermocapillary convection within ellipsoidal fluid droplets, Sec. IV D. According to our results the obtained stationary solutions for the fluid drop with the sticking boundary conditions at the bottom interface and $T_{dn} < T_{up}$ are stable.  This result remains valid  upon crossover from the ellipsoidal droplets to a flat fluid layer. Such a crossover can be made within the formalism of the stream functions by reducing of the droplet ellipticity ratio $\xi_0$ to zero value.
However, the stability analysis  indicated that the system starts to behave differently for  the opposite direction of temperature gradient,  $T_{dn} > T_{up}$.
We have shown that the critical thermocapillary motion (with the positive Ma$_c$  values) develops both  in the ellipsoidal drop and in a flat liquid layer only when the hot fluid volume from the bottom surface with the sticking properties flows in the direction of the cold free surface. In another words, only when
$T_{dn} > T_{up}$  the critical deviations from the stationary dependencies of the stream functions and temperature distribution both  in  the drops and  in the flat layers evolve (compare with \cite{Gershuni1972,Pikina2020}).

The predictions of our hydrodynamic theory can be checked using various experimental set ups. The  experiments can be carried out both in the laboratory at the earth and under microgravity conditions at the International Space Station. The geometry of the experiment with the ellipsoidal fluid drops embedded in smectic films with asymmetric boundary conditions implies that the heat transfer occurs from the hot plate positioned at the upper side of the drop to the cold one placed at its bottom side. This corresponds to the positive direction of the temperature gradient ($T_{up}  >  T_{dn}$), Fig. \ref{FigSt}, and ensures the buoyancy effects related to the convection in the surrounding air are absent.

  The most important quantitative characteristic of the circulating Marangoni flow in the drops that can be directly measured in experiment is the time period, $\Delta t$, i.e. the time interval required for the movement along the closed stream line. The time period can be defined as
\begin{equation}
\Delta t = \, \oint \frac{d l}{v} \ , \
\label{Dt}
\end{equation}
where $v$ is a velocity modulus along the trajectory of the flow and $dl$ is a tangential element of the curved trajectory, which is determined as
$dl_i = \sqrt{(x_{i+1} - x_i)^2  + (z_{i+1} - z_i)^2 }$, where the  set of points $\{(x_i, z_i)\}$, determines the velocity distribution within the drops and, accordingly, the streamlines. The magnitude of the circulation period $\Delta  t$
crucially depends on the values of the Marangoni number, Ma and the drop ellipticity ratio. Because the flow velocity increases upon the Ma increase, the corresponding time period delta $\Delta  t$  diminishes. On another hand, for the fixed values of Ma (i.e. of the  temperature gradient across the drop) the length of the flattened circular trajectory increases upon the ellipticity ratio $\xi_0$ decrease. This leads to increase of the period of circulation $\Delta  t$. The calculated values of $\Delta  t$ for the typical geometrical and material \cite{par} parameters of the fluid drop are shown in the captions to Figs. \ref{sl1}, \ref{sl2-1}. For example, for the velocity patterns shown in Fig. \ref{sl2-1}  the periods of circulation constitute
1.69 s, 1.95 s, 2.575 s and 6.11 s counted off from the center of vortex to its periphery, respectively.
 These time intervals are pretty large and can be registered by tracking the circulatory movement of the properly selected tracers within the drop.

\section{ SUMMARY}

In this work, we present a quantitative description of the Marangoni flows in ellipsoidal isotropic droplets of different origin embedded in free standing smectic films. The convection inside the ellipsoidal fluid drops appears very different from the classical Marangoni convection in the systems with a simple flat geometry. In contrast to the flat fluid films, the mechanical equilibrium within drops is absent due to their curved shape. Because of the nonuniform temperature distribution the tangential Marangoni force activates a fluid flow along its curved interface, making the thermocapillary flow within the drop thresholdless. To describe the vortex formation in the ellipsoidal isotropic droplets we generalized the method of the Stokes stream functions to the case of the curved fluid interfaces. It was shown that the general solution for the stream function can be represented as a sum over the limited amount of the basic functions, which  satisfy the boundary conditions of the problem and reflect the properties of the real physical fields. Moreover, we developed the original operator method for the solution of the  differential equations for the stream functions.

In general, the basic stream functions satisfy  the symmetry of the problem and the  condition of the absence of the fluid flow through the external drop boundary. Formally this corresponds to the symmetrical case. In parallel,  we developed the straight method of the obtaining of the basic set of the stream functions describing the thermocapillary flow in the drop for the asymmetric boundary conditions. In this  case the upper drop interface is connected with the air, while the bottom surface is in contact with the static smectic layers -- so called, sticking or no-slip boundary conditions. To solve such a problem we represented the stream functions and the corresponding velocity components as a set of odd and even functions. The idea was to combine the pairs of such functions to generate the basic set of the stream functions satisfying the sticking conditions at the bottom boundary of the drop.
It is important that the basic stream functions (velocities) derived in this  way   provide a continuous variation through the points of contact between the free and bounded by the smectic layers surfaces of the drop. At the next step, we derived the distribution of the temperature deviations inside the drop and in the ambient air. This allowed us, first, to resolve the Marangony boundary condition, and then to find the general solution for the stream functions and flow velocities, describing the stationary thermocapillary convection inside the drop with account to the actual temperature field within it.
As a result,   the general stream function and velocity fields, as well as the temperature distribution within the ellipsoidal drops, were derived in the stationary regime for the fixed Marangoni numbers as a function of the droplet ellipticity ratio, and for the different values of the heat conductivity of the liquid crystal and air. Additionally, the numerical hydrodynamic calculations of the thermocapillary motion in the ellipsoidal drops with asymmetric boundary conditions were carried out. Both the analytical and numerical simulations describe the axially-symmetric circulatory convection flow induced by the thermocapillary effect at the droplet free surface.

Finally, we note that the developed theory of Marangoni flow in droplets is quite general and thus applies to a wide variety of thermocapillary convection problems in fluid drops of  ellipsoidal form. As the first and foremost task we consider the Marangoni flows in isotropic ellipsoidal droplets suspended on the circular frame. The mechanical stability of such drops is determined by the sticking conditions at the solid bounding frame. Both the isotropic phase of various liquid crystal compounds, as well as the simple liquids of the type of glycerol or silicone oil can be considered for the experimental and theoretical investigations. These droplets have the shape of the spherical segments (circular flat lenses), the height of which can be varied relative to their lateral dimension by changing of the amount of the material. As a second problem we indicate the ellipsoidal nematic droplets spontaneously formed in overheated FSSF.  For this case, the FSSF of appropriate material should be heated above the bulk smectic-nematic transition.  There are also examples of the formation of the fluid droplets of anisometric shape in various colloidal suspensions and among anisotropic fluids placed on the substrate with an ultra low wetting properties. Of special interest are also the thermocapillary processes in phospholipid membranes with various fluid inclusions, which can mimic the reaction of the cell membranes to the small temperature gradients.

\section*{Acknowledgments}
We are grateful to Vladimir V. Lebedev, Efim I. Kats,  Igor  V. Kolokolov,  Sergey S. Vergeles for  fruitful discussions.
We acknowledge support from the Russian Science Foundation (Grant No. 18-12-00108, general theory of Marangoni convection in isotropic drops embedded in free standing films and corresponding  numerical experiments).
The work on the derivation of the  stress tensor and expressions for the tangential forces in ellipsoidal coordinates  and the elaboration of the thermocapillary experiments  was supported by the Ministry of Science and Higher Education within the corresponding State assignments of  FSRC "Crystallography and Photonics" RAS.
The work on the statement of problem and the solving of the problem of the temperature distribution within the ellipsoidal isotropic drops   was supported by the Ministry of Science and Higher Education within the corresponding State assignments N. 0029-2019-0003.

\medskip

\section*{Authorship contribution statement}
E.S.Pikina: conceived of the presented idea, calculated  the Marangoni convection, solved of the problem of the temperature distribution, discussed the results, final manuscript writing.
 M.A. Shishkin: calculated  the Marangoni convection, developed the original operator method for calculation of the stream functions, discussed the results.
 K.S. Kolegov: made the numerical experiment, discussed the  results.
 B.I. Ostrovskii: conceived of the presented idea,  presentation of the results of the calculations, the elaboration of the thermocapillary experiments,  discussed the  results, final manuscript writing.
 S.A. Pikin: conceived of the presented idea, worked on the derivation of the  stress tensor and expressions for the tangential forces in ellipsoidal coordinates, contributed to the calculations, discussed the  results.

 All authors read and agreed on the final text of the paper.

\medskip

  \numberwithin{equation} {section} { \bf
\appendixname{}}

\appendix

\section{Derivation of the basic functions of  Laplace equation  }
 \label{sec:A1}

The  Laplace equation for the temperature distribution   $T$ within the drop in  the oblate spheroid coordinates $\xi, u, \varphi$ reads:
\begin{eqnarray}
   \Delta T  = \frac{1}{h_\xi h_u h_\varphi}\Big(
    \partial_\xi \frac{h_u h_\varphi}{h_\xi}\partial_\xi +
    \partial_u \frac{h_\xi h_\varphi}{h_u}\partial_u
    \nonumber \\
    + \,      \partial_\varphi \frac{h_\xi h_u}{h_\varphi}\partial_\varphi \Big) \, T
         =
    \frac{c}{h_\xi h_u h_\varphi}\Big(
    \partial_\xi (1+\xi^2)\partial_\xi
    \nonumber \\
  +\, \partial_u (1-u^2)\partial_u   + \,  (\xi^2+u^2)\partial_\varphi^2\Big) \, T = 0 \ . \quad
    \label{lapl}
\end{eqnarray}
To solve this equation the separation of the variables is used: $T[\xi, u, \varphi] = \Xi[\xi]U[u]\Phi[\varphi]\,$   \cite{NLebedev65,NLebedev652,Happel}, where
 $\Xi[\xi], U[u], \Phi[\varphi]$  are the  functions of one single variable $\xi, u,$ or $\varphi$, respectively; $\Phi = e^{im\varphi}$ for $m\in \mathbb{Z}$ due to continuity on $\varphi \in \mathbb{T}^1$.
Due to an axial symmetry of the system  $m=0$ , and for the  function  $U$ we obtain the Legendre`s equation. As a result, the constant of separation is $n(n+1)$ and $U=U_n = P_n[u]$, where $P_n[u]$ is a  Legendre polynomial of the first kind (the temperature  $T$  is supposed to be regular for all $u$).
In turn,  for  $\Xi$, we obtain the equation:
\begin{equation}
    \partial_\xi(1+\xi^2)\partial_\xi  \Xi_n - n(n+1) \Xi_n =  0 \ , \
    \label{Xieq}
\end{equation}
 the first solution  of which, $\Xi_n^{(1)}[\xi]$,  is the  Legendre polynomial of the first kind of the imaginary argument  $P_n[I\xi]$ \cite{NLebedev65}.
 The exclusion of the imaginary part leads to the simple transformation rules:
\begin{eqnarray}
    \Xi^{(1)}_{2n}[\xi] \,= \, P_{2n}[i\xi] \ , \qquad
   \label{lei1}  \\
    \Xi^{(1)}_{2n+1}[\xi] \, = \, (-i)\cdot P_{2n+1}[i\xi]  \ . \
    \label{lei2}
\end{eqnarray}
On the basis of  these rules, we can use for $\Xi_n^{(1)}[\xi]$ the transformed recurrent relations for  Legendre polynomials with the  argument  $z=i\xi\,$,
in  particular
\begin{eqnarray}
   (1 -  z^2)\,\frac{d P_n[z]}{d z}  =  n\, P_{n-1}[z] \, - \,n \, z\, P_n[z] \ . \
    \label{lei3}
\end{eqnarray}
The later can be transformed to
\begin{eqnarray}
   (1 +  \xi^2)\,\frac{d \Xi^{(1)}_{2n}[\xi]}{d \xi}  =  \, -\,2\,n\, \Xi^{(1)}_{2n-1}[\xi] \, +\, 2 \, n \, \xi\, \Xi^{(1)}_{2n}[\xi]\ , \qquad
    \label{lei4}
\end{eqnarray}
\begin{eqnarray}
   (1 +  \xi^2)\,\frac{d \Xi^{(1)}_{2n +1}[\xi]}{d \xi}  = \, (2n +1)\, \Xi^{(1)}_{2n}[\xi]
   \nonumber \\
   \, +\,  \,(2n +1) \, \xi\, \Xi^{(1)}_{2n +1 }[\xi] \  \ . \quad
    \label{lei5}
\end{eqnarray}
The equation (\ref{lei5}) is a certain  representation of  the Legendre polynomial of the second kind.
By disposing of the imaginary unit, we obtain directly:
$\Xi_0^{(1)}[\xi] = 1,\; \Xi_0^{(2)}[\xi] = \arctan[\xi]$, $\Xi_1^{(1)}[\xi] = \xi, \; \Xi_1^{(2)}[\xi] = \xi\arctan[\xi] + 1$; other  solutions for $\Xi_n $ ($n>2$) obtained via the recurrent relation
\begin{equation}
\Xi_{n+1} = (-1)^n \frac{2n+1}{n+1}\,\xi\, \Xi_n - \frac{n}{n+1}\,\Xi_{n-1} \ . \
\label{Xin}
\end{equation}

Using the recurrent relation  (\ref{Xin}) we can write the  successive expressions of $\Xi_n[\xi]$ for various $n$:
\begin{eqnarray}
  \Xi_0^{(1)}[\xi] = 1 \ , \ \Xi_0^{(2)}[\xi] = \arctan[\xi]   \ , \
    \nonumber \\
  \Xi_1^{(1)}[\xi] = \xi \ , \ \Xi_1^{(2)}[\xi] = \xi\arctan[\xi] + 1  \ , \
    \nonumber \\
 \Xi_2^{(1)}[\xi] =  - \frac{1}{2} -  \frac{3 \xi^2}{2}
 \ , \
    \nonumber \\
   \Xi_2^{(2)}[\xi]= - \frac{\arctan[\xi]}{2}   -
   \frac{3}{2} \xi \big(1 + \xi\arctan[\xi]\big)
 \ , \
    \nonumber \\
  \Xi_3^{(1)}[\xi] =  -\frac{2 \xi}{3} -
   \frac{5}{3}\, \xi \,\Big(\frac{1}{2} + \frac{3 \xi^2}{2}\Big)
    \ , \quad
           \label{lapl1b}
\end{eqnarray}
and etc.

 To find the temperature  distribution  in the air it is convenient to use the
 linear combinations of
$\Xi_n^{(1,2)}$, damped at $\xi\to+\infty$,  designated below as $\Xi_n^{(a)}$ :
\begin{eqnarray}
\Xi_0^{(a)}  = - \arctan[\xi] + \frac{\pi}{2} \ , \
    \nonumber \\
\Xi_1^{(a)}  =  \xi \arctan[\xi] - \frac{\pi \xi}{2} -  1 \ , \
    \nonumber \\
\Xi_2^{(a)}  =  \frac{3 \xi \left(2 \xi \arctan[\xi] - \pi \xi + 2\right)}{4} + \frac{\arctan[\xi]}{2} - \frac{\pi}{4} \ , \
    \nonumber \\
\Xi_3^{(a)}  =  \frac{5 \xi^{3} \arctan[\xi]}{2} - \frac{5 \pi \xi^{3}}{4} + \frac{5 \xi^{2}}{2} + \frac{3 \xi \arctan[\xi]}{2}
    \nonumber \\
 - \frac{3 \pi \xi}{4} + \frac{2}{3} \ , \qquad
        \quad
\label{lapl2b}
    \end{eqnarray}
and etc.

\bigskip

\section{Ellipsoidal coordinates and differentiation of unit vectors. Boundary conditions  }
 \label{sec:A4}

  \begin{figure}
    \hskip-0.5true cm
   \includegraphics[width=0.7\linewidth]{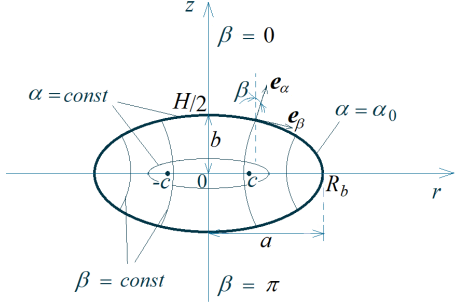}
    \caption{ The parameters characterizing the ellipsoidal drop shape  are represented in discussed oblate spheroid coordinates.   Semiaxes of  ellipsoid, $a=c\cosh[\alpha_0]\equiv R_b$, $b=c\sinh[\alpha_0]\equiv H/2$, $\tanh[\alpha_0]=b/a\ll\,1$ (the latter inequality is possible only for $\alpha_0 < 1$), $c$   is a focus distance (coordinate of the focal point). Here ${\bf e}_{\alpha}$ ,   ${\bf e}_{\beta}$   are the unit vectors in the  discussed oblate spheroidal coordinates in their meridional plane. Note that ${\bf e}_{\alpha}$ is outward normal vector to the oblate spheroidal surface of constant ${\alpha}=\alpha_0$,  unit vector ${\bf e}_{\varphi}$ is the  azimuthal unit vector, oriented beyond the page (sheet) plane, ${\bf e}_{\beta}$ lies in the tangent plane to the oblate spheroid surface and completes the right-handed basis set.
      \label{FigAp02}}
    \end{figure}

There are various methods of introducing of the ellipsoidal coordinates. The conventional approach consists in implementation  of orthogonal coordinates
$\alpha,\beta, \varphi\,$ which are useful to solve certain problems.  The every point of space is   described by a triple of numbers ($\alpha,\beta, \varphi\,$),  which correspond to an unique point in the Cartesian coordinates $(x,y,z)$.
The corresponding
orthogonal   system of surfaces consists of
 oblate spheroids formed by the surfaces of constant  $\alpha\,$ ($\alpha=\alpha_0\,$  is the spheroid of the given boundary),
    one-sheeted  hyperboloids of revolution of constant  $\beta$, and planes of azimuthal angle  $\varphi\,=$ const, \cite{NLebedev65,Happel}, (compare with  Fig. \ref{Figure02}).
 These parameters are related  to the rectangular coordinates by the  following matrix representation \cite{NLebedev65,NLebedev652,Happel}:
\begin{eqnarray}
  \left(\begin{array}{c}
 x \\ y \\ z
 \end{array} \right)
\, =
\,  \left(\begin{array}{c}
 \,c\,\cosh[\alpha]\,\sin[\beta]\,\cos[\varphi]\\
c\,\cosh[\alpha]\,\sin[\beta]\,\sin[\varphi] \\
 c\,\sinh[\alpha]\,\cos[\beta]\,
 \end{array} \right) \  \
 \label{1elc}
 \end{eqnarray}
where the focus distance $c$ plays a role of a scale parameter and
\begin{eqnarray}
0\, \le \alpha < \,\infty\, , \ 0\, \le \beta \le \, \pi \, , \
  -\,\pi\, < \varphi \le \pi \,\, \qquad \qquad  . \  \ \label{1elc1}
\end{eqnarray}
 The corresponding metric coefficients are:
\begin{eqnarray}
h_{\alpha}^2 =  \Big(\frac{\partial x}{\partial \alpha} \Big)^2 + \Big(\frac{\partial y}{\partial \alpha} \Big)^2 + \Big(\frac{\partial z}{\partial \alpha} \Big)^2 \ , \
 \nonumber \\
h_{\beta}^2 =  \Big(\frac{\partial x}{\partial \beta} \Big)^2 + \Big(\frac{\partial y}{\partial \beta} \Big)^2 + \Big(\frac{\partial z}{\partial \beta} \Big)^2 \ , \
 \nonumber \\
h_{\varphi}^2 =  \Big(\frac{\partial x}{\partial \varphi} \Big)^2 + \Big(\frac{\partial y}{\partial \varphi} \Big)^2 + \Big(\frac{\partial z}{\partial \varphi} \Big)^2 \ , \
 \nonumber \\
h_{\alpha} = h_{\beta}  = c\,\sqrt{\cosh^2[\alpha] - \sin^2[\beta]} \, , \nonumber \\
h_{\varphi} = c\,\cosh[\alpha]\,\sin[\beta]\, \qquad \qquad  .
 \  \ \label{1metr}
\end{eqnarray}

To solve the differential equations for the stream functions and to simplify the corresponding boundary conditions   we prefer to use  somewhat different representation of the orthogonal  oblate spheroidal coordinates:  $\,\xi\, = \sinh[\alpha]$,
$u=\cos[\beta] , \, d\beta = -du/\sin[\beta]$ and $\varphi$. The corresponds to
       new oblate spheroidal coordinates $u, \xi,  \varphi\,$ and new right-hand triple of unit vectors
     $({\bf e}_{u}, {\bf e}_{\xi}, {\bf e}_{\varphi})$.  This new representation is related to rectangular coordinates by the  following expressions \cite{Happel}:
\begin{eqnarray}
x  = c\,\sqrt{1+\xi^2}\,\sqrt{1-u^2}\,\cos[\varphi] \, ,\qquad \qquad \ \nonumber \\
  y  = c\,\sqrt{1+\xi^2}\,\sqrt{1-u^2}\,\sin[\varphi] \, ,\qquad \qquad \ \nonumber \\ \  z = c\,u\,\xi\,   . \  \qquad \qquad \ \label{1elc}
\end{eqnarray}
 In turn,
             \begin{eqnarray}
   h_\alpha =   h_\xi \,\sqrt{1+\xi^2} \ , \
    h_\beta  =  h_u\,\sqrt{1-u^2} \ ,   \nonumber \\
    h_\varphi = c\,\sqrt{1+\xi^2}\sqrt{1-u^2}
      \  . \ \label{a1Lamexi}
\end{eqnarray}

  Let us  define also the transformation relation for the differential operator $\nabla$
\begin{eqnarray}
 e^i_{\mu}  \partial_i  \,=\, \frac{1}{h_{\mu}}  \partial_{\mu}
      \ , \  \label{der}
\end{eqnarray}
where    indices $i$, $\mu\,$  correspond  to Cartesian  orthogonal coordinates and to  oblate spheroidal coordinates (\ref{1metr}), (\ref{1elc}), respectively; $e^i_{\mu}$ is the $\mu$-component  of  the  unit vector  in  Cartesian  coordinates.

Following Happel and  Brenner, \cite{Happel},  we write the components of the unit vectors  in oblate spheroidal coordinates and define the rule of differentiation of these  unit vectors
   \begin{eqnarray}
     {\bf e}_{\mu} \, = \,\frac{1}{h_{\mu}} \partial_{\mu} x_i {\bf e}_i
             \, . \ \label{bc5}
      \end{eqnarray}

For certain derivations we need to know the projections of the unit vectors  on the z-axis:
\begin{eqnarray}
(e_z, \hat{\xi}) = \frac{1}{h_\xi} \partial_{\xi}{z} = \frac{c\, u}{h_\xi} \  , \ \label{ez1}\\
(e_z, \hat{u}) = \frac{1}{h_u} \partial_{u}{z} = \frac{c\,\xi}{h_u} \  , \ \label{ez2}
\end{eqnarray}
and on the radial axis:
\begin{eqnarray}
(e_x, \hat{\xi}) = \frac{1}{h_\xi} \partial_{\xi}{x} =\frac{\xi \sqrt{1-u^2}}{\sqrt{\xi^2+u^2}}\,\cos[\varphi] \  , \ \label{ez3} \\
(e_x, \hat{u}) = \frac{1}{h_u} \partial_{u}{x} = \frac{-u\sqrt{1+\xi^2}}{\sqrt{\xi^2+u^2}}\,\cos[\varphi] \ .  \ \label{ez4}
\end{eqnarray}

       The hybrid boundary conditions   for the balance of tangential forces  are given by  the expressions
          \begin{eqnarray}
       \hat{\sigma}_{\mu\,\alpha}\,n_{\alpha} = e^k_{\mu} e^i_{\alpha}\hat{\sigma}_{i k}\,n_{\alpha}
        \nonumber \\
       =\eta_{vi} \, e^k_{\mu} e^i_{\alpha}\, \Big(\partial_i ( e^k_{\nu} v_{\nu})  + \partial_k ( e^i_{\nu} v_{\nu})  \Big)\,n_{\alpha}
        \nonumber \\
        = \, \eta_{vi} \,
       \Big\{ e^i_{\mu}  \frac{\partial_{\alpha}  }{h_{\alpha}} \big( e^i_{\nu} v_{\nu} \big)  +  e^k_{\alpha}  \frac{\partial_{\mu}  }{h_{\mu}} \big( e^k_{\nu} v_{\nu} \big)  \Big\}\,n_{\alpha}
     \nonumber \\
        =  \frac{1}{h_{\mu}}\,\partial_{\mu} \gamma \ ({\hbox{ top surface}}   )
             \, , \ \label{abc4f}
      \end{eqnarray}
       \begin{eqnarray}
       \hat{\sigma}_{\mu\,\alpha}\,n_{\alpha} = 0  \ ({\hbox{ bottom surface}}  )
             \, . \ \label{abc42f}
      \end{eqnarray}
       Because  $e^i_{\mu} e^i_{\alpha}=\delta _{\mu \alpha}$, $e^i_{\mu} e^k_{\mu}=\delta _{i k}$
     and  taking Eq. (\ref{der}) into account, the components of the unit vectors $e^i_{\alpha}$  in  oblate spheroidal coordinates are calculated using Eq. (\ref{bc5}).

     After substitution of the variables  $\xi, u$  to Eqs. (\ref{abc4f}) and (\ref{abc42f}) one obtains
     \begin{eqnarray}
       \sigma^{u\xi} = \eta h_\xi h_u \left[
    \frac{1}{h_u^2}\partial_u \frac{v_\xi}{h_\xi}+
    \frac{1}{h_\xi^2}\partial_\xi \frac{v_u}{h_u}
    \right]   \nonumber \\
     =
    \eta \left[ \frac{\partial_u v_\xi}{h_u}+
    \frac{\partial_\xi v_u}{h_\xi} - \frac{v_u}{h_\xi}\frac{\xi}{\xi^2+u^2} \ , \
    \right]
     \nonumber \\
     = \, \frac{\partial_u \gamma}{h_u} \qquad ({\hbox{ top surface}}   ) \ , \
     \nonumber \\
     \sigma^{u\xi} = 0 \qquad   \ ({\hbox{ bottom surface}}  )
     \ , \
     \nonumber \\
     \sigma^{\varphi \xi} = \eta h_\xi h_\varphi \left[
    \frac{1}{h_\varphi^2}\partial_\varphi \frac{v_\xi}{h_\xi}+
    \frac{1}{h_\xi^2}\partial_\xi \frac{v_\varphi}{h_\varphi}
    \right] = 0 \ . \
    \label{sigmaxi}
\end{eqnarray}
    Taking into account that at the free boundary  $v_\xi = 0$ (at $\xi=\xi_0$),   we obtain
       \begin{eqnarray}
       \mathbf{v} =  \, - \frac{{\bf e}_{u}}{h_\xi h_\varphi}\partial_\xi \psi \, +\, \frac{{\bf e}_{\xi}}{h_u h_\varphi}\partial_u \psi \ , \
       \label{abfu}
              \end{eqnarray}
              where ${\bf e}_{\xi}$ and ${\bf e}_{u}$ - are the unit vectors along the $\xi$ and $u$ axis, respectively,
       i.e.
         \begin{eqnarray}
         v_u =  \frac{1}{c^2\,\sqrt{u^2  +  \xi^2} \sqrt{1 - u^2}}\partial_\xi \psi \ . \
         \label{vu}
              \end{eqnarray}

   It is convenient also to rewrite the continuity equation  using  the variables:
    \begin{eqnarray}
   (\nabla {\mathbf v}) = \frac{1}{\sqrt{1+\xi^2}\sqrt{1-u^2}\,h_\xi h_u h_\varphi}\,
   \nonumber \\
   \times \,
    \Big(
   \sqrt{1+\xi^2}\, \partial_\xi (h_u \sqrt{1-u^2}\, h_\varphi v_\xi)
   \nonumber \\
    + \sqrt{1-u^2} \,\partial_u (h_\xi \sqrt{1+\xi^2} \,h_\varphi v_u) \Big)
    \nonumber \\
     =\,\frac{1}{h_\xi h_u h_\varphi}\,\Big(
   \partial_\xi (h_u  h_\varphi v_\xi)
       + \partial_u (h_\xi h_\varphi v_u) \Big)
                 = \, 0 \  \, . \ \label{aContxi}
             \end{eqnarray}

\bigskip

\begin{widetext}
\section{Some algebraic relations.}
\label{sec:A5}

We derive here some   algebraic relations which are useful in calculation of the $\psi\,$-functions.
\begin{equation}
    \mathcal{F}_n = \int_{-1}^{u}P_n[u']\,d{u'} = \frac{P_{n+1}-P_{n-1}}{2n+1} \ . \
    \label{defF}
\end{equation}

After multiplying of the both sides of Eq. (\ref{defF})  by $u$ we obtain:
\begin{equation}
    u \mathcal{F}_n = \frac{n+2}{2n+1}\mathcal{F}_{n+1} + \frac{n-1}{2n+1}\mathcal{F}_{n-1} \ . \
\end{equation}

After  doing this for a second time we have
\begin{equation}
    u^2 \mathcal{F}_n = \frac{(n+2)(n+3)}{(2n+1)(2n+3)}\mathcal{F}_{n+2} +
    \left[\frac{n(n+2)}{(2n+1)(2n+3)} + \frac{(n-1)(n+1)}{(2n+1)(2n-1)}\right]\mathcal{F}_{n}+
    \frac{(n-1)(n-2)}{(2n+1)(2n-1)}\mathcal{F}_{n-2} \ . \
     \label{Fnre}
\end{equation}

For the derivation of the temperature distribution within a drop it is useful to know the following equation:
\begin{equation}
    u \partial_u \mathcal{F}_n = u P_n = \frac{n+1}{2n+1} P_{n+1} + \frac{n}{2n+1}P_{n-1} \ . \
    \label{uduPn}
\end{equation}

To calculate the certain  boundary conditions the following equations might be useful:
\begin{equation}
    F_{ik} = \int_{-1}^{1} \mathcal{F}_i \mathcal{F}_k \frac{d{u}}{1-u^2} = \delta_{ik} \frac{2}{i(i+1)(2i+1)} \ , \
    \label{Fik}
\end{equation}
\begin{equation}
    \partial_u P_n = \frac{n}{1-u^2}(P_{n-1} - u P_n) = \frac{n}{1-u^2}\frac{n+1}{2n+1}(P_{n-1} -  P_{n+1}) =  - \frac{n(n+1)}{1-u^2}\mathcal{F}_n
    \ . \
\label{duPn}
\end{equation}

\section{Derivation of Streamfuctions}
\label{sec:A6}

We introduce here a number of operators which considerably simplify the calculations of the stream functions.
\begin{align}
    \hat{P} &= \partial_u (1-u^2)\partial_u, \; \hat{P}P_n = -n(n+1)P_n  \ , \
    \label{E1}
    \\
    \hat{\mathcal{F}} &= (1-u^2)\partial_u^2, \; \hat{\mathcal{F}}\mathcal{F}_n =  -n(n+1) \mathcal{F}_n \ , \
    \label{E2}
    \\
    \hat{\Xi} &= \partial_\xi (1+\xi^2)\partial_\xi, \;
    \hat{\Xi}\Xi_n = n(n+1)\Xi_n \ , \
    \label{E3}  \\
    \hat{\mathcal{X}} &= (1+\xi^2)\partial_\xi^2, \;
    \hat{\mathcal{X}} \mathcal{X}_n = n(n+1)\mathcal{X}_n \ , \
    \label{E4}
\end{align}
where the functions $\mathcal{X}_n $ are introduced similarly to $\mathcal{F}_n$,
\begin{equation}
    \mathcal{X}_n = \frac{\Xi_{n+1}-\Xi_{n-1}}{2n+1} \ . \
\label{eqXn}
\end{equation}

Since  $\psi(u=\pm1) = 0$,  any stream function  can be expanded over a set of $\mathcal{F}$, i.e.
 $\psi = \sum_n \mathcal{F}_n[u] g_n[\xi]$, where $g_n[\xi]$ is an unknown function of $\psi$.
The operator $\hat{E}^2$ acts on the monom $\mathcal{F}_n g_n$ as
 \begin{equation}
    \hat{E}^2 \mathcal{F}_n g_n = \frac{\hat{\mathcal{X}} + \hat{\mathcal{F}}}{\xi^2+u^2}(\mathcal{F}_n g_n) =\frac{1}{\xi^2+u^2} \mathcal{F}_n \underbrace{(\hat{\mathcal{X}}g_n -n(n+1)g_n)}_{G_n} \ , \
\end{equation}
where we define an unknown function $G_n[\xi]$. Let us  note that the kernel of the operator $\hat{E}^2$ is  $\{\mathcal{X}_n \mathcal{F}_n\}$ (see Eqs. (\ref{E2}, \ref{E4}).
In turn,  for  the action of the operator $\hat{E}^4$ on the monom $\mathcal{F}_n g_n$ we obtain:
\begin{equation}
    \hat{E}^4 \mathcal{F}_n g_n = \hat{E}^2 \frac{1}{u^2+\xi^2} \mathcal{F}_n G_n = \frac{1}{\xi^2 +u^2}(\hat{\mathcal{X}} + \hat{\mathcal{F}}) \frac{1}{u^2+\xi^2} \mathcal{F}_n G_n \ . \
    \label{1E4e}
\end{equation}

Let us derive the commutator
\begin{equation}
   \Big[\hat{\mathcal{F}}{\frac{1}{\xi^2+u^2}}\Big] = \Big[(1-u^2)\partial_u^2 , {\frac{1}{u^2+\xi^2}}\Big] = (1-u^2)\Big[{\partial_u^2}, {\frac{1}{\xi^2+u^2}}\Big] = (1-u^2)\Big(\frac{-2(\xi^2+u^2)+8u^2}{(\xi^2+u^2)^3} + \frac{-4u}{(\xi^2+u^2)^2}\partial_u \Big) \ , \
 \label{Com}
\end{equation}
with the help of which the required expression (\ref{1E4e}) reads
\begin{equation}
    \hat{E}^4 \mathcal{F}_n g_n = \frac{1}{(\xi^2+u^2)^{3}}\left({6(\xi^2-u^2)+4} -4((1-u^2)u\partial_u + (1+\xi^2)\xi\partial_\xi)+({\xi^2+u^2})(\hat{\mathcal{X}} - n(n+1))\right) \mathcal{F}_n G_n \ . \
    \label{E4e}
\end{equation}
In order to expand the expression (\ref{E4e}) over $\mathcal{F}_m$ we should know how the operator $u(1-u^2)\partial_u$ acts on this function
\begin{eqnarray}
    u(1-u^2)\partial_u \mathcal{F}_n = -\frac{(n+1)(n+2)(n+3)}{(2n+1)(2n+3)}\mathcal{F}_{n+2}
    + \,\left(-\frac{n(n+1)(n+2)}{(2n+1)(2n+3)}+\frac{n(n+1)(n-1)}{(2n+1)(2n-1)}\right)\mathcal{F}_n
  \nonumber \\
    + \, \frac{n(n-1)(n-2)}{(2n+1)(2n-1)}\mathcal{F}_{n-2} \  . \
\end{eqnarray}
Finally, we arrive to
\begin{equation}
(\xi^2+u^2)^{3} \hat{E}^4 \mathcal{F}_n g_n  = \mathcal{F}_{n+2}\,\hat{U}_n G_n + \mathcal{F}_{n}\,\hat{S}_n G_n + \mathcal{F}_{n-2}\,\hat{\mathcal{D}}_n G_n \ , \
 \label{E4u}
\end{equation}
where
\begin{eqnarray}
    \hat{U}_n =  \frac{(n+2)(n+3)}{(2n+1)(2n+3)}( \hat{\mathcal{X}}-(n-2)(n-1)) \ , \
     \label{Up}\\
    \hat{S}_n =  6\xi^2 + 4-4\xi(1+\xi^2)\partial_\xi+\xi^2\big(\hat{\mathcal{X}}-n(n+1)\big)  + \frac{n(n+2)}{(2n+1)(2n+3)}\big(\hat{\mathcal{X}} - (n-2)(n-1)\big) \nonumber \\
      + \frac{(n-1)(n+1)}{(2n+1)(2n-1)} \big(\hat{\mathcal{X}} - (n+2)(n+3)\big) \ , \
      \label{Sn} \\
    \hat{\mathcal{D}}_n = \frac{(n-1)(n-2)}{(2n+1)(2n-1)}\big(\hat{\mathcal{X}}-(n+2)(n+3)\big) \, . \
    \label{Dn}
\end{eqnarray}

Let us consider the  representation $\psi_N^{(p)} = \mathcal{F}_N g_N + \mathcal{F}_{N-2} g_{N-2}$,  ($N > 2$),  for the solutions of Eq. (\ref{big}) for $\psi$,  which  does not include
the kernel of the  operator  $\hat{E}^2$.
Then  it follows  from  the Eqs. (\ref{E4}, \ref{Up}) that for the contribution proportional to $\mathcal{F}_{N+2}$  in Eq. (\ref{E4u}) to be equal to zero, the following equation should fulfilled
\begin{eqnarray}
G_N = \mathcal{X}_{N-2} \ . \
 \label{Gn}
\end{eqnarray}
   In this case to obtain the zero coefficient at $\mathcal{F}_N$ in Eq. (\ref{E4u}) the  following  equality should hold
\begin{eqnarray}
     -\hat{U}_{N-2} G_{N-2} = \hat{S}_N \mathcal{X}_{N-2} \ , \
   \label{eqUn}
         \end{eqnarray}
        In turn,  using Eqs. (\ref{lei1}-\ref{Xin}, \ref{E4},\ref{Sn}), we find
    \begin{eqnarray}
 \hat{S}_N \mathcal{X}_{N-2} =    \Big( 6\xi^2 + 4-4\xi(1+\xi^2)\partial_\xi+\xi^2(\hat{\mathcal{X}}-N(N+1))
     \nonumber \\
     +\, \frac{(N-1)(N+1)}{(2N+1)(2N-1)}(\hat{\mathcal{X}} - (N+2)(N+3)) \Big) \, \mathcal{X}_{N-2}\, =\,
     \frac{4\,N\,(N+1)}{(2N-1)} \mathcal{X}_{N} \ . \
      \label{kFn}
\end{eqnarray}
By means of  the simple algebra,  we obtain from Eq. (\ref{eqUn}), using  Eqs. (\ref{Up}, \ref{kFn}),   \\
  $\,(\hat{\mathcal{X}}- (N-4)(N-3))G_{N-2} = -4(2N-3)\mathcal{X_N} $, which leads   to
\begin{eqnarray}
 \hat{G}_{N-2} =  -\, \mathcal{X}_N  \ . \
 \label{Gnm2}
\end{eqnarray}
After substitution of the expressions for $G_N, G_{N-2}$  ((\ref{Gn},\ref{Gnm2})) in Eqs. (\ref{E4u}-\ref{Dn})
one can see, that multiplier at $\mathcal{F}_{N-2}$ is equal to zero,
$\hat{\mathcal{D}}_{N-2}G_{N-2}=0$, and  $\hat{\mathcal{D}}_{N} G_{N} = - \hat{S}_{N-2} G_{N-2}$. This means that  these terms cancel  out each other in Eq. (\ref{E4u}).

Thus the final representation of the partial solution for  the  stream function  $\psi_{N>2}^{(p)}$
(i.e. $\psi_{N>2}$  without the contribution of the kernel of operator  $\hat{E}^2$) is:
\begin{equation}
    \psi_{N>2}^{(p)} = \mathcal{F}_N\mathcal{X}_{N-2}  + \mathcal{F}_{N-2} \mathcal{X}_{N} \ . \
     \label{fpsol}
\end{equation}

Let us  consider the partial solutions  for  $\psi_{N\leq2}^{(p)}$  separately:
\begin{description}
    \item[$N=1$:] In this case $\psi_1^{(p)} = \mathcal{F}_1 g_1$, in turn, $\hat{\mathcal{D}}_1 =0$, and for  the equality
      $\hat{U}_1 G_1=0$ to be satisfied, the fulfillment of the equality  $\hat{\mathcal{X}}G_1 = 0$ is necessary; at the same time
      $\hat{S}_1 G_1 = 0\implies G_1 \propto \xi, (\hat{\mathcal{X}} -2)g_1 = \xi $, that leads to the relation
    $\underline{g_1 \propto \xi} $.
    \item[$N=2$:]  In this case  $\psi_2^{(p)} = \mathcal{F}_2 g_2$, $\hat{\mathcal{D}}_2 =0$, $\hat{\mathcal{X}} G_2 = 0, \hat{S}_2 G_2=0\implies {G_2\propto 1}$, $(\hat{\mathcal{X}} -3)g_2 = 1$, i.e.  $\underline{g_2 \propto 1}$.
\end{description}

As a result, we arrive to the full solution of Eq. (\ref{big}) for the stream function $\psi$, which is a combination  of  the
 stream functions $\psi_{1,2}^{(p)}, \psi_N^{(p)(1,2)}$  and of the  kernel of  operator $\hat{E}^2$:
\begin{multline}
    \psi = c_1 \xi \mathcal{F}_1 + c_2 \mathcal{F}_2 \\
    +\sum_{N>2} c_N^{(1)} (\mathcal{F}_N\mathcal{X}^{(1)}_{N-2}  + \mathcal{F}_{N-2} \mathcal{X}^{(1)}_{N}) + c_N^{(2)} (\mathcal{F}_N\mathcal{X}^{(2)}_{N-2}  + \mathcal{F}_{N-2} \mathcal{X}^{(2)}_{N}) +\\
    +\sum_{N\ge1} c_{No}^{(1)}\mathcal{X}^{(1)}_N\mathcal{F}_N + c_{N  \mathcal{K}}^{(2)}\mathcal{X}^{(2)}_N\mathcal{F}_N \ . \
    \label{fullsol}
\end{multline}
For the problem under consideration
we are interested  in  the continuously differentiable (smooth)  solutions  for the stream function inside  an oblate spheroid.
 In accordance with the general rule, for such solutions
 the same
evenness  over  $u$  and  $\xi$ should be fulfilled  (in this case the solution is automatically regular at $\xi=u=0$, see Sec. I, A). This leads to
$\{c_1,c_2, c_N^{(2)}, c_{N \mathcal{K}}^{(2)}\} = 0$.

Finally, the necessary smooth  solution of Eq. (\ref{big}) for the stream function  has the  form
\begin{equation}
\psi = \sum_{N>2}\, c_N\, \psi_N\,   \ , \ \label{Psift}
\end{equation}
where  basic functions $\psi_N$ can be written as
\begin{equation}
 \psi_N = \mathcal{F}_N \,\big(\mathcal{X}_{N-2}^{(1)} + c_{N  \mathcal{K}}\, \mathcal{X}_{N}^{(1)}\big) + \mathcal{F}_{N-2} \,\big(\mathcal{X}_{N}^{(1)} + \, \widetilde{c}_{N \mathcal{K}} \,\mathcal{X}_{N-2}^{(1)} \big),
         \label{Psi1t}
\end{equation}
in turn,  the constants $c_{N  \mathcal{K}}, \widetilde{c}_{N \mathcal{K}} $
can be found from the condition  $\psi(\xi=\xi_0)=0$:
\begin{equation}
   c_{N  \mathcal{K}} = - \frac{\mathcal{X}_{N-2}^{(1)}(\xi_0)}{\mathcal{X}_{N}^{(1)}(\xi_0)},\; \widetilde{c}_{N \mathcal{K}}  = - \frac{\mathcal{X}_{N}^{(1)}(\xi_0)}{\mathcal{X}^{(1)}_{N-2}(\xi_0)} = \frac{1}{c_{N  \mathcal{K}}} \ . \
        \label{cNt}
   \end{equation}
The expression (\ref{Psi1t}) is used in our analytical calculations (see, for example, Eq. (\ref{GenS}) in Sec. IY A).

\end{widetext}

\section{Details of calculations of the basic stream functions. General stationary solutions and critical values of Ma}
\label{sec:A7}

\centerline{ \bf 1. Derivation of the  basic set $\{\psi_{j, st}\}$.}

Here we present the details of the derivation of the basic set of the stream functions $\{\psi_{j, st}\}$, for the case of  the sticking boundary conditions  at the bottom drop surface.
In doing so we use the  basic  stream functions $\psi_j[\xi, u]$ and introduce the corresponding  tangential velocity component $v_{u, j}[\xi, u]$  for  odd and even  functions over the variable $u$: $\psi^{odd}_j$, $\psi^{ev}_j$, $v_{u, j}^{odd}$, $v_{u, j}^{ev}$.
To build up  the new basic stream functions which satisfy the sticking boundary condition we are using the following mathematical
trick.
 We start with the expansion of  the partial derivative $ \partial_\xi \psi_l^{odd}[\xi=\xi_0,|u|]$  at  $\xi=\xi_0$  over set of functions $\partial_\xi \psi_j[\xi=\xi_0, u]$
\begin{eqnarray}
    \partial_\xi \psi_l^{odd}[\xi=\xi_0,|u|] = \sum_{j_{even}} c_{lj} \,\partial_\xi \psi_j[\xi=\xi_0, u] \ , \
     \end{eqnarray}
    which leads to equality
     \begin{eqnarray}
    \partial_\xi\underbrace{\big(\psi_l^{odd} + \sum_{j_{even}} c_{jl}\,\psi_j \big)}_{\psi_{l, st}}[\xi=\xi_0] =0 \ \text{ for } u < 0 \ . \
 \label{cst}
\end{eqnarray}
The above equations provide us with a  set of functions $\psi_l^{odd}$  which allows to  generate the full basis $\{\psi_{j, st}\}$.

 The desired  basis  provides equation $\partial_\xi \psi_l^{odd}[|u|]\vert_{\xi=\xi_0} = \hat{O}^{stick}_{i l}\mathcal{F}_i[|u|]$,
where  the matrix  $\hat{O}_{il}^{stick}$  describes  the action of the operator $\partial_\xi $ on  the expansion of the stream  function $\psi_l^{odd}[\xi,u]$ at $\xi=\xi_0$.
  For the implementation of the above procedure it is necessary to obtain  an  expansion of  the symmetrized function $\mathcal{F}_{i}[|u|]$  over a  set  of functions  $\{\mathcal{F}_m\}$:
\begin{equation}
    \mathcal{F}_{i}[|u|] = k_{im}\mathcal{F}_m \ . \
    \label{Fsim}
\end{equation}
The corresponding coefficients $k_{im}$  can  be calculated
  by means of Eq. (\ref{duPn}),  taking  the
orthogonality of the  functions $\mathcal{F}_m$  in the interval $[-1,1]$ into account (see Eq.
(\ref{defF})):
\begin{eqnarray}
k_{im} \,=\, P_i[0](P_{m+1}[0] - P_{m-1}[0])\,
\nonumber \\
\times \, \frac{m(m+1)}{i(i+1)-m(m+1)} \ . \
    \label{kim}
\end{eqnarray}

In order to calculate the coefficients $c_{jl}$  in the expansion of the stream function $\psi^{odd}_l$ over the function $\psi^{even}_j$
it is necessary to invert the matrix  $\hat{O}^{stick}_{m j}$ which leads to the following  equation
\begin{equation}
   \sum_{j_{even}} \hat{O}^{stick}_{m j} c_{jl}  =   \sum_i k_{im}\hat{O}^{stick}_{i l} \ . \
    \label{Fsim1}
\end{equation}
 Eq. (\ref{Fsim1}) can be solved explicitly
 because in accordance with the expression (\ref{GenS})
 the matrix $\hat{O}^{stick}_{m j}$ is a lower triangular matrix (see Eq. (\ref{Psi1t})).

\bigskip

\centerline{ \bf 2. Validation of the condition $n \ne k$ }

The next point that should be clarified is justification of the condition $n \ne k$
 in  Eq. (\ref{Tnk}). Due to the explicit form of the solution for $v_z$
    that for the response to the right part of Eqs. (\ref{Tnin},\ref{Tn}) the condition $n \ne k$ is valid due to the explicit form of the solution for $v_z$
 , as each $\psi_n$ from obtained set $\{\psi_j\}$ has the contributions of the view $\mathcal{F}_n \mathcal{X}_n$ and $\mathcal{F}_n\mathcal{X}_{n-2} + \mathcal{F}_{n-2} \mathcal{X}_n$, see Eqs. (\ref{GenS},\ref{mFn},\ref{mXn},\ref{Xin}-\ref{lapl2b}):
\begin{eqnarray}
       v_z[\psi=\mathcal{F}_n \mathcal{X}_n] = \, - \,  \frac{P_{n+1}\Xi_{n-1} - P_{n-1}\Xi_{n+1}}{2n+1} \ , \qquad
\end{eqnarray}
\begin{eqnarray}
  \hskip-0.6true cm  v_z[\psi=\mathcal{F}_n \mathcal{X}_{n-2} + \mathcal{F}_{n-2} \mathcal{X}_n]
     =\  2 \,\frac{P_{n+1} \Xi_{n-1} - \Xi_{n+1} P_{n-1}}{(2n-3)(2n+1)}  \,
     \nonumber \\
     - \ \frac{2n-1}{(2n+1)(2n-3)}(P_{n+1}\,\Xi_{n-3}- P_{n-3}\,\Xi_{n+1}) \qquad
      \nonumber \\
      -\ 2 \,\frac{P_{n-1}\,\Xi_{n-3} - P_{n-3}\,\Xi_{n-1}}{(2n+1)(2n-3)} \ . \ \qquad
\end{eqnarray}
In another words in the  right part of Eq. (\ref{Tn})  the functions $P_n(u)$ are always multiplied by  $\Xi_k^{(1)}$ with $ n \ne k$.

\bigskip

\centerline{\bf 3. Derivation of the general stream function }

\centerline{\bf through the  basic set  $\psi_{i, st}$}

In order to derive the general stream function we need  $c_{i, st}$ of  the expansion of the full stream function over basic functions $\psi_{i, st}$, see Eq. (\ref{tpsist}).
We are looking for the finite approximation  of this solution.
The series is broken  once the following convergence criterion is
satisfied:   the norm of deviation from the required equation (\ref{Mar2}), given by Eq. (\ref{Normst}),
should be minimal for the derived $N_r$-measured set of $c_{i, st}$, i.e.
\begin{eqnarray}
    E[\{c_{j, st}\}]  = c_{j, st} c_{m, st} \underbrace{\sum_{i,l} \int \hat{O}_{ij}^{free} \mathcal{F}_i \hat{O}_{lm}^{free} \mathcal{F}_l \, \frac{d{u}}{1-u^2}}_{\hat{M}_{jm}}
    \nonumber \\
     -\, 2\, c_{j, st}\, \underbrace{\sum_{i}\int \hat{O}_{ij}^{free} \mathcal{F}_i \, r[u]\, \frac{d{u}}{1-u^2}}_{{\mathcal{R}}_j} \, + \int (r[u])^2 \frac{d{u}}{1-u^2}\
    \nonumber \\
     \to \ \delta E=0 \implies \hat{M}_{mj} c_{j, st}^{(opt)} = \mathcal{R}_m \quad \ . \  \
     \label{Norm1st}
\end{eqnarray}
In this way  the optimal solution  for  a set $\{c_{j, st}\}$ can be calculated with the  given accuracy.

\bigskip

\centerline{\bf 4. Details of calculations   of  the }

\centerline{\bf critical Marangoni number Ma$_c$}

In this subsection  we present the  details of the finding of solution  of  Eq. (\ref{QEP}).
The standard method to obtain  the solution  of Eq. (\ref{QEP})  is  its  reduction to the generalized eigenvalue problem:
\begin{eqnarray}
    \hat{\mathcal{A}}  - \hbox{Ma}_c \hat{\mathcal{B}} = \begin{pmatrix}
    \hat{M_0} & 0\\
    0 & \hat{I}
    \end{pmatrix} - \hbox{Ma}_c \begin{pmatrix}
    \hat{M_1} & -\hat{M_2}\\
    \hat{I} & 0
    \end{pmatrix},
    \noindent \\
     \ \vert{{\mathcal{V}}}\rangle = \begin{pmatrix}
    \vert{c_c}\rangle \\
    \hbox{Ma}_c \vert{c_c}\rangle
    \end{pmatrix} \  \to \ (\hat{\mathcal{A}}  - \hbox{Ma}_c \hat{\mathcal{B}})\vert{{\mathcal{V}}}\rangle =0 \ , \
    \label{matr}
\end{eqnarray}
where  matrixes $\hat{M}^{(0)}, \hat{M}^{(1)}, \hat{M}^{(2)}$  are determined in expression (\ref{qwF}) as  interlinear (footnote) designations,
and matrixes $\hat{\mathcal{A}}, \hat{\mathcal{B}}$ are obtained from   Eq. (\ref{QEP}) through $\hat{M}^{(0)}, \hat{M}^{(1)}, \hat{M}^{(2)}$, and $\vert{{\mathcal{V}}}\rangle $  is the unknown eigenvector of the matrix $(\hat{\mathcal{A}}  - \hbox{Ma}_c \hat{\mathcal{B}})$.
Because  the quadratic eigenvalue problem is well known, there are reliable  methods of finding  its solution \cite{QEP}.


\begin{thebibliography}{76}
\bibitem{crw}   The parameters of the spherical segments of the drop
 are determined from the condition of  minimum of
its  surface energy $F_s=\,2\,\pi\,\gamma\,(R_{cap}^2+(H/2)^2)+2\,\gamma_{\hbox{\tiny A}}\,(S_f-\,\pi\,R_{cap}^2)\,$,  provided that the volume of the droplet $V_{dr}=\,\pi\,(H/2)(R_{cap}^2+(H/2)^2/3)\,+\,\pi\,2h\,R_{cap}^2\,$  is  fixed.
 Here  $S_f$  is the total area of FSSF,  $R_{cap}$  and $(H)$  are the base radius and the height of the drop, respectively, $h$ is a half of the film thickness and  $\gamma$,  $\gamma_{\hbox{\tiny A}}$  are interfacial tensions defined in text.
\bibitem{Ma} The  estimation of Marangoni force for $Ma=1$:

$\qquad d_0\sigma_{Ma} \sim d_0\,\partial\gamma/\partial x \sim    10^{-10}$  J m$^{-2}$

$\ll E_{d}/R_c \sim 10^{-6} \div 10^{-5}$  J m$^{-2}$,

demonstrates that Marangoni force can play  only the  role for further growing and moving of
dislocation loops of above-critical radius to the bounding meniscus of the drop (where  $E_{d}$ is the  excess line energy  of the
 elementary dislocation, $E_{d} \sim 10^{-13}$ J m$^{-\,1}\,$ \cite{Pikina15,Pikina17}).
Note also that the value Ma$=1$ corresponds to the drops of the height $H\sim 10\, \mu$m  to the  temperature gradient $A=10^{-2}$ K$/\mu$m or $A=10^{-5}$ K$/$ nm, i.e. the difference between neighbouring regions is very small: $10^{-4}\div\,10^{-3}$ K and not enough for a long process of thickening.
\end{thebibliography}
\end{document}


\renewcommand{\baselinestretch}{1.2}\normalsize

\title{{\large \bf SUPPLEMENTARY INFORMATION}
 to the manuscript
 "{\large \bf CIRCULATING  MARANGONI  FLOWS  WITHIN  DROPLETS IN   SMECTIC FILMS}"}

\author{ E.S. Pikina$^{1,2,5}$, M.A. Shishkin$^{2,4}$, K.S. Kolegov$^{2,6}$,  B.I. Ostrovskii$^{2,3}$, and S.A. Pikin$^{3}$. }

\affiliation{$^1$ Landau Institute for Theoretical Physics, RAS,
142432, Chernogolovka, Moscow region, Russia \\
$^2$ Institute of Solid State Physics of the RAS, Chernogolovka, Russia \\
$^3$ FSRC "Crystallography and Photonics" of the RAS,
Leninsky pr. 59, 119333 Moscow, Russia\\
$^4$ NRU Higher School of Economics,
101000, Myasnitskaya 20, Moscow, Russia, \\
$^5$ Oil and
Gas Research Institute,
 RAS, Gubkin str. 3, 119333 Moscow, Russia \\
$^6$ Astrakhan State University,
20a Tatischev Str., Astrakhan 414056 Russia
}


\begin{abstract}

\end{abstract}

\maketitle

\section{Straight derivation of the stream functions $\Psi_N$}
 \label{sec:S1}

In order to find the solution of Eq. (33)  in ellipsoidal coordinates in terms of the stream functions  it is important to know explicitly the properties of these functions. According to the definition of the stream function, $\psi=0\,$ along the $z$-axis, i.e. at $\beta=0$ ($u=1$) or $\beta=\pi$ ($u=-1$). The
 solutions for $\psi\,$ can be either symmetrical or asymmetrical with respect to variable $u$. This means that all solutions  should be  proportional  either to
$(1 - u^2)\,$ or to  $u\,(1 - u^2)\,$.

Since  $\psi(u=\pm1) = 0$,  any stream function  can be expanded over a set of $\mathcal{F}$ (see Appendix C, (C1)), i.e.
 $\psi = \sum_n \mathcal{F}_n(u) g_n(\xi)$, where $g_n(\xi)$ is an unknown function of $\psi$.
  According to the properties of Legendre polynomials (see, for example, Eq. (3.47) in \cite{Lebedev2020}),
  if we aim  to find solutions with the accuracy up to $u^4$ or $u^5$ we look for combine the solutions for stream function either symmetrical :
\begin{eqnarray}
    \psi_{ev}  = \int_{-1}^u du P_3[u]  g_3[\xi] + \int_{-1}^u du P_1[u] g_1[\xi]
  \ , \ \label{StrL4}
                  \end{eqnarray}
or,   asymmetrical with respect to  $u$:
\begin{eqnarray}
    \psi_{odd}  =  \int_{-1}^u  du P_4[u]  g_4[\xi] + \int_{-1}^u du  P_2[u] g_2[\xi]
  \ . \ \label{StrL5}
                  \end{eqnarray}
  If we are looking for the solutions  of  higher  accuracy  - up to $u^{2 m}$ or $u^{2 m +1}$ for $m>2$,  we should find the solution in the form of  a sum of linearly independent terms proportional to  integrals over  Legendre polynomials of the higher order:
  \begin{eqnarray}
    \psi_{ev}^{(2 m + 1)}  = \int_{-1}^u  du P_{2 m + 1}[u]  g_{2 m + 1}[\xi]   \quad \
   \nonumber \\
    + ... + \int_{-1}^u du P_3[u]  g_3[\xi] + \int_{-1}^u du P_1[u] g_1[\xi]
  \ , \ \label{StrL2m}
                  \end{eqnarray}
                  or
                  \begin{eqnarray}
    \psi_{odd}^{(2 m)}  = \int_{-1}^u  du P_{2 m}[u]  g_{2 m}[\xi] \quad \
    \nonumber \\
     + ... +  \int_{-1}^u  du P_4[u]  g_4[\xi] + \int_{-1}^u du  P_2[u] g_2[\xi]
  \ . \ \label{StrL3m}
                  \end{eqnarray}
In accordance with Eqs. (\ref{StrL4}-\ref{StrL3m}),  the  substitution for the stream function:  $\psi = f[u]\,g[\xi]$,
can be used. Then, the equation $\,\Delta \psi = 0\,$ takes the form:
\begin{eqnarray}
\hat{\text{E}}^2 \psi = \frac{1}{c^2\,\big(u^2  + \xi^2 \big)} \Big(f(u)\,( 1 + \xi^2)\, \partial_\xi^2 g(\xi)
\nonumber \\
+ \, g(\xi)\,(1-u^2) \,\partial_u^2 f(u) \Big)=  \frac{1}{c^2\,\big(u^2  + \xi^2 \big)}\, \mathrm{Y}[u,\xi]\ . \
 \label{StrFE1}
\end{eqnarray}
Accordingly, Eq. (33) can be rewritten as:
\begin{eqnarray}
    c^4 (\xi^2+u^2)^2 \hat{\text{E}}^4 \psi = ( 1 + \xi^2)\Big\{\big(
    \frac{8 \xi^{2}}{(u^{2} + \xi^{2})^{2}} - \frac{2}{(u^{2} + \xi^{2})}\big)\,\mathrm{Y}
       \nonumber \\
      -\, 2\frac{2 \xi}{(u^{2} + \xi^{2})}\partial_\xi \, \mathrm{Y} +\partial_\xi^2 \, \mathrm{Y}
    \Big\}
    +\, (1-u^2)\Big\{
    \big(    \frac{8 u^{2}}{(u^{2} + \xi^{2})^{2}}
     \nonumber \\
     -\, \frac{2}{(u^{2} + \xi^{2})}\big)\,\mathrm{Y}
      -\, 2 \frac{2 u}{(u^{2} + \xi^{2})}\partial_u \mathrm{Y} + \partial_u^2 \, \mathrm{Y}
    \Big\} \, . \  \ \ \
     \label{StrFE2}
\end{eqnarray}

It is instructive to find  the  solutions of Eq.  (\ref{StrFE1}) for the case (\ref{StrL4}) and  (\ref{StrL5}).
For  $\psi_{odd}$,  Eq. (\ref{StrFE1}) takes the form:
\begin{eqnarray}
 \hat{\text{E}}^2 \psi_{odd} = \frac{u^2-1}{u^2+\xi^2} \Big( \underbrace{\frac{(\xi^2+1)g_1'' - 2g_1}{2}}_{G_1[\xi]} +
  \nonumber \\
 \underbrace{\frac{(1+\xi^2)g_3''-12g_3}{8}}_{G_3[\xi]} (5u^2-1)\Big) \ , \
    \label{StrFE3}
  \end{eqnarray}
in turn, Eq. (\ref{StrFE2}) can be written as
\begin{eqnarray}
\hat{\text{E}}^4 \psi = \frac{u^2-1}{(u^2+\xi^2)^3}\Big\{u^{4}\, \mathcal{F}_4 \,
  +\, u^{2} \mathcal{F}_2\,
    +\,\mathcal{F}_0 \Big\} \, = 0 \, , \ \ \ \
     \label{StrFE4}
\end{eqnarray}
where
  the functions $\mathcal{F}_2, \mathcal{F}_4 $ and $\mathcal{F}_0$ are conjugated with $\{u^4, u^2, 1\}$, respectively, and can be represented as
\begin{eqnarray}
\mathcal{F}_4[\xi] =\, 5 (1 + \xi^{2} )\partial_\xi^2 G_3
 - 10  G_3   =\,0\, \ , \qquad
\label{StrFE15} \\
\mathcal{F}_2[\xi] =\,(1 + \xi^{2})\,\big(- 20 \xi^{3} \partial_\xi G_3 - \partial_\xi^2  G_3 + 5 \xi^{2} \partial_\xi^2  G_3   \big)       \qquad
     \nonumber \\
    - \, 30 \xi^{2}  G_3  - 10  G_3 + (1 + \xi^{2}) \partial_\xi^2 G_1   =\,0\,   , \qquad
\label{StrFE25} \\
\mathcal{F}_0[\xi] =\,   (1 + \xi^{2})\,\big\{ \xi^{2} \big(\partial_\xi^2 G_1 -  \, \partial_\xi^2  G_3\big)
     - 4 \xi \,( \partial_\xi G_1 - \partial_\xi  G_3 )\big\} \,
      \nonumber \\
     +\, 4\, (1 + \xi^{2}) G_1 - 4  G_3  + 6 \xi^{2}  G_3
  =\,0
 \qquad \qquad
    \label{StrFE35}
  \end{eqnarray}
 Each of the expressions (\ref{StrFE15}-\ref{StrFE35}) should be equal to zero  since  $\{u^4, u^2, 1\}$ are linearly independent. This allows us to obtain  the triple set of differential equations.
From Eq. (\ref{StrFE15}) we obtain the equation for $G_3$,:
\begin{eqnarray}
G_3 = \mathbf{c_{1}} \big(1 + \xi^{2}\big) + \mathbf{c_{2}} \big(\xi + (\xi^{2} + 1) \arctan[\xi]\big)\, \ , \quad
    \label{StrFE14}
  \end{eqnarray}
where $\mathbf{c_{1}}$ and $\mathbf{c_{2}}$ are the integration constants. Substituting (\ref{StrFE14}) to  Eq. (\ref{StrFE25}), one obtains
\begin{eqnarray}
G_1 = \mathbf{c_{1}} \xi^{2} \big(5 \xi^{2} + 6\big)  + \mathbf{c_{2}}\, \big(5 \xi^{4} \arctan[\xi]  + \,5 \xi^{3}
  \nonumber \\
  + 6 \xi^{2} \arctan[\xi]
   - \xi + \arctan[\xi]\big)  + \mathbf{c_4} \xi + \mathbf{c_3} \ , \
    \label{StrFE24}
\end{eqnarray}
where $\mathbf{c_{3}}$ and $\mathbf{c_{4}}$ are also  the integration constants.
After substitution of expressions (\ref{StrFE14}, \ref{StrFE24}) to Eq. (\ref{StrFE35}) we obtain $\mathbf{c_{1}} = \mathbf{c_{3}}$.
In turn, according to Eq. (\ref{StrFE3}) we find  the functions $g_1, g_3$ as solutions of the corresponding differential equations
\begin{eqnarray}
    g_3 \,= 4 \,\mathbf{c_1}\, \xi^2\, (1 + \xi^2) +
\frac{\mathbf{c_2}}{2} \,\big( \xi\, (1 + 3 \xi^2) \qquad \qquad
\nonumber \\
 + \,(1 + \xi^2)\, (-1 + 3 \,\xi^2) \,\arctan[\xi]\big) +
 \frac{\mathbf{c_{5}}}{5} \,  \big(1 + 6 \xi^2 + 5 \xi^4\big) \qquad
 \nonumber \\
  +\,
  \frac{5}{16} \,\mathbf{c_{6}}\, \big( \xi (13 + 15 \xi^2) +
    3\, (1 + \xi^2)\, (1 + 5  \xi^2)\,\arctan[\xi]\big)
     \, , \qquad
       \label{g3}
\end{eqnarray}
\begin{eqnarray}
    g_1 =   \mathbf{c_1}  \xi^2 (1 +  \xi^2) + \mathbf{c_2}\big( \xi   (3 + \xi^2) + \, (1 + \xi^2)  (- 3  \qquad
     \nonumber \\
    +\, \xi^2)\arctan[\xi]\big)
 + \frac{ \mathbf{c_4}}{2}\big( (1 + \xi^2) \arctan[\xi] - \xi\big)
  \nonumber \\
   +   \frac{\mathbf{c_7}}{2}\big((1 + \xi^2)\arctan[\xi] + \xi\big)
               +    \mathbf{c_8} (1 + \xi^2) \ . \ \qquad
         \label{g1}
\end{eqnarray}
Because the $g_3$ is found as a solution of  the differential equation of the second order it has two free constant, hence we can choose
$c_5 = c_5 - 4 c_1$, $c_6 = c_6 - 8 c_2/25\,$. Finally,  we obtain for the function $g_3$
\begin{eqnarray}
    g_3[\xi] = - \,\mathbf{c_1}\,\frac{4}{5}\,(\xi^2+1) - \mathbf{c_2}\,\frac{4}{5}\, \big((\xi^2+1)\,\arctan[\xi] + \xi \big)
         \nonumber \\
        + \mathbf{c_{5}}\, \big(5 \xi^{4} + 6 \xi^{2} + 1 \big)  + \mathbf{c_{6}}\,  \Big(15 \xi^{3}  + 13 \xi
         \nonumber \\
   +\, 3\, \big(5 \xi^{4} + 6 \xi^{2} + 1\big) \arctan[\xi]\Big) \ .  \qquad
   \label{g3f}
\end{eqnarray}
In the next step we obtain the function $g_1[\xi]$ using  the corresponding differential equation of the second order
\begin{eqnarray}
    g_1[\xi] = \mathbf{c_{1}}\, \big( \xi^{4} - 2  \xi^{2} - 3  \big) + \mathbf{c_{2}}  \big(\xi^{4} \arctan[\xi]
     \nonumber \\
     +  \xi^{3} - 2  \xi^{2} \arctan[\xi] + 3  \xi - 3  \arctan[\xi] \big)
      \nonumber \\
      + \frac{\mathbf{c_{4}}}{2} \big((\xi^{2}+1) \arctan[\xi] - { \xi}  \big)
      \nonumber \\
      + \mathbf{{c_{7}}}\, \, \big( (\xi^{2}+1) \arctan[\xi] + \xi \big)
       + \mathbf{c_{8}}\, \big(\xi^{2} + 1\big) \ , \quad
       \label{g1f}
\end{eqnarray}
where we account for that the function $g_1[\xi]$ can be obtained with the accuracy to
eigenfunctions $\mathcal{X}_n$ of the operator $\hat{\mathcal{X}} = (1+\xi^2)\partial_\xi^2$  (see Appendix D, Eq. (D.4,D.5)). This operator $\hat{\mathcal{X}}$ appears in equation (\ref{g1f}) for $g_1$,
in turn let designate as $g_{01}[\xi]$ the kernel  of the operator $(\hat{\mathcal{X}} - n(n+1))$), and we choose   $g_{01}[\xi]=  - 3 \mathcal{X}_1= - 3 (1 +  \xi^2)$ and use it at obtain of the expression (\ref{g1f}).

In accordance with the boundary condition for normal velocity component $v_\xi[\xi_0,u]=0$, (see Eq.  (20)) we find  that the conditions $g_1[\xi_0]=0$ and $g_3[\xi_0]=0$ are held. One can see  from Eq. (\ref{g1f}) that  $\mathbf{c_{4}}$ corresponds to linearly  independent solution of Eq. (\ref{g1f}) with $\mathbf{c_{1}}=\mathbf{c_{2}}=0$, thus we can fix $\mathbf{c_{4}}=0$. From the condition of regularity at $(\xi = 0,u=0)$ we obtain an equation  $\mathbf{c_{2}}-10\mathbf{c_{6}}+5\mathbf{{c_{7}}}=0$. From the boundary condition for velocity component $v_x[\xi=0,u]=0$ we write one more equation
$\mathbf{c_{2}}-10\mathbf{c_{6}}=0$, which allows us to obtain
 $\mathbf{{c_{7}}}=0,\; \mathbf{c_{6}}=\mathbf{c_{2}}/10$.

Because the velocities of the fluid flow are continuously differentiable functions, their space derivatives  $\partial v_x/\partial z$ and $\partial v_z/\partial x$ do not contain  singularities.  This means  that vorticity $\vec{\varpi} $,
introduced  by Eq. (29,30),
 doesn't have singularities at  $(\xi = 0,u=0)$. Due to this all coefficients $\mathbf{c_{i}}$  corresponding to terms with  $\arctan[\xi]$  in above equations should be set equal to zero.

Summarizing all above restrictions, we obtain the linear space of solutions for  the set $\{\mathbf{c_{1}}, \mathbf{c_{2}}, \mathbf{c_{4}}\}$:
  \begin{eqnarray}
\mathbf{c_{2}}  =  \mathbf{c_{6}} = \mathbf{c_{4}} = \mathbf{c_{7}} =  \mathbf{c_{8}} = 0 \ , \ \qquad \qquad
  \nonumber \\
  \mathbf{c_{5}} = \frac{1}{10 (5 \xi_{0}^{4} + 6 \xi_{0}^{2} + 1)}\,\big\{8 \mathbf{c_{1}} \big(\xi_{0}^{2} + 1\big)
     \big\}\
    \ . \quad
      \label{ci}
  \end{eqnarray}

Finally,  we obtain
 \begin{eqnarray}
  \psi_{3}^{(1)} =
\frac{1}{2}\big(u^2 - 1\big)\,\Big\{- \, \frac{( {5}\,\xi^4 + {6}\,\xi^2 +{1})}{80}
\nonumber \\
+ \,\, \Big(\frac{\xi^2 +  1}{\xi_0^2 + 1}\Big)\,\frac{({5}\,\xi_0^4 + {6}\,\xi_0^2 + {1})\,}{80}\Big\}
\nonumber \\
  +\,  \frac{(5\,u^4 -  6\,u^2 +  1)}{8}\,\Big\{\frac{\xi^2 + 1}{20}\,
\nonumber \\
 - \, \,\frac{ (\xi_0^2 + 1)}{20}\,\frac{( {5}\,\xi^4 + {6}\,\xi^2 +{1})}{({5}\,\xi_0^4 + {6}\,\xi_0^2 + {1})\,} \Big\} \ . \ \
 \label{psi31}
  \end{eqnarray}

Similarly, one obtains the first two  even stream functions $\psi_{ev}$ from Eq. (\ref{StrL4}):
    \begin{eqnarray}
  \psi_4  =   \frac{u\, (1 -  u^2)}{2} \,  g_2[\xi]    +
   \frac{ u\, (1 -  u^2)\, (7  u^2 -  3)}{8}\, g_4[\xi]  \, , \ \ \
   \label{psi4}
  \end{eqnarray}
  and correspondingly
    \begin{eqnarray}
G_4[\xi]  =   \frac{(1 +  \xi^2)\, g_4''[\xi]  - 20 g4[\xi]}{8}  =
  \nonumber \\
=  (\xi + \xi^3) \mathbf{c_{1}} +
 \frac{1}{2}\, (-2 - 3 \xi^2 - 3 \xi \arctan[\xi]
 \nonumber \\
  -   3 \xi^3 \arctan[\xi]) \mathbf{c_{2}}
 \ , \
   \label{G4}
    \end{eqnarray}
    \begin{eqnarray}
  G_2[\xi] = \frac{(1 +  \xi^2)\, g_2''[\xi]  - 6  g2[\xi]}{2}  =
    \nonumber \\
  = \,-\,\frac{23 \mathbf{c_{2}} \xi^2}{2} + 10 \mathbf{c_{1}} \xi^3 - \,\frac{21 \mathbf{c_{2}} \xi^4}{2} +
  \nonumber \\
  7 \mathbf{c_{1}} \xi^5 - \,\frac{9}{2} \mathbf{c_{2}} \xi \arctan[\xi] -
  15 \mathbf{c_{2}} \xi^3 \arctan[\xi]
  \nonumber \\
   - \frac{21}{2} \mathbf{c_{2}} \xi^5 \arctan[\xi] +
  \mathbf{c_{3}} + \xi \mathbf{c_{4}}
  \ , \
   \label{G2} \\
    \mathbf{c_{4}} = 3 \mathbf{c_{1}}  \, . \qquad  \qquad
  \end{eqnarray}


  After solving of  the equations (\ref{G4}, \ref{G2}) and  subsequent  substitution of  all boundary and regularity conditions we obtain the expression for $\psi_4^{(1)}$.

\bigskip

\section{Numerical simulation}
\subsection{Formulation of the problem} An isotropic liquid droplet in the form of an oblate spheroid formed in the free-standing smectic film (FSSF) is placed inside a cylindrical chamber between two round plates with different temperatures: $T_\mathrm{up}$ is  the top plate temperature and $T_\mathrm{dn}$ is the bottom plate temperature. The drop is in the ambient air at the normal atmospheric pressure. A  thin smectic film of uniform thickness  keeps the droplet in horizontal plane at the same distance from each plate. The edges of the FSSF are connected with a rigid frame which is attached to the side wall of the cylindrical chamber. We assume that the geometry of the droplet and film does not change over time. Since the thickness of the smectic film is much less than the thickness of the isotropic liquid drop, we do not take into account the thermal properties of the FSSF.

The side wall of the chamber is made of the thermal insulating material; therefore, we do not take into account the heat exchange of our  system with the environment. We assume that the smectic film prevents the evaporation of a liquid from the droplet, and the air inside the chamber is dry. Hence, the transfer of heat in the air occurs due to the thermal conductivity. The convection and radiation are not taken into account. The heat transfer in the droplet occurs due to the thermal conductivity and convection. Since the thickness of the drop in the vertical direction is much less than along the horizontal, we do not take into account the flows in the bulk caused by the difference in the density of the liquid due to the spatial inhomogeneity of the temperature field. We consider  only the thermocapillary flow of liquid, which arises due to the dependence of surface tension on temperature. This type of convection within the  liquid is known as the Marangoni flow. The physical and geometric parameters of the problem are presented in Table~\ref{tab:Parameters}. We take into account the linear dependence of surface tension on temperature, $\sigma(T)= \sigma_0 - \sigma'(T - T_0)$, where $\sigma_0$ is the surface tension at the temperature $T=T_0$.
\begin{table} \caption{Problem parameters} \centering
	\begin{tabular}{|p{0.13\linewidth}|p{0.45\linewidth}|p{0.36\linewidth}|} \hline
		Symbol& Parameter& Value [unit of measurement]\\ \hline
		$\kappa_l$& thermal conductivity of liquid &0.12 [W/(mK)]\\
		$\kappa_a$& thermal conductivity of air& 0.026 [W/(mK)]\\
		$c_l$& specific heat capacity of liquid& 2500 [J/(kg~K)]\\
		$c_a$& specific heat capacity of air& 1000 [J/(kg~K)]\\
		$\eta$& dynamic viscosity& $1.4 \times 10^{-2}$ [s~Pa]\\
		$\rho_l$& liquid density& 1200 [kg/m$^3$]\\
		$\rho_a$& air density& 1.2 [kg/m$^3$]\\
		$\sigma'$& $\sigma'$ &$1.0 \times 10^{-4}$ [N/(m~K)]\\
		$h_d$& droplet height &20 [$\mu$m]\\
		$H$& domain height &1 [mm]\\ $R_\mathrm{in}$& droplet radius &100 [$\mu$m]\\ $R_\mathrm{out}$& domain radius &1 [mm]\\
		$T_\mathrm{dn}$& bottom plate temperature &324 [K]\\
		$T_\mathrm{up}$& top plate temperature &334 [K]\\ \hline
	\end{tabular} \label{tab:Parameters}
\end{table}

\subsection{Mathematical model} Let the $z$ axis corresponds to the axis of symmetry of the cylindrical chamber and  of the droplet in the form of an oblate spheroid. The zero of the $z$-coordinate axes is in the center of the drop.  The center of the droplet and the chamber in the horizontal plane corresponds to the radial coordinate $r=0$. Here we use the cylindrical coordinates $(r,\varphi,z)$. The maximum droplet radius in the horizontal plane and the maximum droplet height in the vertical plane are designated as $R_\mathrm{in}$ and $h_d$, respectively. The height and the radius of the cylindrical chamber are  indicated as $H$ and $R_\mathrm{out}$. In the considered problem, the transfer of mass and heat does not depend on the angular coordinate $\varphi$. This allows us to consider our problem as a two-dimensional one in the coordinates $(r,z)$. In this formulation, the geometry of the droplet is described by an ellipse with a semi-major axis is $a=R_\mathrm{in}$, and a semi-minor axis  $b=0.5 h_d$.

\begin{figure}[!htb]
	\includegraphics[width=0.6\linewidth]{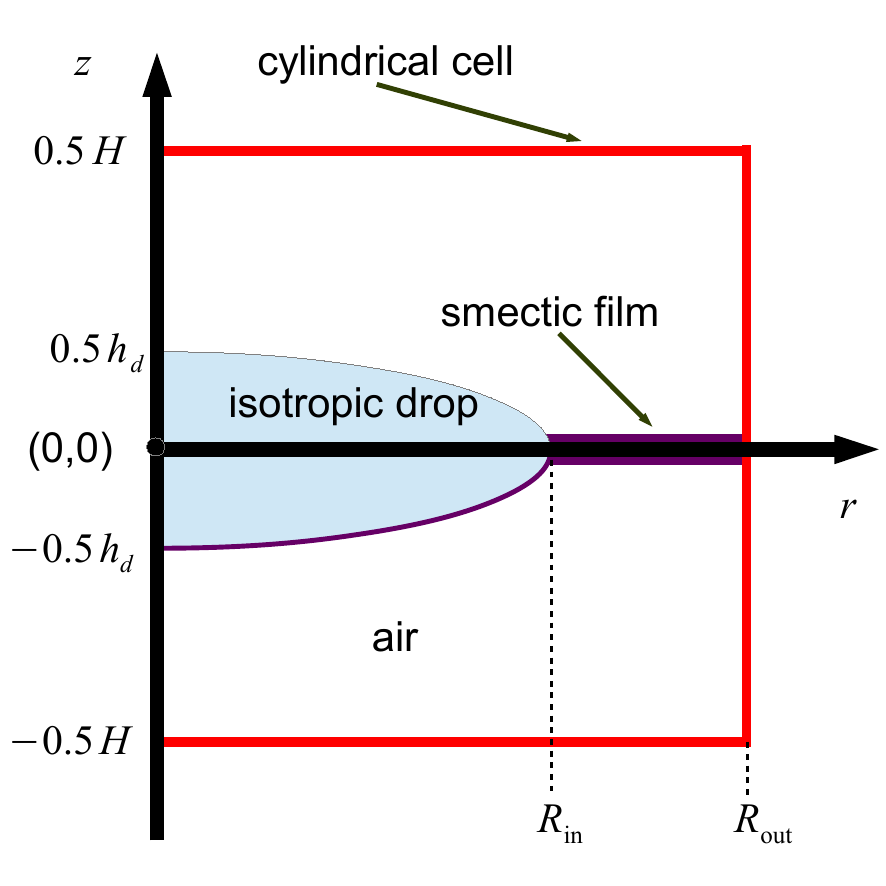}
	\caption{\label{fig:sketch} Sketch for the problem statement:  smectic film on the  bottom drop surface.}
\end{figure}

The hydrodynamics in the droplet in discussed sheme is described as follows. Let us write the Navier--Stokes equations for the incompressible liquid in the cylindrical coordinates \cite{Landau6,Happel}
\begin{multline}\label{eq:NavierStokes_u}
	\frac{\partial u}{\partial t} + u \frac{\partial u}{\partial r} + v \frac{\partial u}{\partial z} = \\ -\frac{1}{\rho_l} \frac{\partial P}{\partial r} + \frac{\eta}{\rho_l}\left( \frac{\partial^2 u}{\partial z^2} + \frac{1}{r}\frac{\partial}{\partial r} \left( r \frac{\partial u}{\partial r} \right)- \frac{u}{r^2}\right),
\end{multline}
\begin{multline}\label{eq:NavierStokes_v}
	\frac{\partial v}{\partial t} + u \frac{\partial v}{\partial r} + v \frac{\partial v}{\partial z} = \\ -\frac{1}{\rho_l} \frac{\partial P}{\partial z} + \frac{\eta}{\rho_l}\left( \frac{\partial^2 v}{\partial z^2} + \frac{1}{r}\frac{\partial}{\partial r} \left( r \frac{\partial v}{\partial r} \right)\right),
\end{multline}
where $t$ is the time, $\rho_l$ is the liquid density, $\eta$ is the dynamic viscosity, $P$ is the pressure, $u$ is the horizontal component of the liquid flow velocity vector, and $v$ is the vertical component of this vector. Let us introduce the stream function $\psi$ that satisfies the relations $\partial \psi/ \partial z = r u$ and $\partial \psi/ \partial r = -r v$. In this case, the equation of continuity is valid by itself, \begin{equation}\label{eq:ContinuityEquation}
	\frac{1}{r} \frac{\partial (ru)}{\partial r} + \frac{\partial v}{\partial z}= \frac{1}{r} \frac{\partial^2 \psi}{\partial r\, \partial z} - \frac{1}{r} \frac{\partial^2 \psi}{\partial z\, \partial r}\equiv 0.
\end{equation}
Introduce also  the vorticity $\omega= \partial u/ \partial z - \partial v/ \partial r$. Let us take the $z$ derivative of Eq. \eqref{eq:NavierStokes_u} and the $r$ derivative of Eq. \eqref{eq:NavierStokes_v}. Then we subtract the resulting equations from each other and perform the transformations. Considering that $$\frac{\partial u}{\partial r} + \frac{\partial v}{\partial z} = -\frac{u}{r}$$ according to the equation \eqref{eq:ContinuityEquation}, we obtain the vorticity transfer equation
\begin{multline}\label{eq:VorticityTransfer}
	\frac{\partial \omega}{\partial t} + u \frac{\partial \omega}{\partial r} + v \frac{\partial \omega}{\partial z} - \frac{u \omega}{r} = \\ \frac{\eta}{\rho_l}\left( \frac{\partial^2 \omega}{\partial z^2} + \frac{1}{r}\frac{\partial}{\partial r} \left( r \frac{\partial \omega}{\partial r} \right)- \frac{\omega}{r^2}\right),
\end{multline}
where $u=(\partial \psi/ \partial z)/r$,  $v=-(\partial \psi/ \partial r)/r$. Rewriting the expression for the vortex $\omega= \partial u/ \partial z - \partial v/ \partial r$ and using the stream function  the following equation is obtained
\begin{equation}\label{eq:StreamFunction}
	\frac{\partial^2 \psi}{\partial r^2}-\frac{1}{r}\frac{\partial \psi}{\partial r}+\frac{\partial^2 \psi}{\partial z^2}=r\omega.
\end{equation}
in such a way we are going from the equations \eqref{eq:NavierStokes_u}--\eqref{eq:ContinuityEquation} to Eqs. \eqref{eq:VorticityTransfer} and \eqref{eq:StreamFunction} in which the pressure $P$ is absent.

The temperature field $T$ in the system is described by the heat transfer equation
\begin{multline}\label{eq:HeatTransfer}
	\frac{\partial T}{\partial t} + \alpha \left( u \frac{\partial T}{\partial r} + v \frac{\partial T}{\partial z}\right) = \\ \frac{1}{c\,\rho}  \left(\frac{\partial}{\partial z}\left(\kappa \frac{\partial T}{\partial z}\right) + \frac{1}{r}\frac{\partial}{\partial r} \left(\kappa\, r \frac{\partial T}{\partial r} \right)\right),
\end{multline}
where
$$\alpha, c, \rho, \kappa= \begin{cases} 1, c_l,\rho_l, \kappa_l:\, h_{-}< z <h_{+};\\ 0, c_a, \rho_a, \kappa_a:\, |z| > h_{+}. \end{cases} $$ Here $c, \rho, \kappa$ are the specific heat capacity, density and thermal conductivity, respectively. The indices $l$ and $a$ indicate the relation of the material parameter either to the liquid or to air. The parameter $\alpha$ allows to turn off the convective term in the air area. As a result, the heat transfer equation \eqref{eq:HeatTransfer} becomes the heat equation at $\alpha=0$. The droplet surface shape $h$ is determined using  canonical equation of the ellipse,
$$h_\pm = \pm b\sqrt{1-\left( \frac{r}{a} \right)^2 }.$$
For the case of a droplet in the form of a lens, we use two spherical segments attached to each other by their bases. From the canonical equation of a circle, we obtain the shape of the droplet surface
$$h_\pm = \pm\sqrt{R^2 - r^2}\mp (R-0.5h_d),$$
where the circle radius
$R=\left(R_\mathrm{in}^2 + (0.5h_d)^2\right)/h_d$. We also consider the case when the shape of the liquid corresponds to the lens with the truncated ends at the point $r=R_\mathrm{in}-\delta r$. For definiteness, let us take the value $\delta r= 0.1 R_\mathrm{in}$.

Finally, we obtain the system of equations that includes the equations \eqref{eq:VorticityTransfer}--\eqref{eq:HeatTransfer}. The model consists of the three equations with the three variables to be find: $\omega$, $\psi$ and $T$.

\subsection{Initial and boundary conditions} \label{sec:InitAndBoundaryCond}
At the initial time, the temperature in the system is uniform, $T(r,z,t=0)=T_\mathrm{up}$, and there are no liquid flows, $\omega(r,z,t=0)=0$. The value of the top plate temperature is the constant, $T(r,z=0.5H,t)=T_\mathrm{up}$. The bottom plate cools down to its stationary temperature $T_\mathrm{dn}$ in a relatively short period of time $t_\mathrm{rel}$, $T(r,z=-0.5H,t)=T_\mathrm{up}+(T_\mathrm{dn}-T_\mathrm{up})U_r(t)$, where
$$U_r(t)= \begin{cases} t/t_\mathrm{rel},\, t\leq t_\mathrm{rel};\\ 1,\, t > t_\mathrm{rel}. \end{cases} $$
Here, the heat relaxation time in the air due to the thermal conductivity is $t_\mathrm{rel}=H^2 c_a \rho_a / \kappa_a$. For the side wall of the chamber we use the condition of the heat flux absence, $\partial T(r=R_\mathrm{out},z,t)/\partial r = 0$. At the <<liquid--air>> boundary, the condition of equality of the temperatures and heat fluxes directed along the normal to this boundary must be satisfied. As indicated earlier, we do not take into account the thin smectic film. The above boundary condition satisfied automatically in the numerical method we use, so it is not described here in detail. At $r=0$, due to the axial symmetry of the system, the conditions
$\partial T/ \partial r = 0, \omega=0, \psi=0\,$ are held. At the boundary <<liquid--air>> we use the condition $\psi(r,z=h_{\pm},t)=0$, which follows from the impermeability condition.
The normal vector $\mathbf{n}$ is directed outside the area with the liquid phase. The tangent vector $\boldsymbol \tau$ points to the right when viewed along the normal direction.
For Marangoni flow to occur, it is necessary to fulfill the condition on the free surface of the droplet
\cite{Barash2009}
$$\omega= \frac{1}{\eta}\frac{d\sigma}{d\tau}+2v_\tau \frac{d\gamma}{d\tau},$$
where $d\sigma/ d\tau= -\sigma'\partial T/ \partial \tau$ is the derivative of the surface tension along the tangent to the surface, $v_\tau$ is the tangential flow velocity, and $\gamma$ is the angle between tangent to the droplet surface and abscissa. We express the tangential velocity as $$v_\tau= u \cos \gamma - v \sin \gamma,$$ where
$$\cos \gamma = \frac{\sgn z}{\sqrt{1+(dh/dr)^2}},$$ $$\sin \gamma = \frac{|dh/dr|}{\sqrt{1+(dh/dr)^2}}.$$
In the paper~\cite{Barash2009}, another useful expression has been derived,
$$\frac{d\gamma}{d\tau}= \left|\frac{d^2 h}{dr^2}\right| \left(1+\left(\frac{dh}{dr}\right)^2\right)^{-3/2}.$$
The tangential  derivative of the temperature is found as
$$\frac{\partial T}{\partial \tau} = \frac{\partial T}{\partial r} \cos \gamma - \frac{\partial T}{\partial z} \sin \gamma.$$

In the case when the surface of the droplet is covered with a layer of smectic, the no-slip condition is fulfilled, since the smectic is assumed to be immobile. This results in $v_\tau = 0$. We discuss the implementation of this condition at one of the drop interfaces in more detail in the next section.

\subsection{Non-dimensional formulation of the problem} Let us introduce the scaled quantities: characteristic length $l_c = H$, characteristic velocity $v_c = \eta/(\rho_l l_c)$, characteristic temperature $T_c = T_\mathrm{up}$, characteristic thermal conductivity $\kappa _c = \kappa_a$ and characteristic time $t_c = l_c/v_c$. Further we use the relations: $r= \tilde r l_c$, $z= \tilde z l_c$, $t= \tilde t t_c$, $u= \tilde u v_c$, $v= \tilde v v_c$, $\omega= \tilde \omega v_c/l_c$, $T= \tilde T T_c$, $\kappa_{l,a}= \tilde \kappa_{l,a} \kappa_c$, $\psi= \tilde \psi v_c l_c^2$. Here the tilde symbol indicates a dimensionless quantity. Now, in all formulas, we use the dimensionless quantities, and omit the tilde symbol for  the brevity sake. The heat transfer equation \eqref{eq:HeatTransfer} in the dimensionless form reads
\begin{multline*}
	\frac{\partial T}{\partial t} + \alpha \left( u \frac{\partial T}{\partial r} + v \frac{\partial T}{\partial z}\right) = \\ \frac{1}{\mathrm{Pe}} \left(\frac{\partial}{\partial z}\left(\kappa \frac{\partial T}{\partial z}\right) + \frac{1}{r}\frac{\partial}{\partial r} \left(\kappa\, r \frac{\partial T}{\partial r} \right)\right),
\end{multline*}
where Peclet number is written as $$\mathrm{Pe}= \begin{cases} l_c^2 c_l \rho_l/(t_c \kappa_c),\, h_{-}< z <h_{+};\\ l_c^2 c_a \rho_a/(t_c \kappa_c),\, |z| > h_{+}. \end{cases} $$
The boundary condition for the vorticity on the free liquid surface is now written as
\begin{equation}\label{eq:MarangoniCondition}
	\omega= 2 v_\tau \frac{d\gamma}{d\tau} - \mathrm{Ma^\prime} \frac{dT}{d\tau},
\end{equation}
where the modified Marangoni number is $\mathrm{Ma^\prime}= \sigma' T_c \rho_l l_c / \eta^2\approx 200$. The use of the upper plate temperature as a characteristic scale is a forced necessity associated with computational difficulties.  For comparison, the Marangoni number from the analytical part of the work is $\mathrm{Ma}=\sigma' h_d^2 (\Delta T/ H) c_l \rho_l/ (\kappa_l \eta)\approx 1$, where $\Delta T= |T_\mathrm{up}-T_\mathrm{dn}|$. For this reason, it is more convenient to make a comparison using the parameter $\Delta T/ H$. In the numerical calculations have been carried out, the parameter is $\Delta T/ H \approx 10^4$ K/m.

On the boundary where the surface of the droplet is covered with a smectic, we use the condition
\begin{equation}\label{eq:NoSlipCondition}
	\frac{\partial \omega}{\partial n}= -p\, \mathrm{rot}_\tau \psi,
\end{equation}
where the penalty parameter $p$~is some relatively large number. The derivative with respect to the normal is recorded on the left. The condition~\eqref{eq:NoSlipCondition} leads to the situation when $v_\tau \approx 0$ with a properly chosen value of the parameter $p$. This allows us to mimic the no-slip condition. The dimensionless form of the equations~\eqref{eq:VorticityTransfer}, \eqref{eq:StreamFunction} and the other conditions (see the section~\ref{sec:InitAndBoundaryCond}) does not differ from their dimensional form.

\subsection{Numerical method} 
\begin{figure}[!htb] \center
	\includegraphics[width=0.5\linewidth]{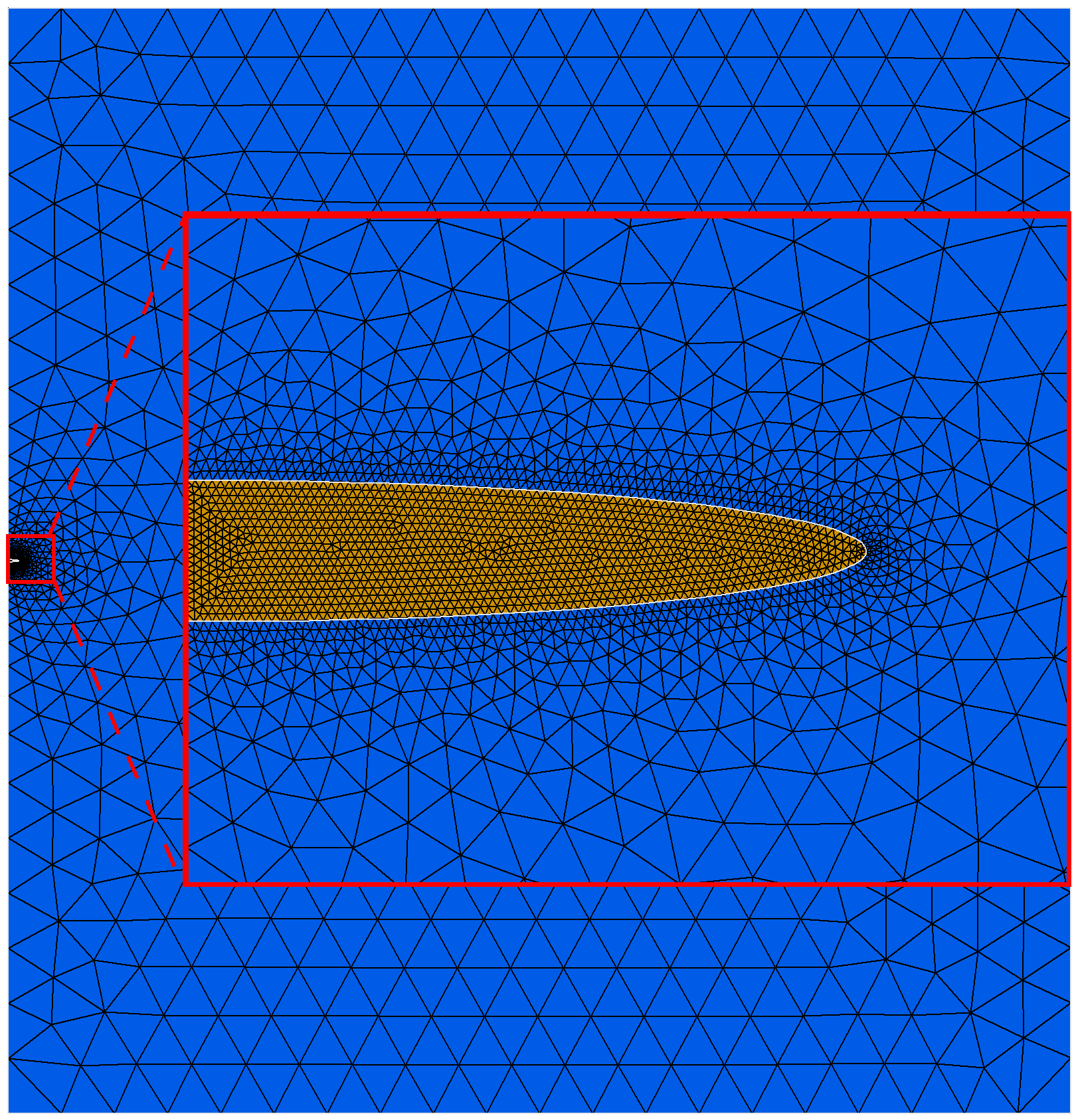} \caption{\label{fig:mesh} Discretization of the spatial domain (the mesh used).}
\end{figure}

The problem (Eqs.~\eqref{eq:VorticityTransfer}--\eqref{eq:HeatTransfer} and conditions from the section~\ref{sec:InitAndBoundaryCond} in dimensionless forms) was solved using the commercial package FlexPDE Professional Version 7.18/W64 3D~\cite{Liu2018}, the algorithm of which was implemented on the basis of the Galerkin finite element method. At each time step the modified iterative Newton-Raphson method is used~\cite{Yamamoto2006}. The size of each time step in the program is determined automatically in order to minimize the calculation error. The mesh convergence of the problem solution is checked. The size of the mesh cells in the area of the droplet periodically decreased until the maximum flow velocity ceased to change. We settled on the mesh that included 2,844 nodes and 5,550 cells (Fig.~\ref{fig:mesh}). For parallel computations, 8 cores were used (our tests showed that this is the optimal number for this problem) on the DEPO Race G790S machine (Intel Xeon CPU E5 2690-v3 2.6 GHz, two CPUs on a dual socket motherboard). The duration of one calculation was approximately 8 minutes. For comparison, the calculation with the use of one core lasted about half an hour. The size of the occupied memory (RAM) was about 120 Mbyte.
